%% file: main.tex
\documentclass[a4paper,11pt]{article}
\pdfoutput=1 

\usepackage{jheppub} 

\usepackage[T1]{fontenc} 

\title{\boldmath An approach to constraining the Higgs width at the LHC and HL-LHC}

\author[a]{Philip Coleman Harris}
\author[a]{Dylan Sheldon Rankin}
\author[b]{Cristina Mantilla Suarez}

\affiliation[a]{Massachusetts Institute of Technology, Cambridge, MA 02139, USA}
\affiliation[b]{Johns Hopkins University, Charles Street, Baltimore, MD 21218, USA}

\emailAdd{pharris@mit.edu,drankin@mit.edu,cmantil1@jhu.edu}

\abstract{Despite the discovery of the Higgs boson decay in five
  separate channels many parameters of the Higgs boson remain largely
  unconstrained. In this paper, we present a new approach to
  constraining the Higgs total width by requiring the Higgs to be
  resolved as a single high p$_T$ jet and measuring the inclusive
  Higgs boson cross section. To measure the inclusive Higgs boson
  cross section, we rely on new approaches from machine learning and a modified jet reconstruction. 
  This approach is found to be complementary to the existing off-shell width measurement and, with the full HL-LHC luminosity, is capable of yielding similar sensitivity to the off-shell projections. 
  We outline the theoretical and experimental limitations and present a path towards making this approach a truly model-independent measurement of the Higgs boson total width.
}

\begin{document} 
\maketitle
\flushbottom

\section{Introduction}
\label{sec:intro}
\input{intro.tex}

\section{Constraining the Higgs Width at the LHC}
\label{sec:measurement}
\input{measurement.tex}

\section{Simulation Setup}                                                                                                                                                      
\label{sec:simulation}                                                                                                                                                            
\input{simulation.tex} 

\section{Measurement Strategy} 
\label{sec:strategy}
\input{strategy.tex}

\section{Achievable accuracies at HL-LHC}
\label{sec:results}
\input{results.tex}

\section{Width calculation and Model Bias}
\label{sec:bias}
\input{bias.tex}

\section{Summary and Conclusions}
\label{sec:conclusion}
\input{conclusion.tex}

\acknowledgments

We thank Petar Maksimovic for his continual support. We also thank
Nhan Tran, Matt Schwartz, and Ian Moult for helpful comments. 
Additionally, we thank members of the
DASZLE collaboration. Part of this work was done on the google
cloud using resources from the MIT Quest for Intelligence. P. Harris
and D. Rankin thank MIT Laboratory for Nuclear Science and the MIT
Physics department. C. Mantilla thanks the Fermilab LHC Physics Center 
Graduate Scholars program. Lastly, we thank members of the 2019 Boost
conference. 

\bibliographystyle{JHEP}
\bibliography{main}

\appendix
\input{appendix.tex}
\end{document}

%% file: intro.tex
Over the past several years, the research output from the LHC has been both fruitful and broad. 
Despite no discovery of beyond the standard model physics, a wealth of measurements have significantly improved our understanding of physics at the TeV energy scale. 
The large collision intensity and the vast amount of collected data have allowed for unprecedented precision in measurements of the standard model, such as b-physics properties and W boson helicity, among others. 
With twenty times more additional data expected in the coming years, the LHC will continue to break records. 
Most recently, the sensitivity of all hadronic decays of vector bosons in the mass range of 10--80~GeV has surpassed constraints from LEP and other colliders~\cite{Sirunyan:2017dnz,Sirunyan:2017nvi,Sirunyan:2019sgo}.
This success comes even though the mass of these bosons are in the precision measurement regime targeted specifically by LEP.
In this paper, we propose a measurement that has not been previously considered at the LHC:
the direct measurement of the SM Higgs boson total width through boosted Higgs boson decays.

The total width of the SM Higgs boson for a mass of 125.1~GeV is predicted to be $\Gamma_{SM}=4.2$~MeV. 
Following the discovery of the Higgs boson~\cite{Chatrchyan:2012xdj,Aad:2012tfa}, the Higgs boson width has been measured using two different approaches.
The first is through a direct measurement of the Higgs mass line-shape using the resonant Higgs decays to diphoton~\cite{PhysRevD.90.052004} and four lepton final states~\cite{Sirunyan:2017exp}.
While this approach is directly sensitive to the Higgs boson width, it is heavily limited by systematic uncertainties from lepton and photon detector resolution. 
The current precision on $\Gamma_{h}$ using this approach is $1.1$~GeV, equivalent to $270\times\Gamma_{SM}$~\cite{Sirunyan:2017exp}.
The second approach involves the use of interference of gluon fusion production of the Higgs boson with gluon fusion production of diboson production, measured using several diboson final states; most recently, the four lepton final state~\cite{Sirunyan:2019twz}.
This interference results in a modification in the high mass distribution of the diboson mass spectrum, yielding a constraint on the width. 
The extraction of the Higgs width in this manner requires a knowledge of the interference pattern, which intrinsically implies a standard model-like behavior of the product of the couplings $g_{ggh}$ and $g_{VVh}$ across a large mass range. 
Further details of this approach and its model-dependent limitations are discussed extensively in the literature~\cite{Kauer:2012hd,Caola:2013yja,Campbell:2013una,Campbell:2013wga,Englert:2014ffa,Englert:2015bwa,Campbell:2014gua,PhysRevD.89.053011,PhysRevD.90.053003,Martin:2012xc,Martin:2013ula,deFlorian:2013psa,Dixon:2013haa}. 
The current best measurement for the total width using this approach is $\Gamma_{h} = 3.2^{+2.8}_{-2.2}$~MeV, while the expected constraint based on simulation is $\Gamma_{h} = 4.1^{+5.0}_{-4.0}$~MeV~\cite{Sirunyan:2019twz}.
Projections for measuring the Higgs width in the four lepton channel alone at the HL-LHC, with assumptions similar to the ones mentioned above, suggest that $\Gamma_{h}$ can be constrained with a precision of $\Gamma_{h} = 4.2^{+1.5}_{-2.1}$~MeV (ATLAS) and $\Gamma_{h} = 4.1^{+1.0}_{-1.1}$~MeV (CMS)~\cite{Cepeda:2019klc}.

\begin{figure}[tbp]
\centering
\includegraphics[width=.45\textwidth]{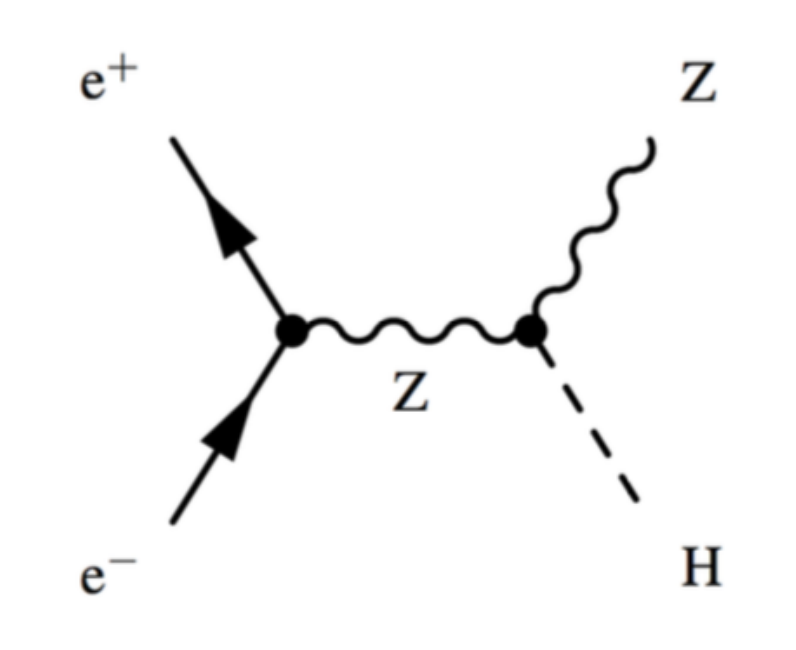}
\caption{Feynman diagram for the Higgsstrahlung process given by $e^{+}e^{-}\rightarrow Zh$ that dominates the Higgs boson production at $\sqrt{s}$=$250$~GeV in an electron linear collider.}
\label{fig:eeZH}
\end{figure}

Model-independent measurements of the Higgs boson total width are possible through the use of lepton-colliders. 
With a muon collider, the width can be probed through the direct production of $\mu^{+}\mu^{-}\rightarrow h$ by a precise scan of the center of mass energy about the Higgs boson total mass~\cite{Conway:2013lca}.
With an electron--positron collider, the Higgs boson width can be measured through the Higgs boson recoil approach whereby one measures the inclusive Higgs boson cross section~\cite{Han:2013kya,Thomson:2015jda}.
In this method, a Higgs boson is produced through the ZH production mode (Fig.~\ref{fig:eeZH}). 
The recoiling Z boson is identified and through conservation of energy of the collision a missing mass can be computed.
The inclusive $Z$+Higgs boson cross section, $\sigma(e^{+}e^{-} \rightarrow Zh)$, can then be deduced from the missing mass distribution. 

From the inclusive $\sigma(e^{+}e^{-} \rightarrow Zh)$ cross section, the Higgs boson width can be determined as follows. 
Noting that the total $Zh$ cross section is proportional to the square of the coupling between the Higgs and Z bosons, $g^2_{hZZ}$:
\begin{eqnarray}
\sigma(e^{+}e^{-} \rightarrow Zh) & \propto  &  g^2_{hZZ},
\end{eqnarray}
and that the cross sections for the exclusive final-state decays $h \rightarrow XX$ can be expressed as:
\begin{eqnarray}
\sigma(Zh (\rightarrow XX)) = \sigma(e^{+}e^{-} \rightarrow Zh) \times \rm{BR} (h \rightarrow XX) & \propto  &  g^2_{hZZ} \frac{g^2_{hXX}}{\Gamma_{h}},
\end{eqnarray}

we can obtain the Higgs boson total width, by directly measuring the total cross section of $\sigma(e^{+}e^{-} \rightarrow Zh)$ and correcting it by the branching ratio of $h \rightarrow ZZ$:
\begin{eqnarray}
\Gamma_{h}  & \propto  &  g^2_{hZZ} \frac{g^2_{hZZ}}{\sigma(e^{+}e^{-} \rightarrow Zh (\rightarrow ZZ))}\\
                     & \propto  &  \frac{\left(\sigma(e^{+}e^{-} \rightarrow Zh)\right)^{2}}{\sigma(e^{+}e^{-} \rightarrow Zh (\rightarrow ZZ))} \\
                     & \propto  &  \frac{\sigma(e^{+}e^{-} \rightarrow Zh)}{\rm{BR} (h \rightarrow ZZ)} \\
                     & \propto  &  \frac{g^2_{hZZ}}{\rm{BR} (h \rightarrow ZZ)}
\end{eqnarray}

Note that we could have chosen another final state, e.g. $h \rightarrow XX$. In that case we would have:
\begin{eqnarray} 
\Gamma_{h} & \propto  &  \frac{g^2_{hXX}}{\rm{BR} (h \rightarrow XX)}    
\end{eqnarray} 

In other words, the inverse of the branching ratio would provide a direct measurement of the Higgs total width, provided the coupling $g_{hXX}$, or $g_{XX}$, agrees with the standard model.
The key to this measurement is the computation of the inclusive Higgs cross section from the recoil mass distribution, which allows us to remove the degeneracy of specific measurements in a final state.
When a specific final state is measured, we are only sensitive to $g^2_{XX} g^2_{ZZ}/\Gamma_{h}$ and so both $g_{XX}$ and $g_{ZZ}$ can scale in such a way as to hide the width measurement.
A measurement of the inclusive cross section for a specific decay yields the partial width $\Gamma_{X}$ of that specific process and, as a consequence, breaks the degeneracy present in a measurement of a specific final state. In summary, to measure the width one would need e.g. the following ratio:
\begin{eqnarray}
\Gamma_{h} & \propto  &  \frac{[\sigma(e^{+}e^{-} \rightarrow Zh)]^2}{\sigma(e^{+}e^{-} \rightarrow Zh (\rightarrow ZZ))}
\end{eqnarray}

\begin{figure}[tbp]
\centering
\includegraphics[width=.45\textwidth]{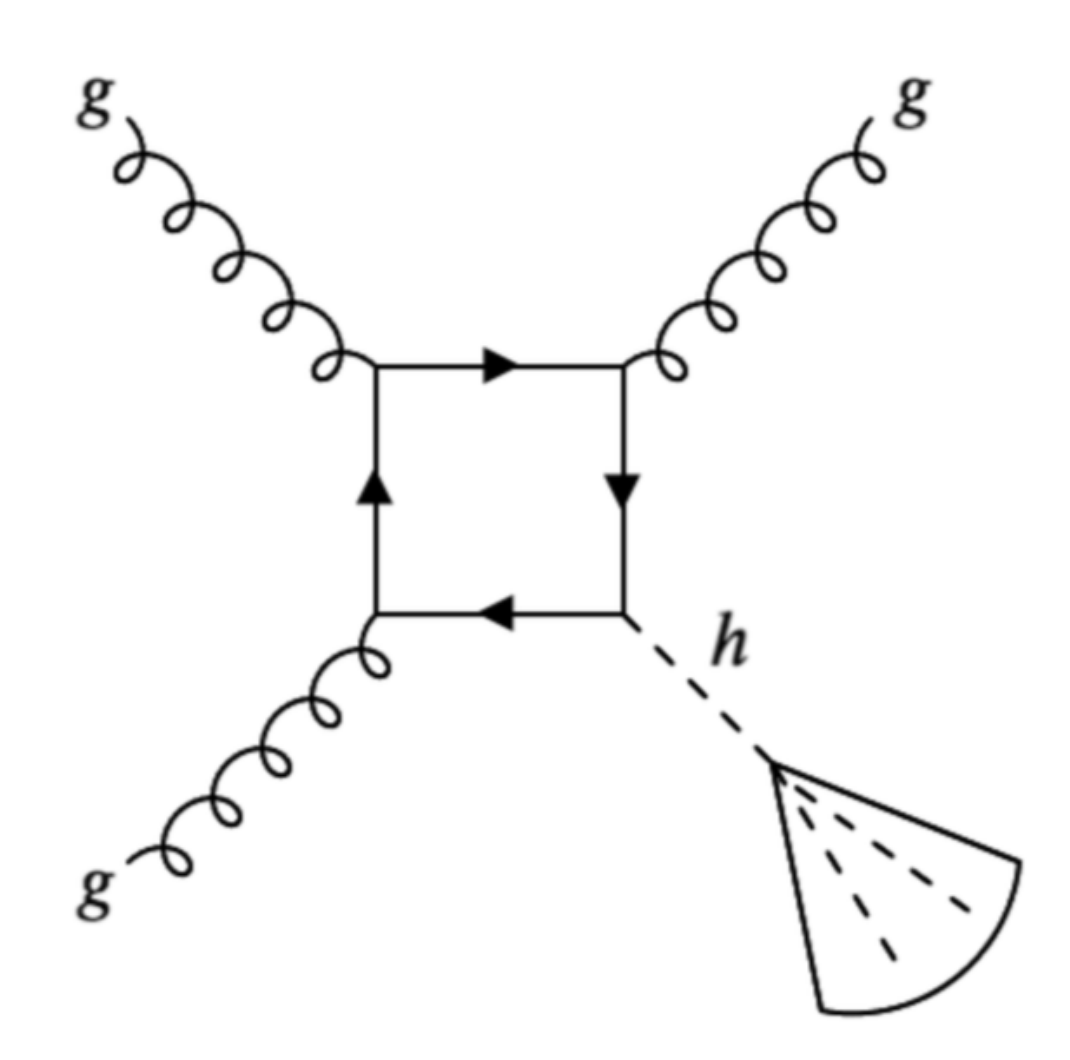}
\caption{\label{fig:ggH} Sample Feynman diagram contributing to the gluon fusion Higgs production process at the LHC where the Higgs is recoiling against a jet.}
\end{figure}

In this paper, we present a new approach to measure the Higgs boson total width at the LHC and HL-LHC by measuring the inclusive Higgs cross section under a set of assumptions. 
To do this, we start from an analogy of the recoil measurement as used at a lepton collider and we make two changes. 
First, in place of a recoiling Z boson, we study a Higgs + jet(s) topology, as shown in Fig.~\ref{fig:ggH}. 
Second, we assume that the recoiling jet(s) give sufficiently high energy to the Higgs boson such that its decay products all fall into a single cone. 
We then reconstruct the decay products as a single jet and extract the Higgs boson signal from this jet by cutting as minimally as possible on the decay components. 

\begin{figure}[tbp]
\centering
\includegraphics[width=.75\textwidth]{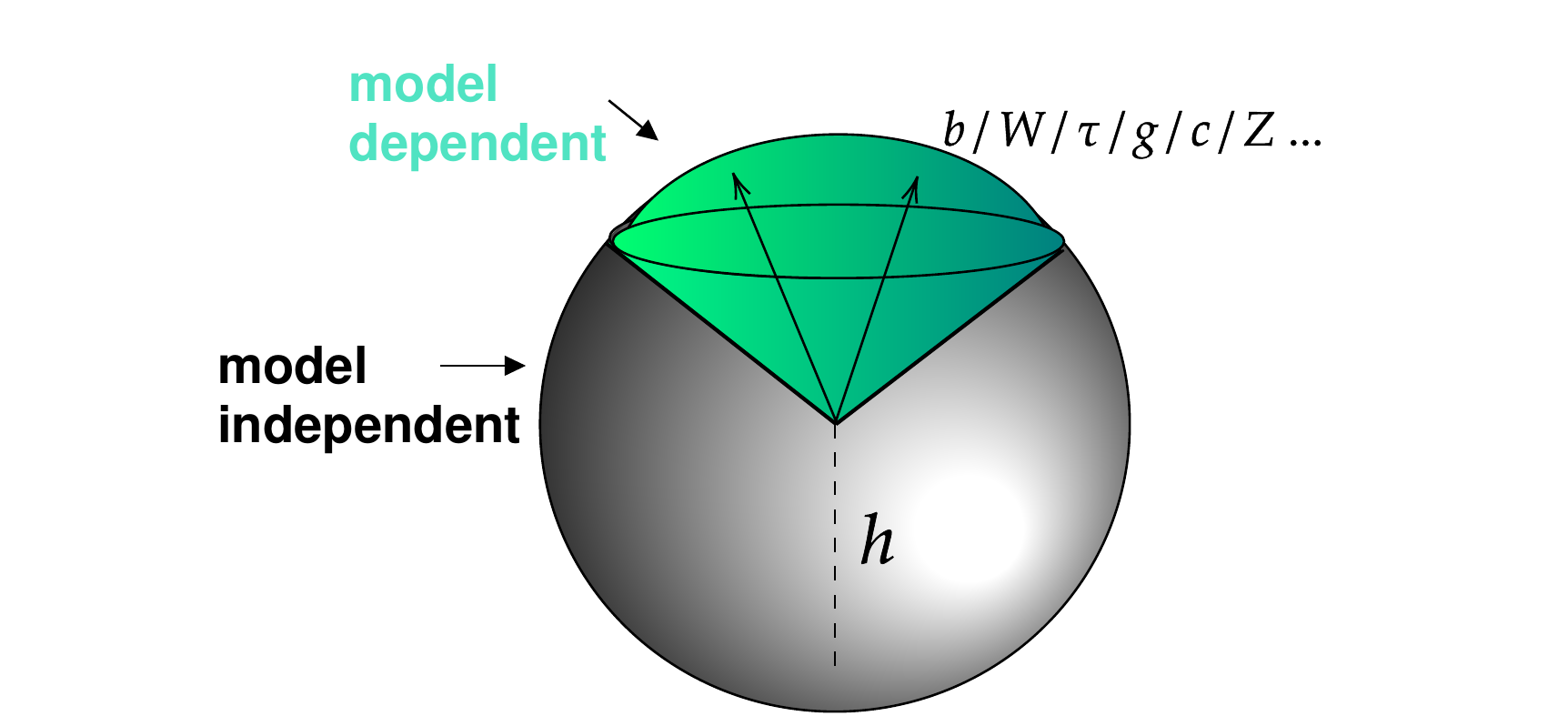}
\caption{\label{fig:cone} Diagram depicting the model dependence behind the inclusive Higgs cross section measurement. The measurement assumes that the Higgs boson transverse momentum is sufficiently high such that all its decay products fall within a jet cone. If we only attempt to isolate the Higgs jet cone without any knowledge of its decay and effectively dispose of the gray area, then the result is model-independent. If, however, we attempt to extract the Higgs boson signal with a biased selection on its decay products that would partially discard the green jet cone area, we introduce a model dependence.
}
\end{figure}

Model dependence enters this measurement when we attempt to extract the Higgs boson from the reconstructed jet cone, as illustrated in Fig.~\ref{fig:cone}. 
The extent of this dependence varies by the level of purity, selection, and sensitivity which is desired. 
Additionally, the model dependence is further complicated by the capture of invisible and partially visible decays, which we will elaborate on further in the text. 

This paper is structured as follows. 
In Sec.~\ref{sec:measurement}, we discuss the formalism to constrain the Higgs width by measuring the inclusive Higgs cross section. 
We describe the simulation setup used for this study in Sec.~\ref{sec:simulation}. 
In Sec.~\ref{sec:strategy}, we discuss the measurement strategy. 
We first quantify the potential of a measurement of the inclusive cross section with a baseline selection and translate this to an upper bound on the total Higgs width at the LHC and HL-LHC. 
We present the expected uncertainties with proton-proton collision pseudo-data corresponding to integrated luminosities of 36~fb$^{-1}$ and 3000~fb$^{-1}$.
We further explore identification and reconstruction techniques to enhance the boosted Higgs signal over the dominant multijet background and discuss their impact on the expected HL-LHC cross section bounds.
We then present the results and limitations of this measurement in Sec.~\ref{sec:results}.
The translation of the inclusive Higgs cross section results to width constraints and its model assumptions are presented in Sec.~\ref{sec:bias}.
Finally, we gather our conclusions and discussion of future work in Sec.~\ref{sec:conclusion}. 

%% file: measurement.tex
To understand how the width measurement is performed, we start with the statement that we can \emph{measure the inclusive cross section of boosted Higgs production}. 
Taking this measurement as a measurement of the gluon fusion production (we will remove this assumption later), then the total cross section can be written as proportional to the gluon coupling,
\begin{eqnarray}
\sigma(gg \rightarrow h) & \propto & \tilde{g}^2_{gg}(p_T)
\end{eqnarray}
where we have written the gluon coupling as $\tilde{g}_{gg}(p_T)$ to make it clear that this is really an effective coupling which is dependent on the $p_{T}$.

Once we have measured $\sigma(gg \rightarrow h)$, which is the heart of this paper, we need to consider three other measurements:
\begin{itemize}
\item{the gluon fusion measurement of the boosted Higgs boson to b-quarks: $\sigma(gg \rightarrow h \rightarrow b\bar{b})$,}
\item{the W boson associated production cross section of the Higgs boson to b-quarks: $\sigma(W+h (\rightarrow b\bar{b}))$, and }
\item{the production cross section of the Higgs boson to W bosons through associated production: $\sigma(W+h (\rightarrow WW))$}
\end{itemize}

We choose these exclusive cross section measurements since the LHC experiments have already explored these final states and have put initial bounds on their production. 
In the following, we will demonstrate how these measurements help to constrain $\Gamma_h$.

The Higgs to b-quark coupling can be written for two production modes: gluon fusion ($gg \rightarrow h$) and W boson associated production ($W+h$), though for the latter we could also write this as weak boson fusion production ($WBF+h$). Following the narrow width assumption, the cross sections can be expressed as:
\begin{eqnarray}
\label{eq:hbb}
\sigma(gg \rightarrow h \rightarrow b\bar{b}) & \propto & \frac{\tilde{g}^2_{gg}(p_T) g^2_{b\bar{b}}}{\Gamma_{h}}, {\rm and } \\
\sigma(W+h \rightarrow b\bar{b}) & \propto & \frac{g^2_{WW} g^2_{b\bar{b}}}{\Gamma_{h}}, 
\end{eqnarray}

and their ratio yields:
\begin{eqnarray}
\frac{\sigma(W+h \rightarrow b\bar{b})}{\sigma(gg \rightarrow h \rightarrow b\bar{b})} & \propto & \frac{g^2_{WW}}{\tilde{g}^2_{gg}(p_T)}
\end{eqnarray}

Multiplying this by the inclusive cross section, we get:
\begin{eqnarray}
\label{eq:ratio}
\sigma(gg \rightarrow h) \times \frac{\sigma(W+h \rightarrow b\bar{b})}{\sigma(gg \rightarrow h \rightarrow b\bar{b})} & \propto & g^2_{WW}
\end{eqnarray}

Moreover, for the W boson decay by W associated production, $W+h \rightarrow  WW$, the cross section is proportional to the W boson coupling over the width:
\begin{eqnarray}
\sigma(W+h \rightarrow WW) & \propto & \frac{g^4_{WW} }{\Gamma_{h}}
\end{eqnarray}

Thus, we can take $\Gamma_{h}$ and square the ratio from \ref{eq:ratio} to write the total Higgs boson width as:
\begin{eqnarray}
\Gamma_{h} & \propto & \frac{1}{\sigma(W+h \rightarrow WW)} \times \left(\sigma(gg \rightarrow h) \times \frac{\sigma(W+h \rightarrow b\bar{b})}{\sigma(gg \rightarrow h \rightarrow b\bar{b})}\right)^2
\end{eqnarray} 

Given this proportionality, we can perform the measurement of the Higgs boson width by computing a scale factor $\mu_{\Gamma} = \mu_{h} / \mu_{SM}$ defined as:
\begin{eqnarray}
\label{eq:unc}
\mu_{\Gamma} & = & \mu_{gg \rightarrow h}^2 \frac{\mu_{Wh \rightarrow b\bar{b}}^2}{\mu_{ggh \rightarrow b\bar{b}}^2~\mu_{W+h \rightarrow WW}} {\rm, with~uncertainty} \\
\delta\mu_{\Gamma}^2 & = & 4\delta\mu_{gg \rightarrow h}^2+\delta\mu_{W+h \rightarrow WW}^2+4\delta\mu_{W+h \rightarrow b\bar{b}}^2+4\delta\mu_{ggh \rightarrow b\bar{b}}^2 ,
\end{eqnarray}
Where $\delta\mu$ signifies the uncertainty on the respective scale factor. 
As a consequence, to get the width we need to measure gluon fusion Higgs to b-quark production, either $W+h$ or $WBF+h$ decaying to b-quarks, either $W+h$ or $WBF+h$ decaying to W bosons, and the inclusive Higgs boson cross section. 
Ideally, all measurements should be done in the same phase space so that scale dependence of the coupling factorizes; this is never fully possible.
This factorization has the additional benefit that it does not rely on the precise form of the effective coupling and thus is unaffected by the presence of new physics contributions to the loop.

If we consider the ultimate precision of the LHC data for $W+h \rightarrow \bar{b}b$~\cite{Sirunyan:2018kst,Aaboud:2018zhk},~$W+h \rightarrow WW$~\cite{Sirunyan:2018egh,Aaboud:2019nan}, and $gg \rightarrow h \rightarrow \bar{b}b$~\cite{Sirunyan:2017dgc,ATLAS-CONF-2018-052}, we find that with 36~fb$^{-1}$ 1$\sigma$ uncertainty of 20\%, 18\% and 80\%, respectively. 

To estimate bounds at HL-LHC we utilize projections on the uncertainties to a full $3~\textrm{ab}^{-1}$ dataset~\cite{CMS:2018qgz}. 
For the $W+h \rightarrow \bar{b}b$ and $W+h \rightarrow WW$ measurements we assume 1$\sigma$ uncertainties of~$9\%$ and $5\%$, respectively.
For the gluon fusion Higgs to b-quark production~($\mu_{ggh \rightarrow \bar{b}b}$), we assume that this uncertainty scales directly with the $\sigma(gg \rightarrow h)$ result and assume that the uncertainty is 25\% the current inclusive Higgs boson uncertainty.
The combined uncertainty on all terms, excluding $\delta\mu_{ggh}$, would give an uncertainty of $\delta\mu_{\Gamma}= 25\%$ on the width at $3~\textrm{ab}^{-1}$. 
Thus, for an inclusive measurement with uncertainties above 12\%$\times\sigma_{SM}$ the uncertainty on the inclusive cross section becomes the largest source of error. Finally, we believe the $\delta\mu_{\Gamma}= 25\%$ limit originating from all measurements excluding the exclusive Higgs boson cross-section will likely improve through the use of more advanced measurements and a simultaneous fit on all Higgs boson couplings. 

Another more direct measurement of the width can be performed provided the high energy $h \rightarrow WW$ channel is measured. 
In this instance, we can write the width and uncertainty given by:
\begin{eqnarray}
\mu_{\Gamma} & = & \mu_{ggh}^2 \frac{\mu_{W+h \rightarrow WW}}{\mu^2_{ggh \rightarrow WW}} {,\rm~with~uncertainty} \\
\delta\mu_{\Gamma}^2 & = & 4\delta\mu_{ggh}^2+\delta\mu_{W+h \rightarrow WW}^2+4\delta\mu_{ggh \rightarrow WW}^2
\end{eqnarray}

With a measurement of $h \rightarrow WW$ boson production in the electron+muon final state for a Higgs boson $p_{T} > 350$~GeV with the full LHC luminosity, we find a projected uncertainty would be comparable to that of the combined $h\rightarrow\gamma\gamma$ and $h\rightarrow ZZ$ results in the same $p_{T}$ region yielding an uncertainty of roughly 10\%~\cite{Cepeda:2019klc}. 
We believe that this uncertainty can be further reduced through the addition of the hadronic channels. 
As a consequence, the dominant uncertainty is again the measurement of the inclusive Higgs boson cross-section at high $p_{T}$, which will be the focus of the rest of this paper. 

Lastly, we mention that the use of gluon-fusion production in the above result, $\mu_{ggh}$, was a choice. All production modes are possible, provided they are later factorized in the measurement through the direct measurement of either $h \rightarrow b\bar{b}$ or $h \rightarrow WW$ final states in a single jet, using the same phase space as the inclusive measurement. 

In summary, we aim to measure the inclusive Higgs cross-section in the boosted regime. 
This ensures that the Higgs boson has sufficiently high energy such that all its decay products fall into a single large-radius jet.
Through the use of an explicit measurement of the Higgs boson to either b-quarks or W-bosons in, preferably, the same kinematic regime we can then translate this measurement to a direct measurement of the Higgs boson total width. 
In this paper, we will focus on the jet~$+$~ISR final state, since this gives us the largest amount of high energy jets. 
However, the production mode of the Higgs is not a critical element of the inclusive cross-section measurement and can be factored out of the width translation. 
For future studies, we would like to explicitly investigate boosted Higgs boson production in the Z, W, and WBF final states, where the Higgs boson purity is enhanced.

%% file: simulation.tex
Samples of simulated background events are generated using Monte Carlo generators and processed through a simplified detector simulation representative of current and future detector concepts. Background events from W+jets and Z+jets processes, where the W and Z decay into quark anti-quark pairs, as well as multijet events from the dominant QCD background, are simulated with the leading-order (LO) mode of MadGraph5 aMC@NLO v5.2.2.2~\cite{Alwall:2007fs,Alwall:2014hca}. We use the POWHEG 2.0~\cite{Frixione:2007vw,Alioli:2010xd} generator at next-to-leading order (NLO) precision to model the $t\bar{t}$ process. 

The Higgs boson signal samples are produced assuming m$_h=125$~GeV. For the gluon fusion production mode, events are generated at one-loop order ({\fontfamily{qcr}\selectfont g g $>$ h j [QCD]} and {\fontfamily{qcr}\selectfont g g $>$ h j j [QCD]}), which corresponds to the leading contribution with the finite top mass included. We account for the overlap between the real emission from the matrix element and the parton shower with the MLM algorithm~\cite{Alwall:2007fs}. After MLM merging is applied, the overall normalization of the Higgs sample is found to differ considerably with existing benchmarks. We re-weight the Higgs yield by a factor of 5 to match the N$^{3}$LO normalized inclusive Powheg Higgs distribution at 400~GeV~\cite{Sirunyan:2017dgc,Bagnaschi:2011tu}; this distribution is the standard distribution used as the benchmark in nearly all CMS Higgs boson analyses from 2012-2017. Despite agreement at 400~GeV with the inclusive Powheg Higgs boson distribution, our $p_T$ spectrum is found to be considerably softer at higher values of transverse momentum. Smaller contributions to the Higgs boson signal from the V$+h$ associated production, the vector fusion production mode (VBF) and the $t\bar{t}h$ process are generated with POWHEG 2.0.

We interface these generators with Pythia 8.212~\cite{Sjostrand:2014zea} for parton showering with the Monash 2013 tune~\cite{Skands:2014pea} 
and the parton distribution function (PDF) set NNPDF3.0~\cite{Ball:2014uwa}. 
Cross sections are computed for a center of mass energy of 13 TeV, but conclusions drawn from this study will hold, if not  improve, at 14 TeV 
since the cross section will change at the 10\% level for signals and backgrounds.

We employ a custom detector simulation that employs particle-flow-based reconstruction, similar to that used for the CMS~\cite{Sirunyan:2017ulk} and ATLAS~\cite{Aaboud:2017aca} detectors at the LHC, that reproduces the main resolution effects relevant for jet reconstruction. We first categorize the generated particles into charged particles, photons (including $\pi^0 \rightarrow \gamma\gamma$), and neutral hadrons. We simulate tracking inefficiencies to reconstruct highly-collimated particles, due to the limited granularity of tracker detectors, by treating charged particles with momenta above a threshold, $p_{T,track}^{\rm max}=220$~GeV, as neutral hadrons. The threshold is chosen such that it matches the jet mass resolution of the current CMS detector at high momenta~\cite{CMS-PAS-JME-14-002} and increased by a factor of 2 as suggested by studies for the HL-LHC Phase-II upgrade of the CMS tracker with higher pixel granularity~\cite{Collaboration:2272264}. The generated neutral hadrons are then discretized to simulate the spatial resolution of the electromagnetic ($\sigma^{\eta,\phi} = 0.017$) and hadronic ($\sigma^{\eta,\phi} = 0.02$) calorimeters. 

The missing transverse energy (MET), is an experimental proxy for the transverse momentum carried by undetected particles and thus a signature of neutrino production.
Experimentally, it is defined as the imbalance of momentum of all detected particles in the transverse plane. 
In the simulation, we define it from the smeared components of the genuine MET.
The ``true'' MET is taken from the vector sum of the momenta of all the generated neutrinos.
Its components $p_{x(y)}$ are then smeared by the $\sum E_{T}$ resolution (60\%), where $\sum E_{T}$ is the scalar sum of the transverse momentum of all the visible final state particles.
The $\sum E_{T}$ resolution matches the resolutions measured experimentally from the hadronic recoil in Z boson events~\cite{CMS-PAS-JME-18-001,Aaboud:2018tkc}.

We cluster the detector-simulated particles into large cone jets using the anti-$k_T$ algorithm~\cite{Cacciari:2008gp} with a distance parameter ($R$) of 0.8.
We assume that the jet is reconstructed with a pileup mitigating algorithm such as PUPPI~\cite{Bertolini:2014bba} to ensure there is a minimal impact from pileup on this analysis; we do not actively apply PUPPI since pileup simulation is not added.
Lastly, we scale the top quark yield by a factor of 0.16 consistent with a b-jet veto using a working point of 60\%.
Additional constraints are possible on the top quark background, but these are not considered further in this study.

We apply the ``modified mass drop tagger'' algorithm~\cite{Dasgupta:2013ihk,Larkoski:2014wba}, also known as soft-drop (SD), with angular exponent $\beta = 0$ and soft cutoff threshold $z_{cut}=0.1$.
This grooming procedure removes soft and wide-angle radiation from the jet, and as a consequence, it reduces the mass for quark and gluon jets and improves the signal mass resolution.
To match CMS public results, we further smear the jet mass for signal and single parton jets by $4$~GeV.
The jet mass is hereinafter referred to as the soft-drop mass ($m_{SD}$) or mMDT.

While we present results using simulation similar to the CMS detector configuration, we expect the study to be representative of the current and future performance of both ATLAS and CMS experiments.

%% file: strategy.tex
Since our goal is to measure the inclusive Higgs boson cross section, we ultimately aim to select Higgs bosons using only event and jet properties common for all its decays.
In practice, we first utilize a basic selection that mimics the selection used in the searches for a low mass spin-one $Z^{\prime}$ resonance decaying into quarks performed at the LHC by the CMS and ATLAS experiments~\cite{Sirunyan:2017dnz,Aaboud:2018zba}.
This allows us to validate our method against the published results, and ensure consistency. 
After we have established a baseline result using this selection we then consider modifications designed specifically to improve the measurement of $\Gamma_h$.

\subsection{Baseline selection and extraction method}
\label{sec:baseline}

The main requirement in this analysis is for the Higgs to be produced at high $p_{T}$ such that its decay products are collimated and reconstructed into one single large radius jet.
Beyond this selection, we have three main handles to enhance the Higgs boson signal and isolate it from the dominant multijet background: the jet $p_{T}$, the substructure of the jet, and the jet mass.

Since the Higgs boson production $p_{T}$ spectrum is harder than the background when a fixed mass window is considered, a higher jet $p_{T}$ selection will have a higher level of purity. 
This requirement is independent of the Higgs boson decay and does not add any further model dependence.
Our baseline selection requires a jet with $p_{T}>500$~GeV, consistent with the current trigger thresholds at LHC experiments. 

The second feature that can be used to purify the event selection is the substructure of the jet. 
SM Higgs bosons decay symmetrically into two objects, which can subsequently decay into other objects, as is the case for $h \rightarrow WW^*$ and $h \rightarrow ZZ^*$ decays.
Our baseline selection requires the leading $p_{T}$ jet to be consistent with a two-prong decay. 
In particular, we use the observable $(\tau_{2}/\tau_{1})^{\rm DDT}$~\cite{Thaler:2010tr,Dolen:2016kst}, tuned to have a low background efficiency of 6\% on the QCD multijet background.
Since some decays, for example $h \rightarrow WW^*$ and $h \rightarrow ZZ^*$, do not always result in jets consistent with two prongs, this requirement can lead to a significant background reduction while introducing additional model assumptions.
Ideally, we would require this selection to be independent of the Higgs jet internal decays. 
We will elaborate on improvements to this in Sec.~\ref{subsec:tagging}. 

The third and most powerful parameter that we can use to perform this measurement is the jet mass. 
For visible decays of the Higgs boson, the mass is a very effective way to discriminate the Higgs signal from the background. 
However, it has the limitation that invisibly decaying particles are not part of the mass reconstruction. 
We start by considering the leading $p_{T}$ jet groomed mass, but we will discuss possible solutions to recover the invisible components of the Higgs decays in Sec.~\ref{subsec:mass}.

The uncertainty on the Higgs boson total cross section is extracted from a likelihood fit that treats the Higgs boson as a signal. 
The procedure follows a CMS-like analysis and is described in detail in the Appendix~\ref{app:fit}.
The key challenge of this procedure is the estimation of the main QCD background, which has a nontrivial jet mass shape that is difficult to model parametrically and depends on the jet $p_T$.
In this analysis we consider two simplified approaches: we either take the shape from the simulation and perform a template fit, or we model the shape using a Bernstein polynomial. 
The number of assumed polynomial parameters and its impact on the signal strength are further discussed in the Appendix~\ref{app:extrap_order}.

\subsubsection{Baseline sensitivity for $\Gamma_h$}
\begin{figure}[tbp]
\centering
\includegraphics[width=.7\textwidth]{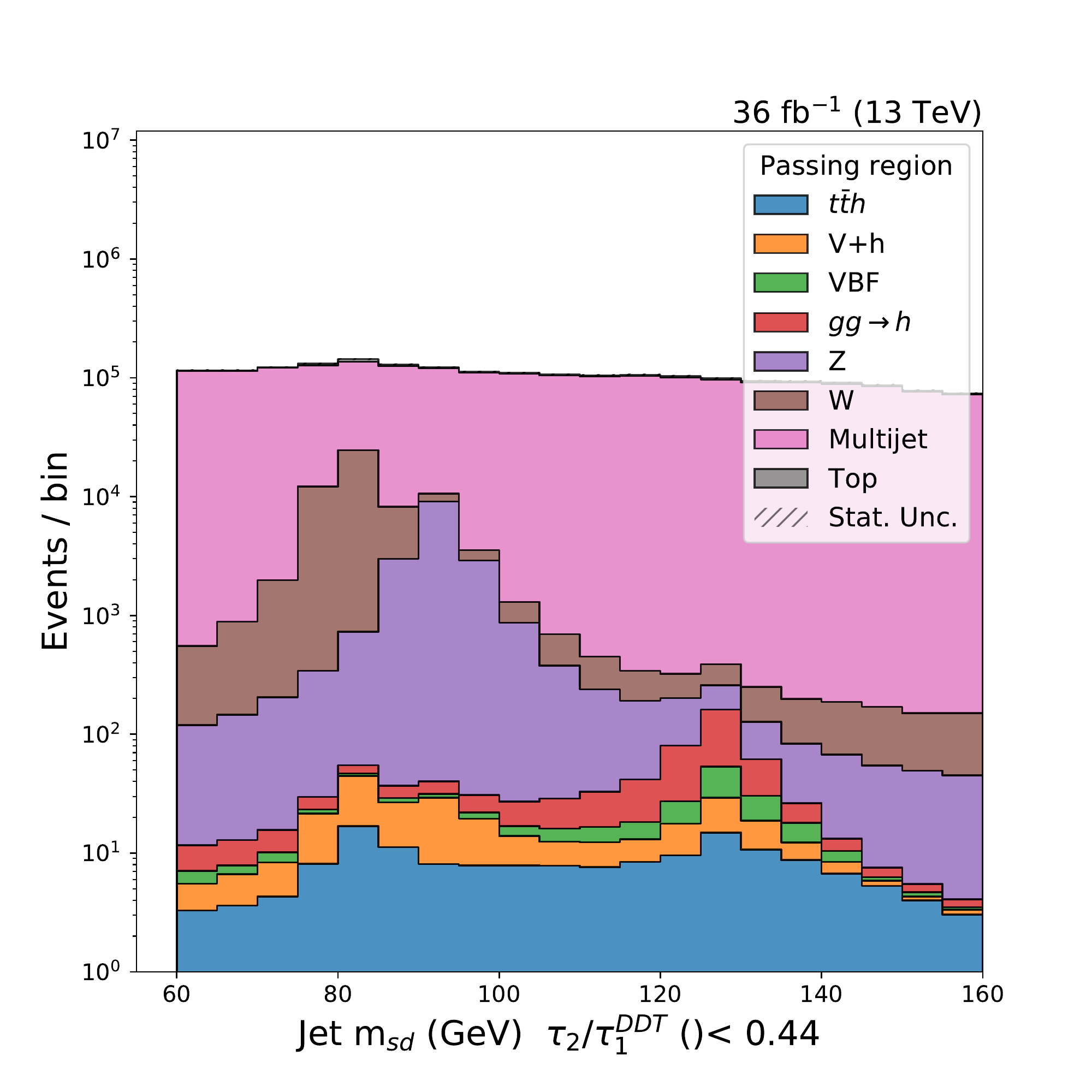}
\caption{Jet mass distribution after a $(\tau_{2}/\tau_{1})^{\rm DDT}$ selection tuned to keep 6\% of the multijet background. 
The main background processes, multijet,~$W$,~$Z$ and $t\bar{t}$, and the signal contributions are shown scaled to 36~fb$^{-1}$.
The distribution is smoothly falling given the decorrelation procedure applied to the $\tau_{2}/\tau_{1}$ observable.
}
\label{fig:t21mass}
\end{figure}

Using the baseline selection given above we can estimate our initial sensitivity with a fit to the mass distribution.
The mass distribution used, after the $(\tau_{2}/\tau_{1})^{\rm DDT}$ selection, is shown in Fig.~\ref{fig:t21mass}.
As a test of the validity of the signal extraction procedure, we first use a $Z^{\prime}$ signal simulation scaled to 36~fb$^{-1}$.
We find a 95\% C.L. upper bound on the $Z^{\prime}$ coupling to quarks of $g_{q}^{\prime} < 0.082$ for a template fit and $g_{q}^{\prime} < 0.100$ for the polynomial fit.
The observed best fit values from the current CMS and ATLAS results at 36~fb$^{-1}$~\cite{Sirunyan:2017nvi,Aaboud:2018zba} give a 95\% C.L. coupling bound of $0.090$, about 25\% better in total cross section than our toy analysis. This improvement is expected given that the actual result corresponds to an analysis where the observable $N_{2}^{\rm DDT}$~\cite{Moult:2016cvt} is used, instead of $(\tau_{2}/\tau_{1})^{\rm DDT}$, and a significantly more sophisticated background method is utilized for the signal extraction.

Next, we translate this to a measurement of the inclusive Higgs cross section using an inclusive Higgs gluon fusion signal and the same $(\tau_{2}/\tau_{1})^{\rm DDT}$ selection.
With a polynomial fit, we find a sensitivity of 1$\sigma$ on $\delta\sigma_{ggh}= 6.6\times \sigma_{SM}$. 
This was further checked by noting that gluon fusion Higgs boson production matches Z$^{\prime}$ production almost identically across p$_{T}$ when $g_q=0.04$.
After including all the other Higgs boson production modes in the signal extraction, the sensitivity of the polynomial fit is improved to $\delta\sigma_{ggh} = 3.6 \times \sigma_{SM}$.
We can translate this bound to one on the Higgs boson total width of $\Gamma_h < 7.8~\times~\Gamma_{SM} = 32$~MeV, using the uncertainty propagation outlined in eq.~\ref{eq:unc}.

The current expected constraint for the Higgs boson total width at 1$\sigma$ from off-shell measurements at the LHC is $5$~MeV from CMS~\cite{Sirunyan:2019twz} using $80~\textrm{fb}^{-1}$ of data and approximately $10$~MeV from ATLAS~\cite{Aaboud:2018puo} using $36~\textrm{fb}^{-1}$ of data.
Thus with the baseline selection, we find that the off-shell measurements are more sensitive.
Including Higgs events with a leading jet of $400 < p_{T} < 500$~GeV allows our estimated  limit to be reduced from $\delta\sigma=3.6\times \sigma_{SM}$ to $\delta\sigma=2.2 \times \sigma_{SM}~(\Gamma_{h} < 19~$~MeV$)$.
This reduction could be obtained by an improved trigger or through the use of scouting as will be discussed further below.

Going from $36~\textrm{fb}^{-1}$ to a full $3~\textrm{ab}^{-1}$ yields a factor of 100 in the amount of data present.
If one were to scale the luminosity, this would increase the statistical precision of the measurement by a factor of 10.
Consequently, a 10~MeV measurement would now be reduced to 1~MeV.
Scaling the samples up to an integrated luminosity of $3~\textrm{ab}^{-1}$ the uncertainty on the total cross section goes to $\delta\sigma_{ggh} = 0.55\times \sigma_{SM}$ using a polynomial fit.
An overall scaling of the result utilizing $\sqrt{\mathcal{L}}$ would give $\delta\sigma=0.24\times \sigma_{SM}$ indicating a reduction of sensitivity resulting from the impact of the systematic uncertainties. 
 
Both ATLAS and CMS have released $3~\textrm{ab}^{-1}$ projections for the off-shell width measurement.
The CMS result finds a final result of $0.26\times\Gamma_{SM}$, whereas the ATLAS result finds a limit of $0.4\times\Gamma_{SM}$. 
This CMS result is roughly $0.24\times\Gamma_{SM}$ when systematic uncertainties are removed from the limit computation, indicating limited room for potential additional improvements.
For future comparison, we translate the off-shell projections to an inclusive cross section measurement uncertainty, using eq.~\ref{eq:unc}. The translated measurements yield an upper bound of $0.09\times \sigma_{SM}$ (CMS) and $0.17\times \sigma_{SM}$ (ATLAS). 

\subsubsection{Per-decay sensitivity}
\label{sec:perdecay}
To understand the relative sensitivity of each decay mode channel, we compute the individual cross section for the dominant SM decay modes and correct it by the SM branching ratio.
This translates the relative uncertainty to bound the inclusive cross section and helps to determine which decay channel can limit the sensitivity of this measurement. 

\begin{figure}[tbp]
\centering
\includegraphics[width=.45\textwidth]{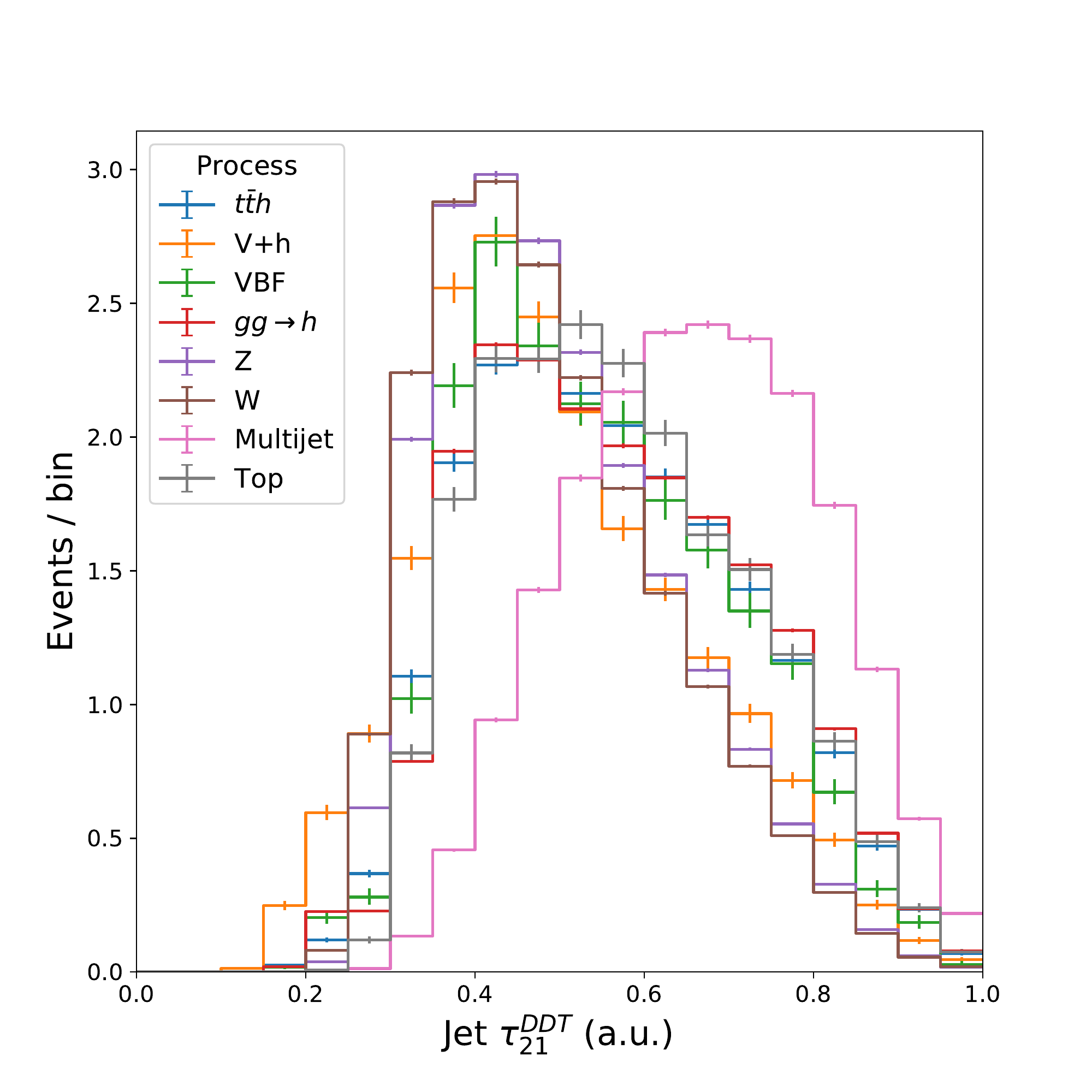} 
\includegraphics[width=.45\textwidth]{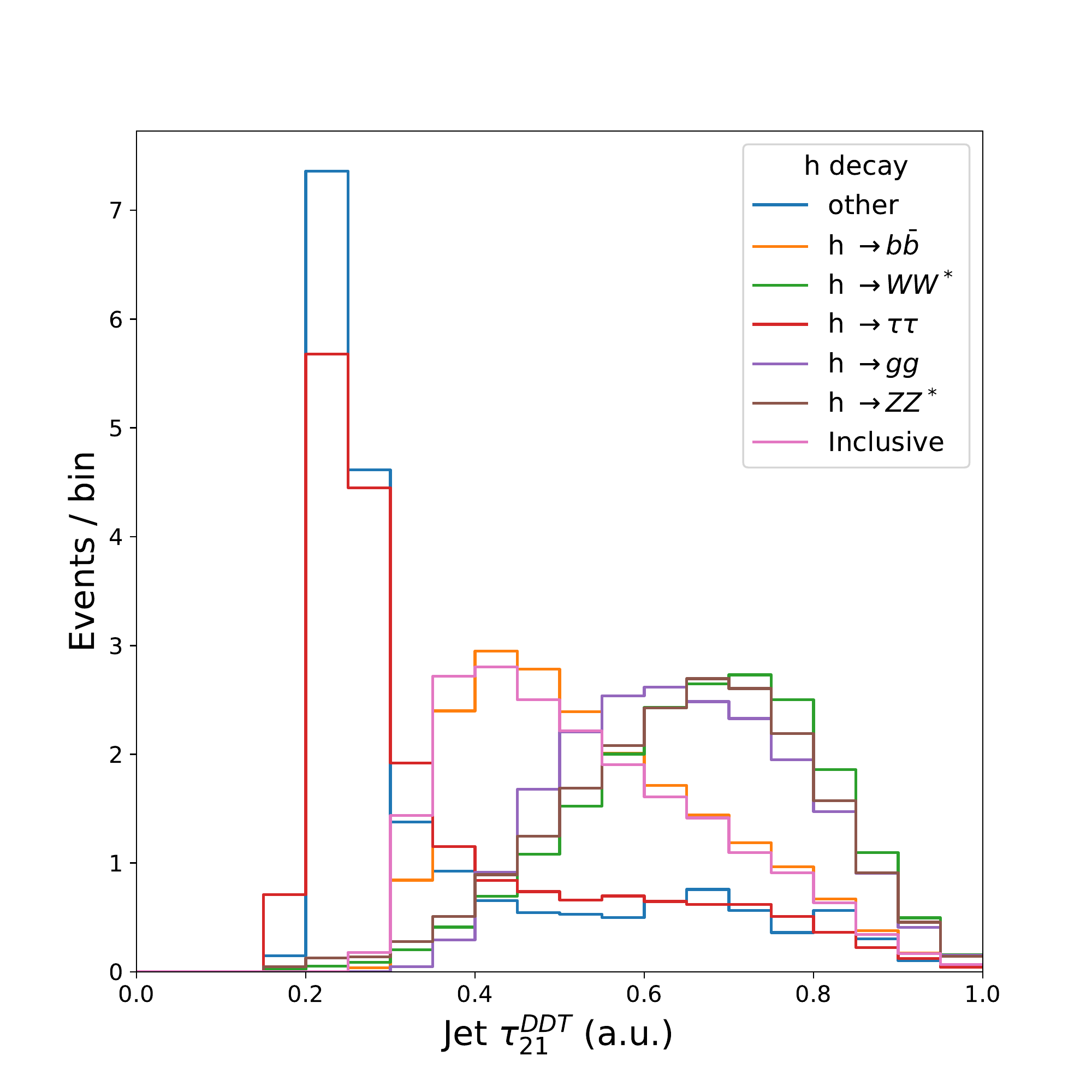}
\caption{\label{fig:t21} $(\tau_{2}/\tau_{1})^{DDT}$ observable shown for the inclusive Higgs signal vs other backgrounds (left) and for different SM decays (right).}
\end{figure}

At $3~\textrm{ab}^{-1}$ we expect a 1$\sigma$ limit of $\sigma_{ggh} < 0.26 \times \sigma_{\rm~SM}$ for a template fit after a $(\tau_{2}/\tau_{1})^{\rm DDT}$ selection.
The $h \rightarrow \bar{b}b$ and $h \rightarrow \bar{c}c$ decay channels have a similar and slightly better sensitivity of $0.16 \times \sigma_{\rm~SM}$ and $0.13 \times \sigma_{\rm~SM}$ given their two-pronged decay.
However, the $h \rightarrow \bar{\tau}\tau$ sensitivity worsens to $0.44 \times \sigma_{\rm~SM}$ in spite of its two-pronged decay.
This is due to the broadening of the jet mass from the neutrinos present in the final state.
We also find that the gluon and EW boson decays of the Higgs have much worse sensitivity given their different radiation patterns and the fact that their $(\tau_{2}/\tau_{1})^{\rm DDT}$ distribution is background-like.
We find a limit of $0.7 \times \sigma_{\rm~SM}$ for $h \rightarrow gg$ decays, where the main signal contribution is from the $W+h$ production mode, discussed later.
For the four or three-pronged decays $h \rightarrow WW^*$ and $h \rightarrow ZZ^*$, we find a limit of $2.4 \times \sigma_{\rm~SM}$ and $1.4 \times \sigma_{\rm~SM}$.
We show the distributions of the jet $\tau_{2}/\tau_{1}^{\rm DDT}$ for different Higgs decay modes and backgrounds in Fig.~\ref{fig:t21}.
The discrimination power against QCD background for different Higgs decays is available in Fig.~\ref{fig:roc_byvar} in the Appendix~\ref{app:performance}.
Potential modifications to the method to improve the performance across all decay modes are discussed in the next sections.

\subsection{Improved selection and reconstruction}
\label{subsec:higgsrec}
To increase the sensitivity to $\sigma_{ggh}$, we have explored various improvements to the Higgs jet reconstruction and the identification of the inclusive Higgs boson signal.
Our simulation framework and signal extraction procedure are the same as those indicated in the previous sections.

\subsubsection{Higgs jet selection}
\label{subsec:ptsel}
One obvious improvement in sensitivity can come from a reduction in the jet $p_{T}$ requirement.
The baseline jet $p_{T}$ threshold is driven by the data rate of the high-level trigger.
Currently, the lowest-threshold unprescaled large radius jet triggers plateau at 100\% efficiency for jets with $p_{T} > 500$~GeV in both the CMS and ATLAS experiments~\cite{atlastrigger,Khachatryan:2016bia,CMS-DP-2018-040}. 
However, novel trigger strategies, such as trigger-level analyses, that record only partial event information, could allow for a threshold as low as $p_{T} > 300-400$~GeV.
Additionally, both ATLAS and CMS are pursuing advanced trigger strategies with higher rates that would allow for low jet threshold events to be either saved offline or scrutinized within the online triggering system~\cite{Aad:1602235,Collaboration:2285584,Collaboration:2283192}.
Thus, in this study, we assume that advances in the capabilities of the trigger system will allow a $p_{T} > 400$~GeV threshold across most of the LHC and HL-LHC running.

\subsubsection{Higgs jet reconstruction}
\label{subsec:mass}
Lorentz invariance implies that the Higgs decay products are all contained within the jet cone, several decays of the SM Higgs boson will lead to final states with neutrinos that escape the detector.
The only method to identify these decays is to utilize the missing transverse energy (MET).
Adding the MET to the jet will recover the lost energy and produce a better estimate of the true Higgs boson properties.
Thus, the first improvement to the jet reconstruction is to require our Higgs jet to be the leading jet in $p_T^{\rm{jet+MET}}$ in the event.
To perform the vector addition of the jet and MET, we assume the missing energy vector is aligned with the jet axis.

A complication to this procedure is that the MET resolution in events with high energy jets is quite poor; Higgs decays without neutrinos in the final state can still produce over 100~GeV of MET.
Since this artificial MET would worsen the mass resolution when it is added to the jet, we first perform a dedicated regression for the true MET and utilize the regressed MET as our default MET calculation.
The regression is designed to remain model independent and to eliminate artificial MET as efficiently as possible.
Additional details on the regression and its performance can be found in Sec.~\ref{app:met_reg} in the Appendix.

Figure~\ref{fig:mass_res_regression} shows a comparison between the mMDT mass for the Higgs jet and the jet mass as computed using the jet - regressed MET combination.

\begin{figure}[tbp]
\centering
\includegraphics[width=.45\textwidth]{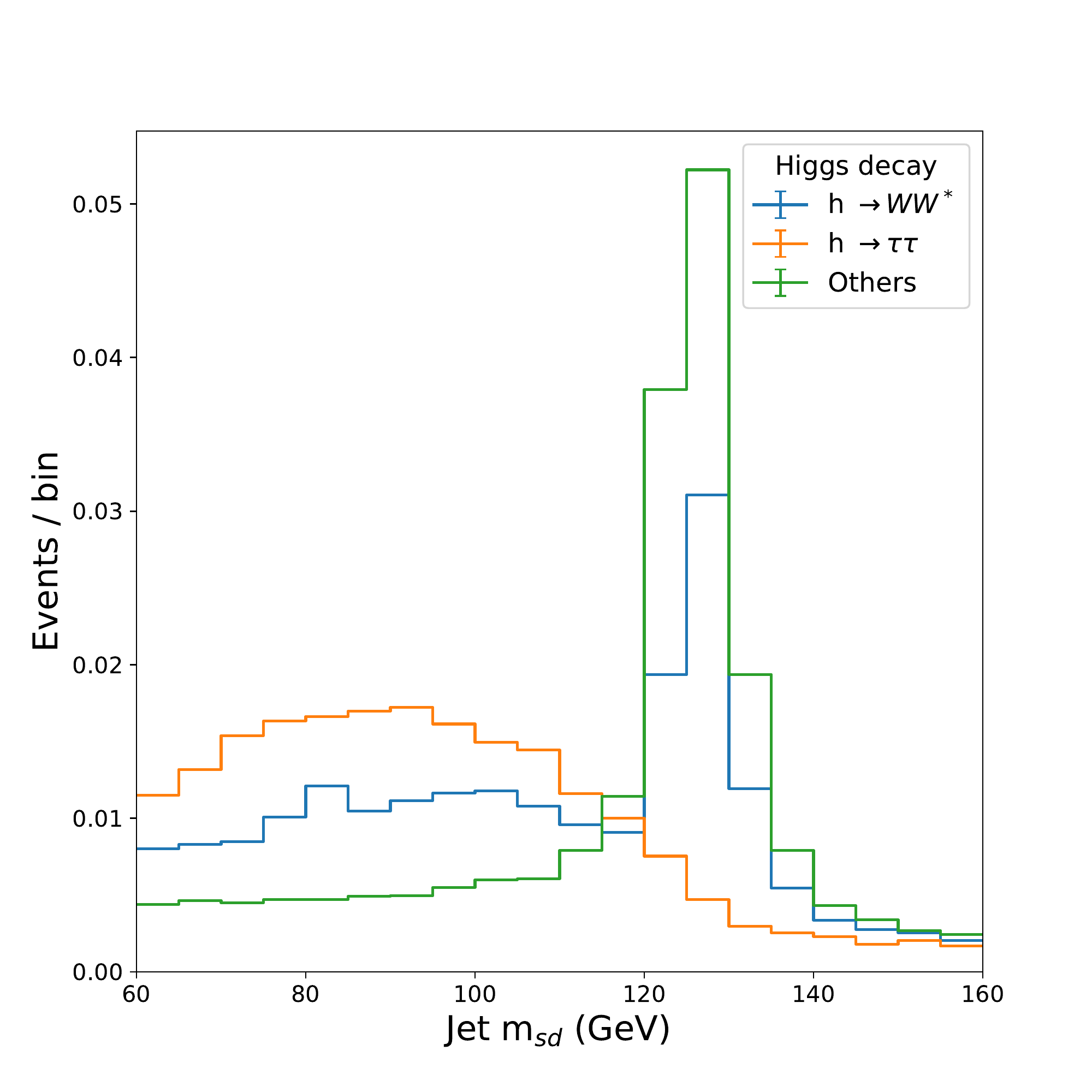}
\includegraphics[width=.45\textwidth]{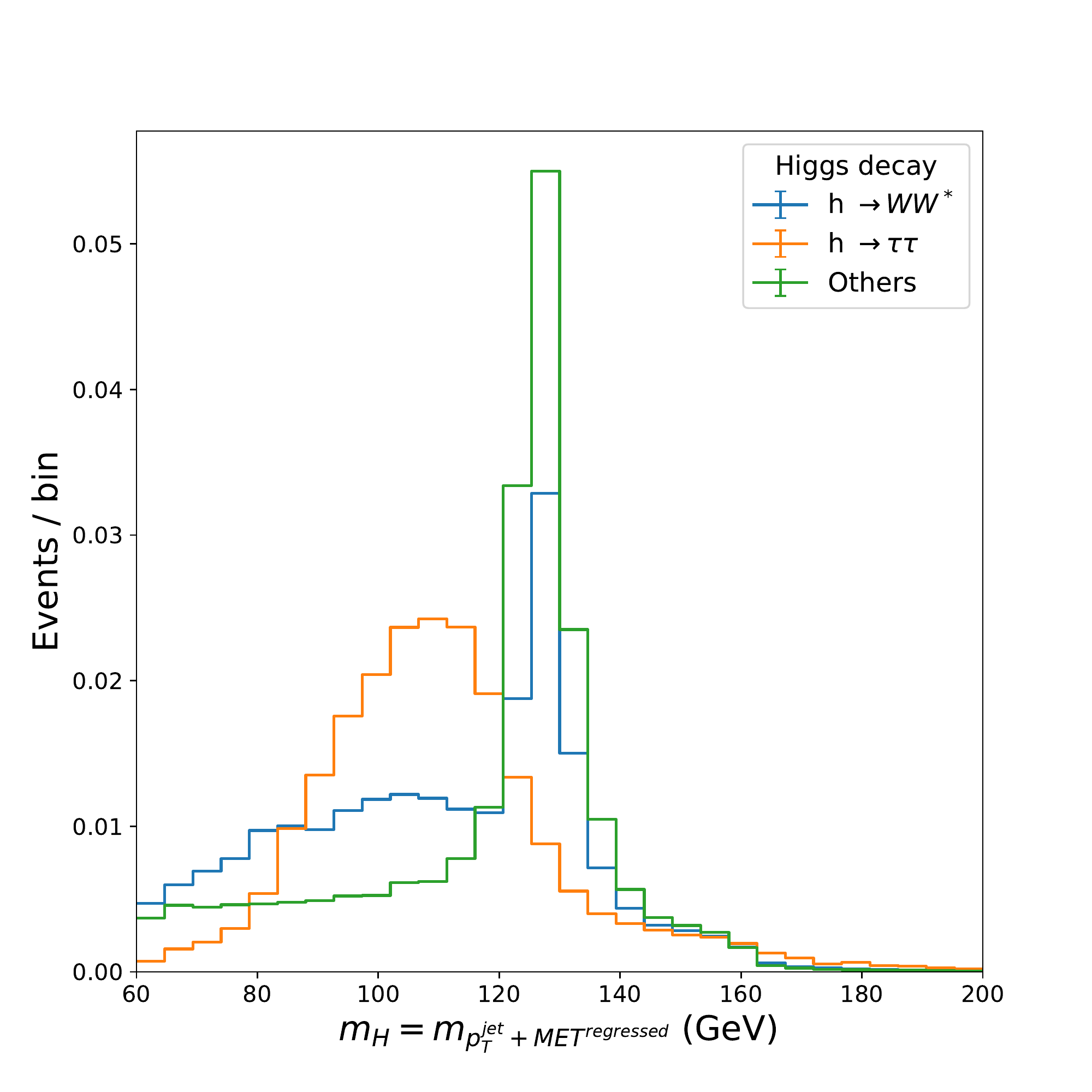}
\caption{\label{fig:mass_res_regression} Comparison of the performance of the SD mass ($m_{SD}$) and the reconstructed Higgs mass $m(\rm jet+MET)$. 
Those Higgs decays without genuine MET (ex. $h \rightarrow gg$) are unaffected by the new mass reconstruction, those with minimal genuine MET (ex. $h \rightarrow b\bar{b}$) are improved slightly, and those with large genuine MET (ex. $h \rightarrow \tau\tau$) are greatly improved.}
\end{figure}

\subsubsection{Higgs jet tagging}
\label{subsec:tagging}
The dominant multijet background can be further suppressed by using information that captures the internal structure of the jet.
While most SM Higgs boson decays result in a jet consistent with either two (ex. $h \rightarrow q\bar{q}$), three (ex. $h \rightarrow qql\nu$)  or four (ex. $h \rightarrow qqqq$) prongs, the multijet background consists primarily of jets consistent with one prong.
However, the ``N-prong'' definition is ambiguous and can introduce further bias to our inclusive $\sigma_{h}$ measurement.
In particular, background like decays of the Higgs boson, e.g. SM $h \rightarrow gg$ decays, resulting in a jet with a radiation pattern very similar to that from the multijet background differing only in the underlying object color structure.
Here, we will discuss several variables and techniques to discriminate signal from the background and our attempts on minimizing this bias. 
Our signal events are taken from the inclusive gluon fusion Higgs signal sample and the Higgs signal jets are selected as discussed in the previous section.
Jets from QCD multijet events with similar p$_T$ are used to define a sample of fake boosted Higgs candidates.

\paragraph{Neural network with jet constituents}
We exploit particle level information by constructing a deep neural network that employs jet particles.
The architecture and model are detailed in the Sec.~\ref{app:performance} of the Appendix.
For brevity, we refer to this network as the GRU classifier~\cite{Louppe:2017ipp,DBLP:journals/corr/ChungGCB14}
When compared to $\tau_2/\tau_1^{\rm DDT}$  the background rejection power of this algorithm significantly improves the Higgs boson signal for a fixed signal efficiency well beyond the critical point where $\epsilon_{S}=\sqrt{\epsilon_{B}}$, as seen in Fig.~\ref{fig:rocs_all}.

\paragraph{Jet mass ratios}
Jet grooming techniques are designed to remove both the soft and collinear radiation from jets.
Unfortunately, these methods make it difficult to distinguish a decay like $h\rightarrow gg$ from quark and gluon jets when considering a fixed mass window.
In contrast, collinear drop observables, recently introduced in~\cite{Chien:2019osu}, are a tool to isolate only the soft radiation in a jet.
The purpose of these variables is to retain components of the soft radiation while removing collinear radiation.
Such observables can be exploited for a study of the color radiation pattern of the particle initiating a jet.
Thus they could provide a handle to isolate the color singlet Higgs jet, without added assumptions of its decay.
In this paper, we test the performance of the ratio of the ungroomed mass of the jet vs the groomed mass:
\begin{equation}
m_{\rm jet}/m_{\rm groomed~jet}
\end{equation}
with different algorithms: trimming ($m_{trim}$)~\cite{Krohn:2009th}, soft-drop ($m_{SD}^{\beta=1},m_{SD}^{\beta=2}$)~\cite{Larkoski:2014wba} and recursive soft-drop $m_{rSD}$~\cite{Dreyer:2018tjj}.
These simplified observables can test how much radiation is removed away by the grooming algorithm: for background QCD jets the mass-ratio distribution is slightly harder than for signal Higgs jets since more soft radiation is removed.
We combine the information from these observables into a simple dense network, designed to classify jets as Higgs-like or multijet-like.
In the following, we refer to this observable as the ``Mass DNN'' or mass-ratios network.
We observe that its discrimination slightly increases with the $p_T$ of the jet.
The performance and distributions of these discriminators can be found in the Appendix~\ref{app:performance}.
Other color-singlet jet isolation techniques have been suggested in the literature, such as~\cite{Chien:2018dfn,Chien:2017xrb}, but not tested in this paper.

\subsubsection{Jet mass correlation}
Since our main observable is the Higgs jet mass, we would ideally like the discriminants we use do not correlate with this variable.
If any such correlation is present, a selection on the tagger will distort the background jet mass distribution depending on the jet $p_T$.
This can bias the way the jet mass is exploited in the analysis to extract the signal.

To decorrelate our discriminators we design a transformation for each variable X to X$^{\rm DDT}$, where ``DDT'' stands for designed decorrelated tagger~\cite{Dolen:2016kst,Sirunyan:2017nvi}.
The transformation $X^{\rm DDT} = X - X^{Y\%}$ effectively varies the Y quantile of the X distribution ($X^{Y\%}$) as a function of the jet mass and $p_T$.
The quantile map $X^{Y\%}(m_{\rm mMDT},p_T)$ is built from multijet events for different bins of the jet $\rho = \rm{log}(m_{\rm mMDT}^2/p_{T}^{2})$ and the jet $p_T$.
The decorrelated selection $X^{\rm DDT} < 0$, equivalent to $X < X^{Y\%}$, keeps a constant $Y\%$ of background events in simulation, irrespective of the jet mass and $p_T$.
This method has little or no impact on the performance of this variable and, by construction, keeps the multijet mass distribution smoothly falling.
The only pitfall of this approach is the smoothness of the quantile map used for the transformation, which depends on the number of background events available in the simulation.
To overcome this limitation, the $X^{Y\%}$ distribution can be smoothed through many procedures e.g. by the use of kernel estimates~\cite{Sirunyan:2017nvi}.

In our study, we choose to build DDT transformations for each of our discriminators, including the outputs of the GRU and mass-ratios network.
More information on alternative methods that were studied for decorrelating neural networks can be found in Sec.~\ref{app:jmass_adv} in the Appendix.

\subsection{Improved sensitivity for $\sigma_{h}$}
\begin{figure}[tbp]
\centering
\includegraphics[width=.7\textwidth]{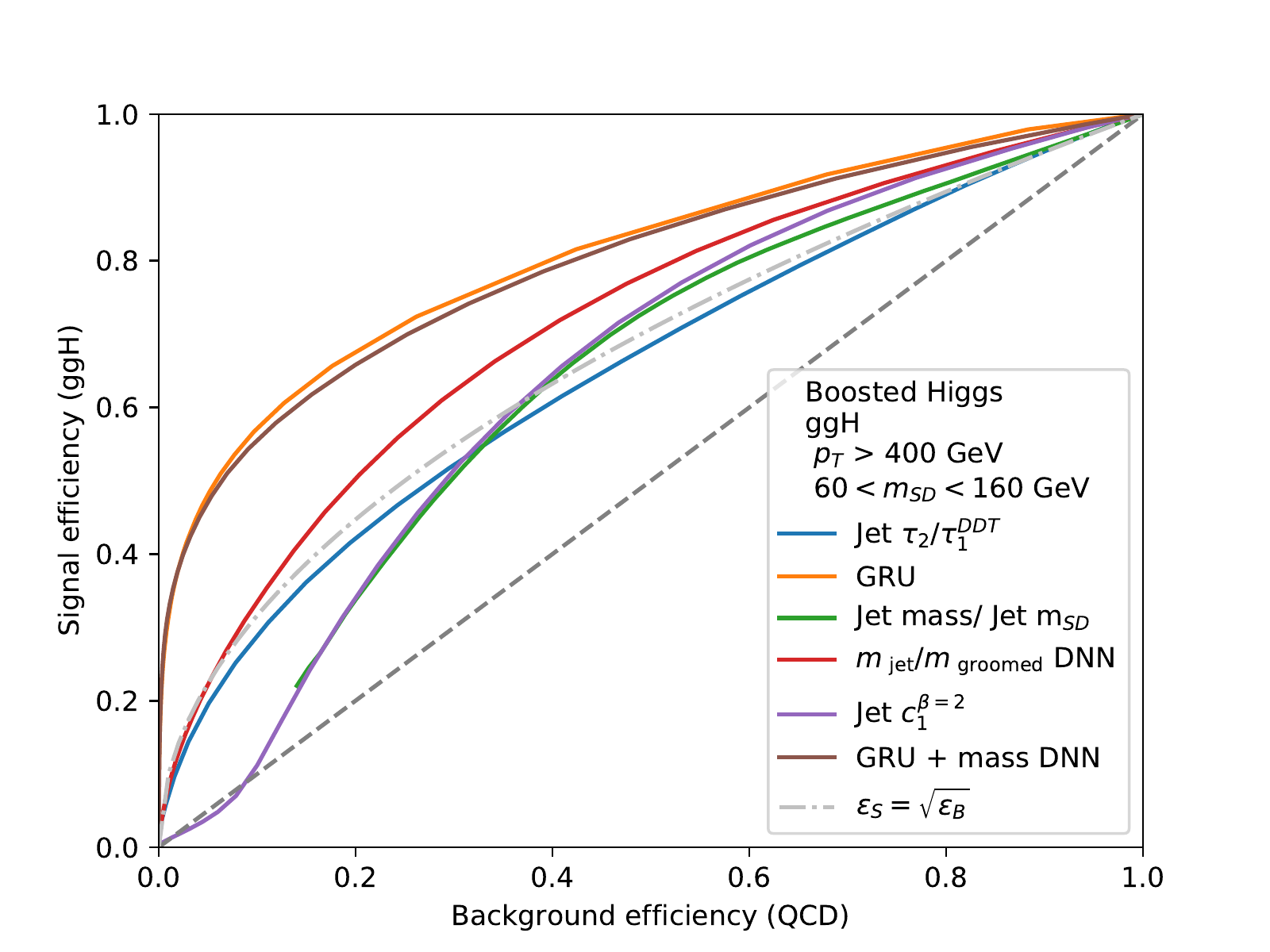}
\caption{\label{fig:rocs_all} Comparison of the performance for the different algorithms used to tag SM decays of the boosted Higgs boson.}
\end{figure}

In Fig.~\ref{fig:rocs_all} we summarize the performance of the tagging approaches considered above in terms of multijet background rejection vs Higgs jet selection efficiency.
The GRU and mass-ratios networks show improved performance compared to the baseline $(\tau_2/\tau_1)^{\rm DDT}$ selection.
We additionally consider a combination of the GRU and mass-ratios networks, constructed by averaging the two discriminants.
We show detailed performance plots such as the discrimination of background-like signals ($h\rightarrow gg$), in Appendix~\ref{app:performance}.

%% file: results.tex
Following the same procedure described in Section~\ref{sec:baseline} we extract a 1$\sigma$ uncertainty on the inclusive Higgs boson cross section by fitting the reconstructed mass distribution. For this fit, we scale the Monte Carlo events to the point that they reach a total integrated luminosity of 3~ab$^{-1}$. All background distributions are smoothed, or, as is the case of the QCD multijet background, a polynomial is fit to the background and used to generate a smoothed background template. 
We assume that the same polynomial order of four, used for 35~fb$^{-1}$, is sufficient for modeling the QCD multijet background; this method will be used to define the upper bound in the ensuing results.
This assumption is addressed in Sec.~\ref{app:extrap_order} in the Appendix.
We note that in recent years there has been a campaign to improve jet variable calculations spanning effects that include higher-order resummation~\cite{Chien:2012ur,Balsiger:2019tne,Chay:2018pvp,Kang:2018jwa,Idilbi:2016hoa,Frye:2016aiz,Marzani:2017mva,Liu:2014oog,Chien:2015cka,Moult:2015aoa,Jouttenus:2013hs,Bauer:2000yr,Bauer:2001yt,Kolodrubetz:2016dzb}, higher orders in perturbative QCD~\cite{Dasgupta:2013via,Dasgupta:2013ihk,Larkoski:2013eya,Dasgupta:2012hg}, and improved Monte Carlo tuning~\cite{Wobisch:1998wt,Dasgupta:2015yua,Reichelt:2017hts,Bellm:2017bvx,Bellm:2016rhh,Sirunyan:2018asm,Skands:2014pea,Sjostrand:2014zea,Buckley:2019kjt,Gras:2017jty}. We strongly encourage this development, and we believe this is the most critical aspect to ensure the sensitivity of this result is preserved.

\subsection{Understanding the mass distribution}
For the signal extraction, a fit of the reconstructed Higgs mass is performed in bins of jet p$_{T}$ following a selection on the discriminator. At this moment, we do not attempt to use the inverted selection or other variations as a control region to model the signal background shape. We instead rely on a polynomial fit of the mass distribution for the QCD multijet background and MC templates for the $W$, $Z$, and top-quark backgrounds. To perform a polynomial fit of the background, the mass distribution needs to have a shape that allows for a sufficient number of sideband events, both below and above the Higgs jet mass peak, to allow for a reliable constraint on the QCD multijet background. 

\begin{figure}[tbp]
\centering
\includegraphics[width=.45\textwidth]{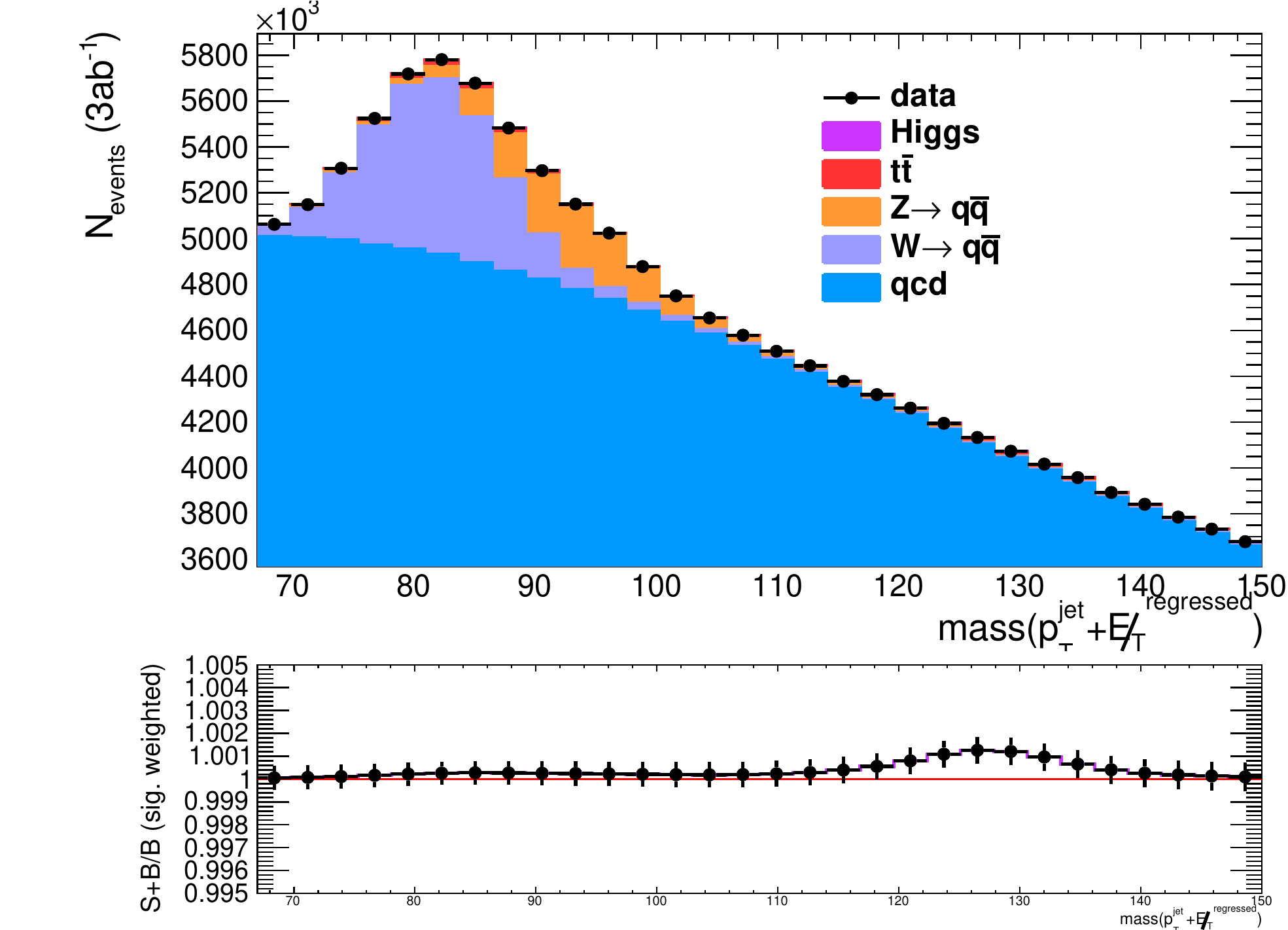}\\
\includegraphics[width=.45\textwidth]{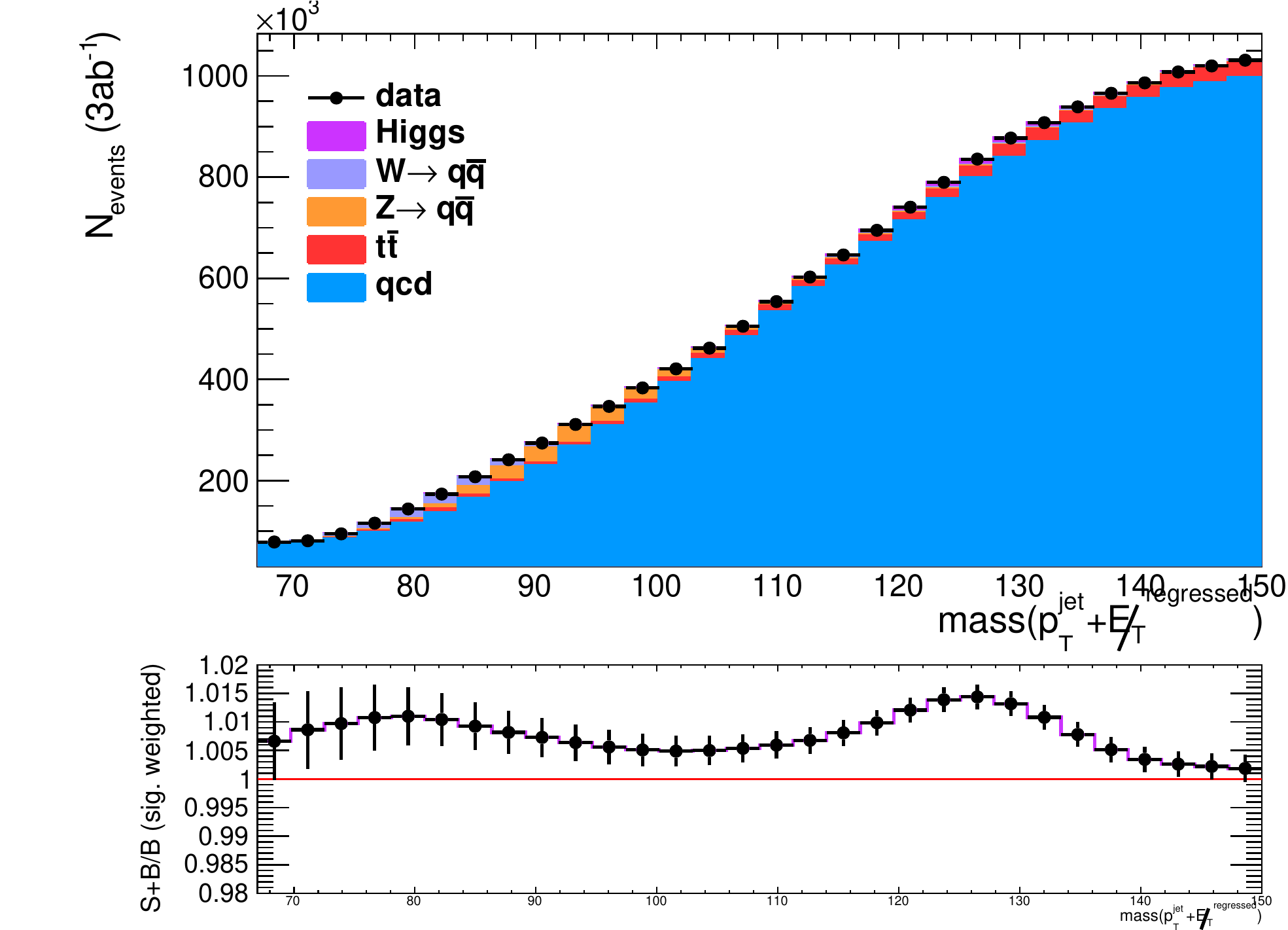}
\includegraphics[width=.45\textwidth]{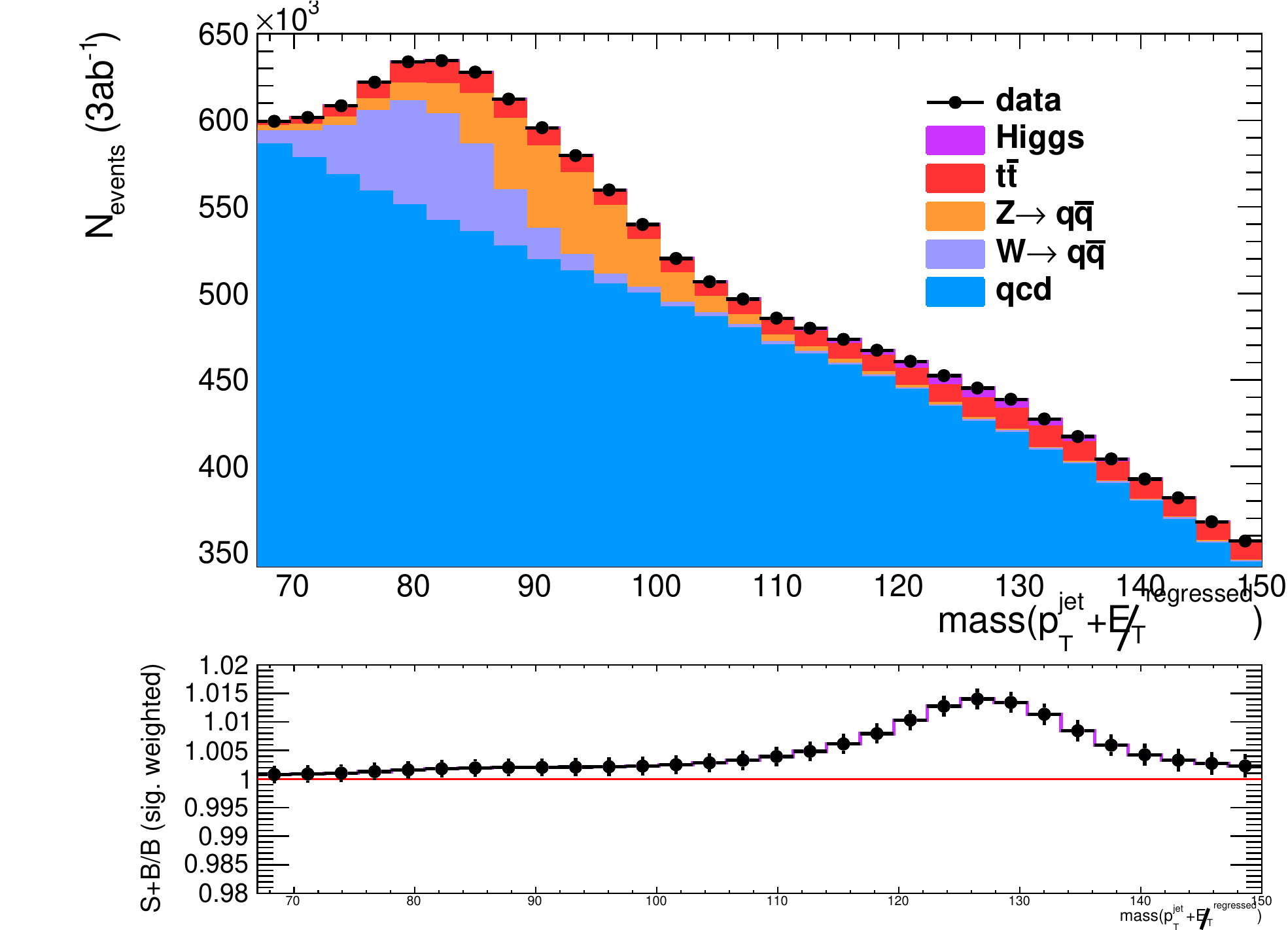}\\
\includegraphics[width=.45\textwidth]{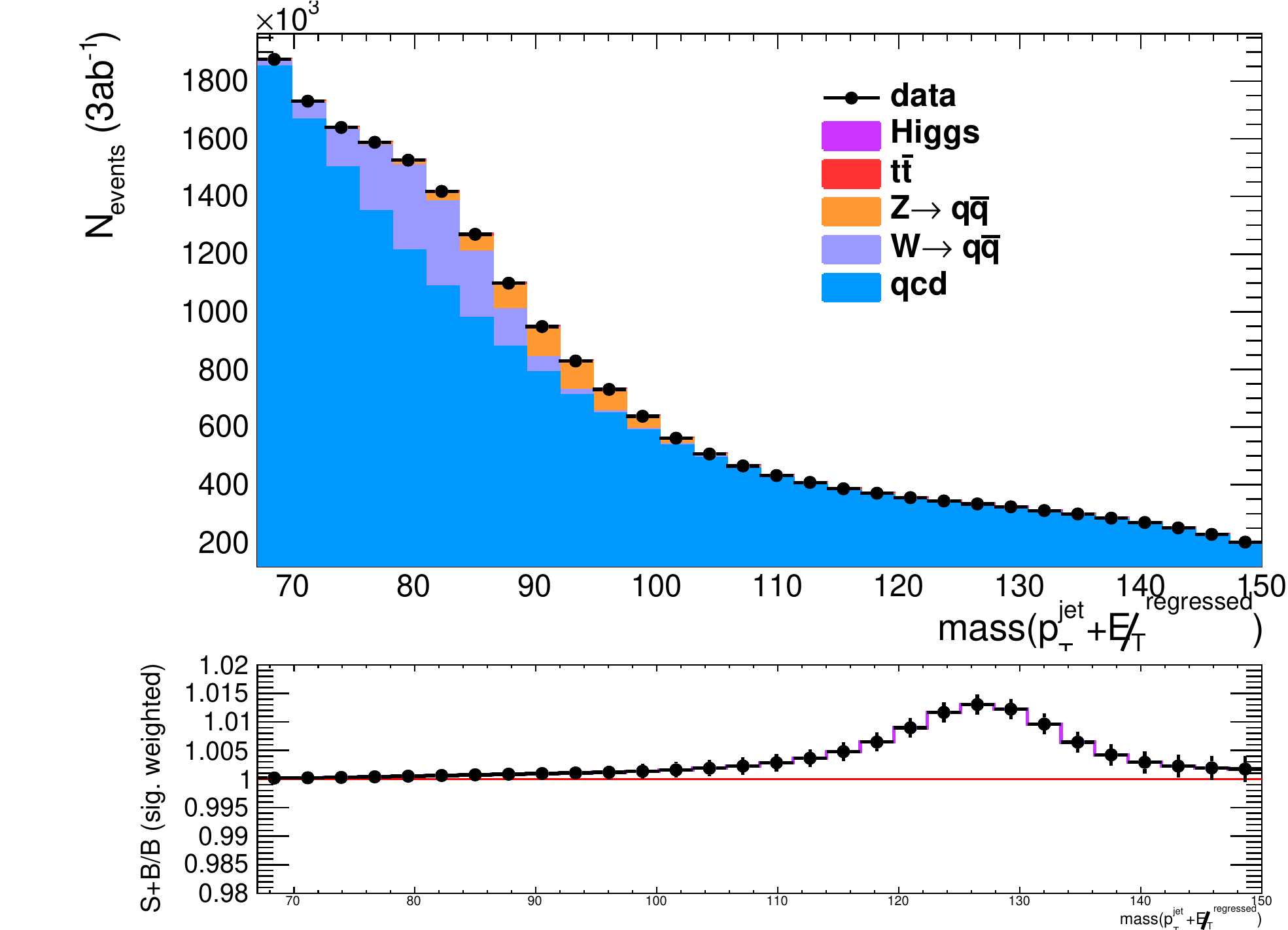}
\includegraphics[width=.45\textwidth]{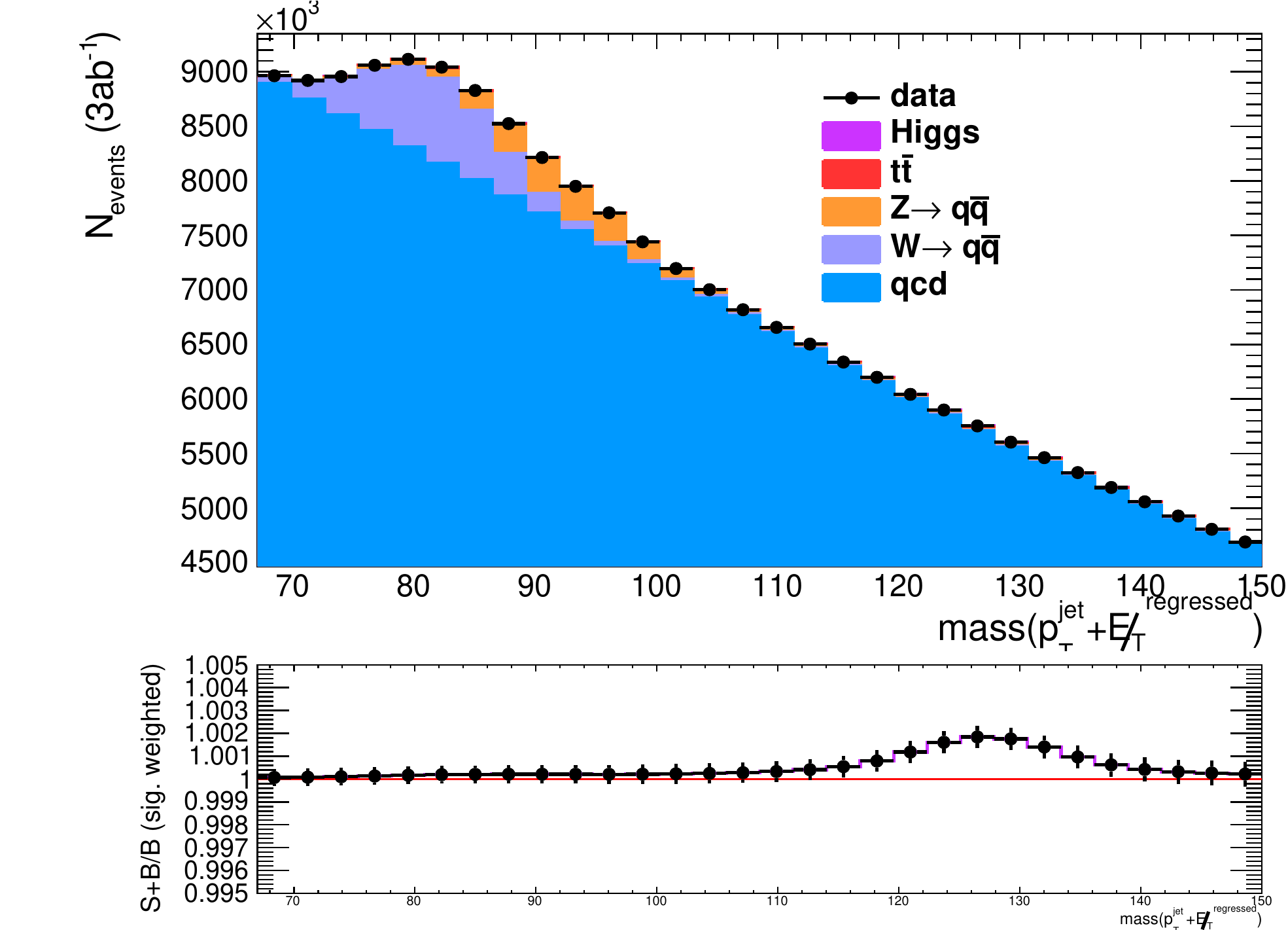}
\caption{\label{fig:massgru} Combined mass distribution of the inclusive Higgs signal extraction for different tagger selections. The top figure shows the smoothly falling mass distribution after a $\tau_2/\tau_1^{\rm DDT}$ selection. In the middle-left a tight GRU selection is shown distorting the mass distribution. Its decorrelated version, in the middle-right, the GRU$^{\rm DDT}$ keeps the mass distribution smoothly falling. The cuts on both discriminants correspond to 1\% of QCD multijet efficiency. Additionally, combined mass distributions selecting on the combined GRU$^{\rm DDT}+\rm{mass-DNN}^{\rm DDT}$ for the 1\% and 10\% QCD multijet efficiency are shown in the bottom-left and bottom-right respectively. The different background efficiency has an effect on the falling shape of the mass distribution. For all cases, the overall signal extraction is performed in $p_{T}$ categories and the final mass distribution is a sum of the $p_{T}$ categories weighted by the individual category significances. The Higgs signal (violet) is small and not visible, but its significance is illustrated in the bottom panel.}
\end{figure}

Figure~\ref{fig:massgru} (top) shows the mass distribution for the baseline selection on the two-pronged observable $(\tau_{2}/\tau_{1})^{\rm~DDT}$. Overall, the mass distribution after this selection is relatively flat and leaves a large peak for the $W$, $Z$ bosons, allowing for a candle to help calibrate the selection. We first consider the mass distribution of a selection that replaces the two-prong observable with the nominal GRU training. The post-fit mass distribution after a tight selection on this variable that leaves 1\% of the background is shown in Fig.~\ref{fig:massgru} (middle-left). The decorrelated version of the GRU, GRU$^{\rm DDT}$, is also shown in Fig.\ref{fig:massgru} (middle-right). This decorrelation does not change the rejection power of this variable significantly and keeps the mass distribution smoothly falling. As a result of it, the $V+h$ and $tt+h$ contributions are less significant in the signal extraction; use of these other processes can introduce additional model dependencies and is better performed through an explicit selection on the rest of the event. 

In light of considering the best signal extraction, we show the mass distribution after cutting at the 1\% QCD multijet background efficiency on a discriminator defined as the average of both the GRU$^{\rm DDT}$ and mass ratio discriminator. The mass distribution of this sample is smoothly falling over the mass range used in this analysis. In the following, we will consider the GRU$^{\rm DDT}+\rm{mass-DNN}^{\rm DDT}$ average discriminator as our benchmark. Additionally, in light of the difficulty to separate Higgs to di-gluon final states from the background, we will consider a second looser working point of this same discriminator where we cut on this discriminator at a 10\% background efficiency. Both of these mass distributions are shown in Fig.~\ref{fig:massgru} (bottom). Note that in each case the DDT computation is performed at the desired background working point to ensure a smoothly falling background. 

\subsection{Comparing results of different approaches}
When comparing the results across different discriminators, we first consider quoting an upper limit on the standard model inclusive Higgs boson cross section assuming branching ratios for the different decays to be consistent with the standard model expectation. For the GRU based discriminators, we consider a tight working point of 1\% background efficiency. For the mass-ratios network, we consider a working point consistent with a 10\% QCD background efficiency. This choice of 10\% represents the lowest background efficiency working point allowed that does not significantly degrade the sensitivity to the Higgs to gluon final state. Lastly, for the combined GRU$+$mass-DNN discriminator, we consider both 1\% and 10\% working points. 

\begin{figure}[tbp]
\centering
\includegraphics[width=.75\textwidth]{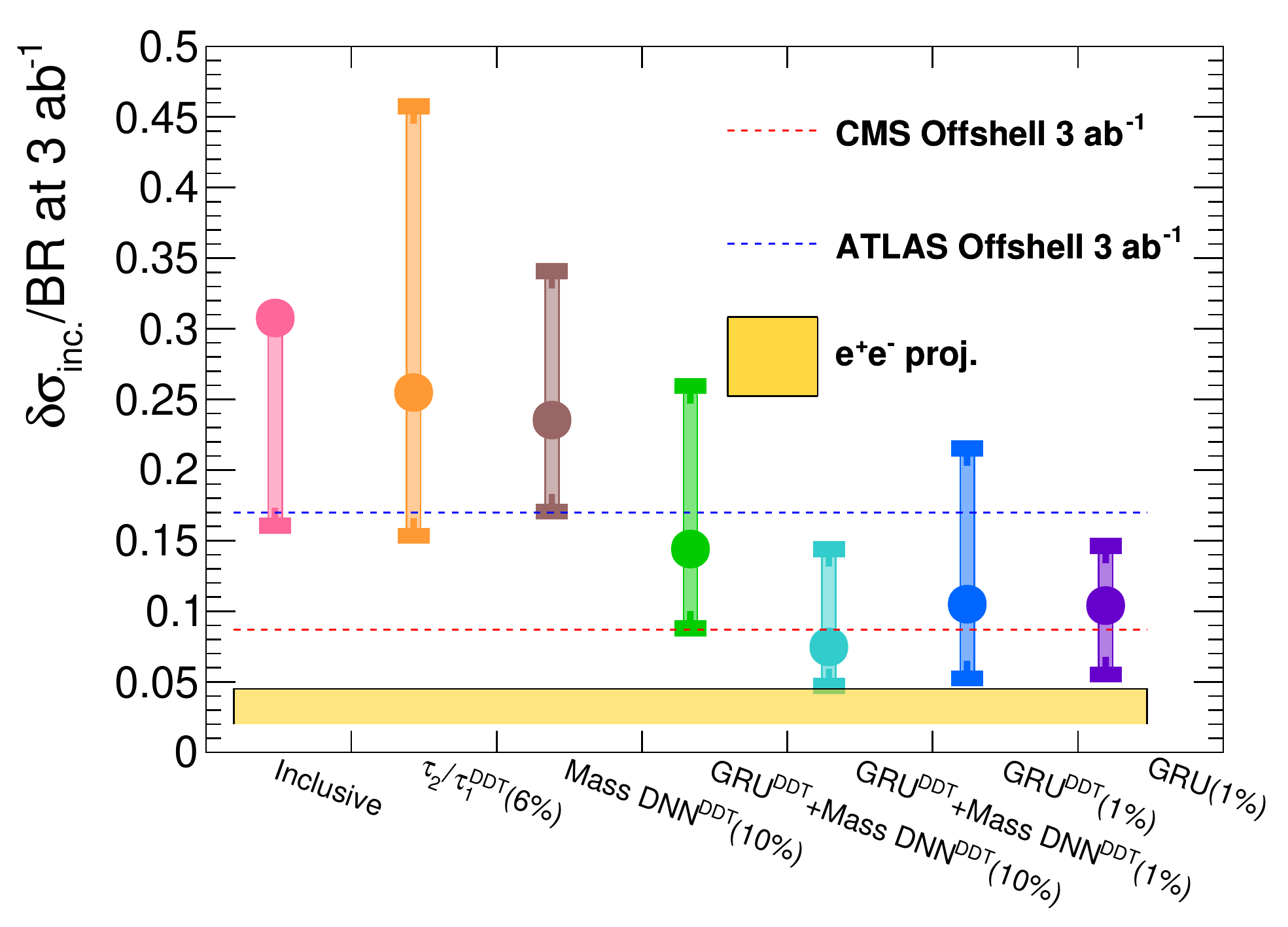}
\caption{\label{fig:allinc}  Estimated 1$\sigma$ sensitivity for selections on various working points and discriminators extrapolated to 3~ab$^{-1}$. For each point, the upper bound of the band indicates the sensitivity using a fourth-order Bernstein polynomial to fit the background. The lower bound of the band shows the result when no systematic uncertainties are included in the calculation and the point indicates the performance when a template fit is utilized for the signal extraction. The inclusive bin has a tadpole-like point indicating that the Bernstein polynomial fit did not converge.}
\end{figure}

Figure~\ref{fig:allinc} shows the results of the inclusive limit using various discriminators and working points. First, we consider the limit when analysis on the inclusive distribution of the reconstructed Higgs mass is performed. With a large number of events and the relatively small signal size, we cannot guarantee convergence of the signal fit when a polynomial form is used for the QCD multijet background, so we just quote the template sensitivity and the sensitivity without systematics (middle point and lower error bar). Following the application of a cut on $(\tau_{2}/\tau_{1})^{\rm DDT}$ an improvement of roughly 20\% is present in the template fit, and the polynomial fit converges. The mass-ratios DNN slightly improves the sensitivity of the analysis. Finally, the application of the combined GRU$^{\rm DDT}$ and mass ratios discriminator gives a full factor of 2 improvements from the inclusive result when a 10\% working point is used. This result is improved by another factor of two when restricting the selection to the 1\% working point. Removal of the mass ratios degrades the sensitivity; it particularly degrades the sensitivity of the polynomial fit as a result of the increased $t\bar{t}$ background. Our benchmark result, using the polynomial fit with the combined mass ratios and GRU$^{\rm DDT}$ discriminator, gives us a measurement of the inclusive cross section 1$\sigma$ uncertainty of $\delta\sigma=0.14\times\sigma_{\rm SM}$. 

Additionally, in Figure~\ref{fig:allinc} we quote the projected extrapolations of the width for both the ATLAS and CMS off-shell interference analyses. This is done by taking the benchmark numbers used in eq.~\ref{eq:unc} and subtracting the width uncertainty with the projected uncertainties on the various Higgs boson projection modes; if the projected uncertainties are reduced then the lines from the off-shell width measurements will increase. Lastly, we show a band for the quoted uncertainties from electron colliders. The various projected results range from 2\% to 4.5\%~\cite{Ruan:2014xxa,Thomson:2015jda,Han:2013kya,Blondel:2018aan,Blondel:2019vdq}. 

\subsection{Results across decay modes}
To establish the model independence of this measurement, we consider the sensitivity across decay modes. In the ideal scenario, the tagger will be independent across decay modes. Consequently, the sensitivity for all Higgs boson final states will be roughly the same and the quoted bound on inclusive Higgs decays will translate to any final state of the Higgs no matter the decay. Variables, such as the mass ratios, which attempt to isolate the Higgs boson through the universal property or the Higgs boson as a color singlet approach to this paradigm. The more model-dependent particle-based taggers, which look broadly at jet internal structure have the potential to improve discrimination at the cost of added model dependence. Furthermore, the reconstructed Higgs mass resolution can worsen for some of the decay modes leading to a worse sensitivity.

We use the standard model Higgs boson decays as a proxy for the sensitivity to all Higgs boson decays.  Additionally, to quote a sensitivity directly applicable to the inclusive Higgs cross section bound we quote the branching ratio corrected sensitivity. This is effectively equivalent to the bound on the Inclusive Higgs boson cross section provided the Higgs boson were to decay uniquely to the selected final state. A detailed study for each of the taggers considered in this analysis is available in the Appendix~\ref{app:decay}. For the two-pronged $\tau_{2}/\tau_{1}^{\rm DDT}$ selection we find a degraded sensitivity to gluon, $W$ boson, and $Z$ boson final states, as already discussed in Sec.~\ref{sec:perdecay}. The results from the GRU and GRU$^{\rm DDT}$ selections show a larger uniformity present amongst the different decay modes, except for the $h \rightarrow gg$ decays, where the sensitivity significantly degrades for the GRU$^{\rm DDT}$.

\begin{figure}[tbp]
\centering
\includegraphics[width=.7\textwidth]{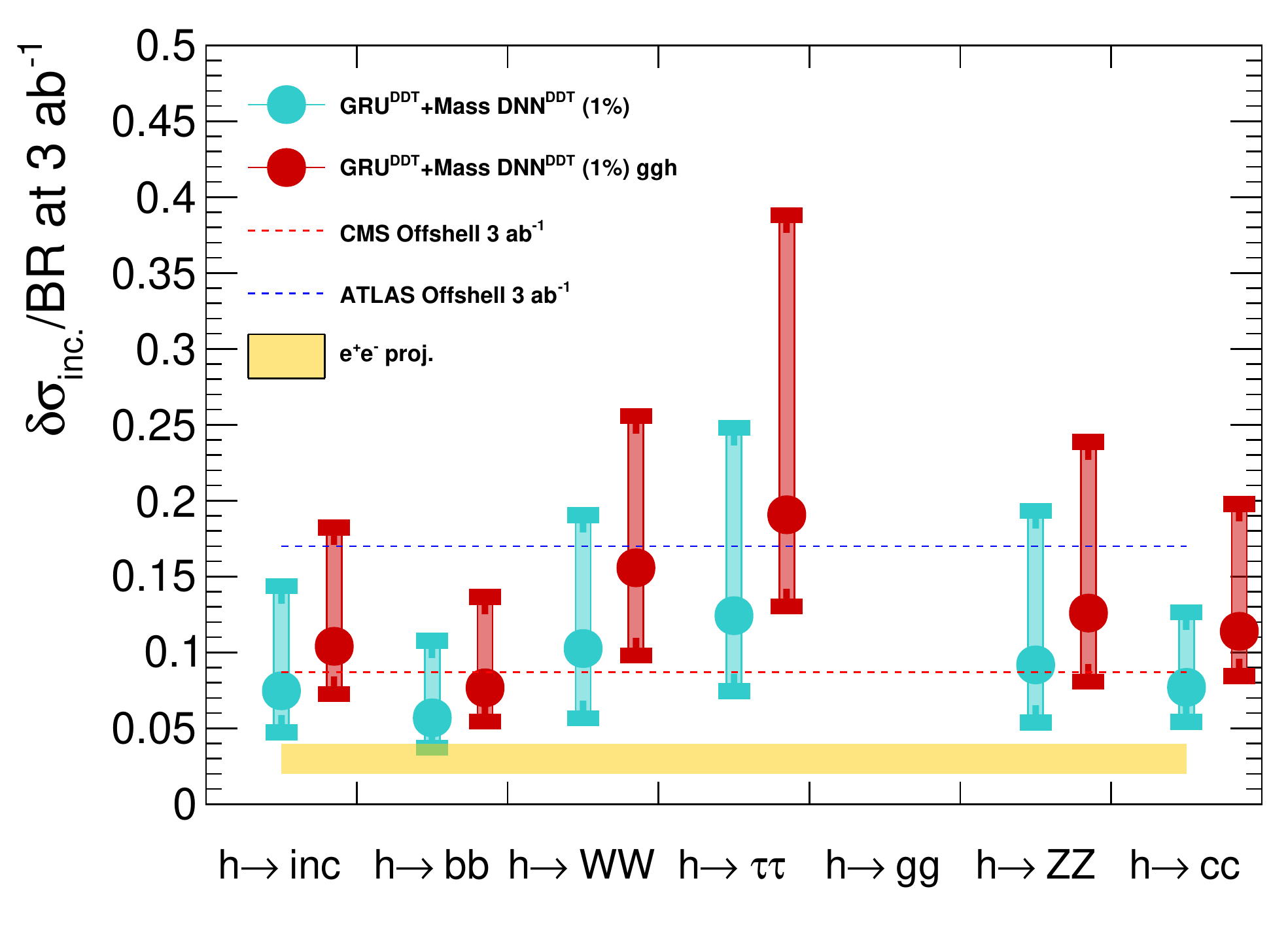}
\caption{\label{fig:finalresult} Estimated 1$\sigma$ sensitivity for the average discriminator from the GRU$^{\rm DDT}$ and mass-ratios$^{\rm DDT}$ networks. The results are shown for an inclusive Higgs boson signal only (red) and including other production modes in the signal extraction. For each point, the upper bound of the band indicates the sensitivity using a fourth-order Bernstein polynomial to fit the background. The lower bound of the band shows the result when no systematic uncertainties are included in the calculation and the point indicates the performance when a template fit is utilized for the signal extraction. Missing points imply the limit is outside the quoted sensitivity range in the plot. In particular, for $h\rightarrow gg$ decays the limit using a polynomial fit reduces to $1 \times \sigma_{SM}$ and to $0.54 \times \sigma_{SM}$ using a template fit. For a $ggh$ signal only, the template fit sensitivity is $0.9 \times \sigma_{SM}$.}
\end{figure}

In Fig.~\ref{fig:finalresult} we show the sensitivity for our benchmark discriminator: the combined GRU$^{\rm DDT}$ and mass ratios DNN.
Here we select signal events with a 1\% QCD background working point.
With this discriminator, we find sensitivity comparable to the ATLAS and CMS off-shell projections for all final states except the di-gluon final state.
Final states with significant missing energy, such as W boson decay, and $\tau$ lepton decay are found to be slightly worse than the inclusive result.
However, the overall degradation is not large (30\% in the worst case).
These final states can be further improved through the explicit use of missing energy in the selection; a topic that goes beyond the scope of this paper.
The di-gluon sensitivity rapidly degrades when a tighter working point is used on the discriminator. 
Consequently, the sensitivity falls outside of the plot; extending the range would give a result of the inclusive cross section at roughly $1.0 \times\sigma_{SM}$.

The rapid degradation of the Higgs boson decay to the di-gluon final state allows us to consider an approach whereby we consider two categories of measurement: a first category consisting of a tight selection on the average discriminator from the GRU$^{\rm DDT}$ and mass ratios DNN$^{\rm DDT}$, and a second looser category obtained by a selection on the 10\% decorrelated version of the same variable (the fraction of events that pass the tight selection and fail the loose selection is $<10^{-3}$ for signal and $<0.03$ for background). In this way, we obtain the bounds shown in Fig.~\ref{fig:finalresult}, and we obtain bounds present at the 10\% bound for the gluon working point. The latter bounds are available in Fig.~\ref{fig:massratiosresult} in the Appendix. For a di-gluon signature we obtain a sensitivity of $0.41 \times\sigma_{SM}$.

By isolating the different decay modes of the Higgs boson, we have effectively opened up Pandora's box and started to peer inside. Looking at the performance amongst each decay mode explicitly highlights the sensitivity towards Higgs boson-like decays. We see that across many final states, we achieve a projected sensitivity comparable to that of other approaches. However, for decays that most closely mimic the QCD multijet background, in particular, a Higgs boson that decays to the di-gluon final state, the ability to measure the equivalent Higgs boson total width becomes worse but remains possible. The resulting question is thus, can we claim an upper-bound on the potential of measurement of the Higgs boson total width through the di-gluon final state? The QCD multijet background in this region largely originates from that of a single gluon that subsequently splits to gluons itself, yielding a di-gluon final state. Other spin-0 combinations of quarks and gluons can potentially occur in extended Higgs boson models~\cite{Curtin:2013fra}. However, in all cases, these models come with additional features such as intermediate bosons. Consequently, we argue that the di-gluon bound can be viewed as an upper bound towards this approach to the measurement of the Higgs boson total width. 

From the above results, we finally quote two bounds on the Higgs boson inclusive cross section. First, a model-dependent bound that assumes, roughly, a standard model admixture of decays. For this bound, we quote the 1~$\sigma$ sensitivity of this process using the polynomial fit and get a result of $0.14 \times \sigma_{SM}$. Given the variation over decay modes, we find that this bound can be applied to final states with varying admixtures of heavy vector-bosons, $\tau$ leptons, and quarks. Secondly, we quote bounds on all Higgs boson decays by taking the worst bound covering all final states. In this case, we quote the Higgs to di-gluon bound found in the GRU$^{\rm DDT}$ and mass ratios DNN$^{\rm DDT}$ selection at the 10\% working point. For this bound, we find a result of $0.41 \times \sigma_{SM}$. 

%% file: bias.tex
The bounds on the inclusive cross section can be translated to upper bounds on the Higgs width using eq.~\ref{eq:unc}. This translation has three main assumptions that can lead to bias in our measurement:

\begin{itemize}
\item All objects from the Higgs boson decay, both with missing energy signatures and visible signatures, occur within a single cone in the event. For missing energy signatures we approximate their decay by using an NN regression. 
\item We have not considered all possible BSM signatures of Higgs boson decays. In place, we have treated all SM Higgs boson decays as a proxy for all signatures. 
\item The Higgs jet selection can introduce a bias by only selecting SM-like Higgs decays.
\end{itemize}

In the following, we discuss these assumptions and our attempts to minimize their biases before translating our results into $\Gamma_h$ constraints.

\subsection{Dealing with MET signatures}
\label{subsec:met}
Higgs decay products can escape detection if neutrinos or other non-interacting particles, such as dark matter, are present in the final state. Amongst SM Higgs boson decays, both $h\rightarrow \tau\tau$ and $h \rightarrow WW^{*}$ decays have invisible signatures. In our analysis, we have attempted to recover these signals by including the missing energy to the jet reconstruction. We have taken two steps in this direction. The first is to modify the leading jet selection in the event and select a composite object instead, defined as the leading $p_T^{\rm{jet+MET}}$ jet in the event. The second consists in adding the regressed MET to the jet, before computing the mass. For the regressed MET direction, we assume the missing energy is collinear to the visible reconstructed components of the jet. We have not studied the modified mass reconstruction in detail and we emphasize further studies can be done to completely recover the direction of the MET. However, a clear improvement is present (Fig.~\ref{fig:mass_res_regression}), which further leads to comparable sensitivity to these final states when compared with a benchmark inclusive Higgs boson selection. While we have only explicitly shown this modified reconstruction helps recover $h\rightarrow \tau\tau$ and $h \rightarrow WW^{*}$ sensitivity, we believe this approach is broadly applicable to all semi-visible decays, both SM based visible decays and beyond SM decays. We leave a more detailed study of MET signatures for future papers. 

In the instance where the Higgs boson decays completely invisibly this analysis is not applicable anymore. Fully invisible searches for Higgs boson decay have been performed by both ATLAS and CMS~\cite{Sirunyan:2018owy,Aaboud:2019rtt}. The equivalent search for an invisibly decaying Higgs boson would occur with a single jet recoiling against missing energy. The search for Higgs to invisible in the single jet final state is currently sensitive at the 1$\sigma$ level at $\sigma = 0.3 \times \sigma_{SM}$. Projected results will thus be significantly better than the inclusive cross section results quoted here. When the additional production channels are used, projected results on the Higgs boson to invisible bounds range at the HL-LHC range from roughly 1\% to 4\%~\cite{Curtin:2018mvb,Cepeda:2019klc}. The reason that this analysis is more sensitive to Higgs boson production largely originates from the significantly reduced background present after a MET selection. Loosened triggers also contribute to improved sensitivity, but these are found to not be as important. 

\begin{figure}[tbp]
\centering
\includegraphics[width=\textwidth]{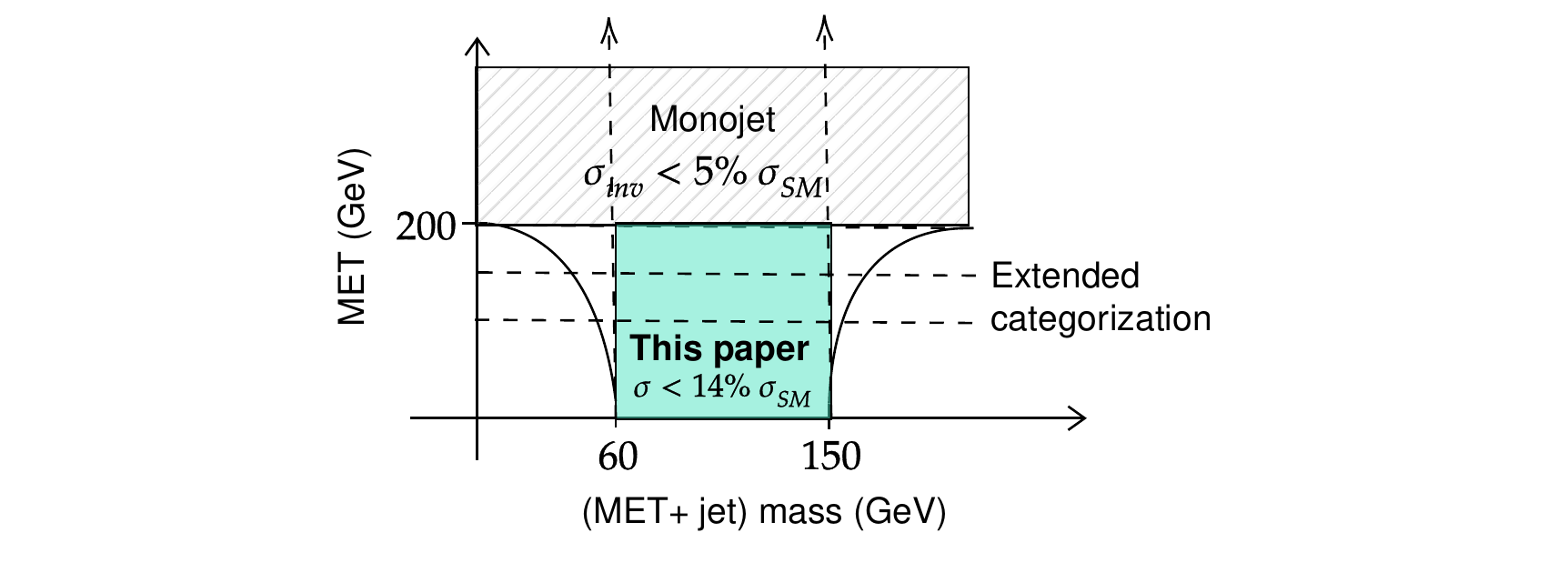}
\caption{\label{fig:metdiag} Diagram depicting a possible extension of the current analysis by adding a missing energy category. The current analysis covers a phase space where the mass of the $\rm{jet+MET}$ object is between 50 and 150~GeV. Adding categories in MET where we progressively loosen the mass window and jet selection can allow covering signatures with semi-visible and completely invisible decays of the Higgs.}
\end{figure}

The excellent sensitivity for a missing energy selection and the existing incorporation of semi-visible decays motivate an analysis strategy whereby we both categorize the analysis in MET and with each MET category we progressively loosen the mass window and jet selection to account for the fact that the background is significantly reduced. In this way, this current analysis can be extended in a natural way to completely cover missing energy signatures that are both semi-visible and completely invisible. This is illustrated in Fig.~\ref{fig:metdiag}. With this extended analysis, sensitivity to MET based events should improve and thus the quoted $h\rightarrow \tau\tau$ and $h \rightarrow WW^{*}$ can be viewed as upper bounds. 

\subsection{Gluons and hard to find signatures}
\label{subsec:gluons}
Another concern about this approach follows from our ability to identify background-like decays. In our current study, we have made a first attempt at isolating $h \rightarrow gg$ decays by exploiting the color-singlet nature of the Higgs boson through the mass-ratios DNN selection. If we can isolate background-like signals by using a universal property of the Higgs boson then we can argue that other non-SM signatures that look like background could be isolated with this approach. Thus the cross section bounds quoted in the last section from the di-gluon final states can be viewed as an upper bound. 

For this paper, we argue that SM Higgs boson decays cover a broad range of signatures that can serve as a proxy for all visible signatures. However, this remains to be shown. While many BSM signatures are either two-pronged or four-pronged like SM Higgs boson decays, other signatures can be more complicated. As an example, Long-lived particle (LLP) decays of the Higgs boson can be viewed as a representative BSM signature. These decays are often two pronged. However, depending on their lifetime, LLPs can escape the detector leaving semi-visible, or fully invisible signatures. While it remains to be proven, we argue the bounding case of Higgs to di-gluons can be viewed as a worst-case scenario. 

\subsection{Uniformity of sensitivity across signatures}
When considering the ultimate sensitivity, we have considered properties largely independent of the decay, such as the jet mass and mass ratios, and, with the case of the GRU tagger, we have considered the internal properties of the jet. When constructing the GRU tagger, we have used SM Higgs boson decays. As a consequence, the tagger itself is designed for SM-only decays of the Higgs boson. However, the large breadth of topology and the fact that the boson of origin is the Higgs boson, indicate that the network could be retrained with additional signatures to incorporate these beyond SM decays. In future iterations of this approach, we aim to directly test BSM decays of the Higgs boson first by using the existing tagger and secondly through training of these additional final states in the tagger. 

\subsection{Width calculation}
\label{subsec:width}
Under the set of assumptions detailed above we can translate our upper bounds on the cross section to bounds on the partial width of the SM Higgs boson to visible and semi-visible decays $\Gamma_h$.

By using a similar selection to the Z$^{\prime}$ analysis we are able to exclude $\delta\sigma = 3.6 \times \sigma_{SM}$ with 35~$\textrm{fb}^{-1}$, which translates to an upper bound on the SM Higgs width of $\Gamma_h < 7.8~\times~\Gamma_{SM} = 32$~MeV.

For a full HL-LHC dataset of 3~ab$^{-1}$ we take the improved tagger selection of the average discriminator from the GRU$^{\rm DDT}$ and mass ratios DNN$^{\rm DDT}$ as our baseline, which gives us a constraint on $\delta\sigma = 0.14 \times \sigma_{SM}$. This translates to an upper bound on the SM Higgs width of $\Gamma_h < 0.33~\times~\Gamma_{SM} = 1.4$~MeV.

Our upper bound on the cross section of the Higgs boson to di-gluon final states is $\delta\sigma = 0.41 \times \sigma_{SM}$, which can be translated to $\Gamma_h < 0.83~\times~\Gamma_{SM} = 3.5$~MeV.

\subsection{Additional assumptions}
\label{subsec:add}

Throughout this analysis, we have made several assumptions for the analysis strategy and signal extraction. Most notably, we have assumed that a signal extraction technique with similar sensitivity to the current Z$^{\prime}$ analysis can be carried to the full dataset without accruing a large amount of additional systematic uncertainties. We believe that preservation of this sensitivity is an important avenue for future work. In particular, minimization of the number of free parameters has previously been performed through the use of more accurate Monte Carlo simulation, and the inclusion of higher-order QCD and electroweak calculations. 

To perform this analysis, we assume the signal efficiency uncertainty in data can be constrained to 5\%. For a 14\% measurement on inclusive Higgs boson production, this uncertainty would need to be constrained to a level significantly below 14\%. While, we do not claim to have resolved this issue there are many handles within the data, such as the W and Z boson resonant peaks, and the Higgs boson itself, that can be used to constrain the Higgs boson selection. Furthermore, we have assumed a uniform signal efficiency across decay modes. It is not clear if this assumption will hold for di-gluon and four-quark final states that could degrade the efficiency measurement. We leave the efficiency study for future papers.

Lastly, to demonstrate the performance, we have utilized a toy simulation to emulate the performance at the LHC. The gains reported with the tagging approaches here would need to be validated with actual LHC simulation and data. Recent developments in boosted object tagging with LHC simulation and data have shown that comparable levels of improvements with similar NN architectures~\cite{CMS-PAS-JME-18-002}. We do not believe this to be a critical issue.

%% file: conclusion.tex
In summary, we have performed a study of the sensitivity of the LHC to the inclusive Higgs boson cross section. For a varying set of assumptions, we can extract a direct measurement of the Higgs boson total width. To minimize the assumptions, we exploit a deep neural network tagger and a dedicated mass reconstruction. In the instance of a standard model Higgs, we are currently able to exclude a total width on the Higgs boson of 32~MeV with 35~$\textrm{fb}^{-1}$. This is roughly 30 times better than the current direct measurement of the Higgs boson total width and on a similar scale with the current off-shell measurements. With the full HL-LHC dataset of 3~ab$^{-1}$ and an improved tagger, we find that we can exclude the Higgs boson total cross section of 0.14~$\times \sigma_{SM}$. This translates to a measurement of the total width with an uncertainty of $\delta\Gamma_h < 1.4$~MeV. Considering the worst possible sensitivity over all decay modes, we find that with the di-gluon final state we obtain an uncertainty on the total cross section of 0.41~$\times \sigma_{SM}$ and corresponding width measurement of $\delta\Gamma_h < 3.5$~MeV. These results are comparable to the ATLAS and CMS projected off-shell measurements of the Higgs boson total width, which are found to be 1.6~MeV and 1~MeV respectively. 

The inclusive measurement without systematics is found to be 0.05~$\times \sigma_{SM}$ when considering SM Higgs boson decays, and 0.19~$\times \sigma_{SM}$ when considering the Higgs to the di-gluon final state. These two numbers are about 3 times better than the quoted result utilizing the full background extraction techniques. A large improvement when systematic uncertainties are removed is a strong indication that better background estimates, selection strategies, and signal extraction methods can lead to additional significant improvements in the quoted Higgs boson width measurement result. One particular source of an improvement would be additional tags on the production mode targeting high $p_{T}$, vector-boson fusion,  vector boson associated Higgs boson production, and $t\bar{t}~+$~Higgs boson production; these processes contribute to roughly 20\% of the total amount of Higgs boson events. Isolation of specific production modes can lead to improved signal purity with minimal model dependence on the Higgs boson final state. We believe this is a very fruitful way to approach future studies of the Higgs boson total width, and can lead to improved bounds. 

\begin{figure}[tbp]
\centering
\includegraphics[width=\textwidth]{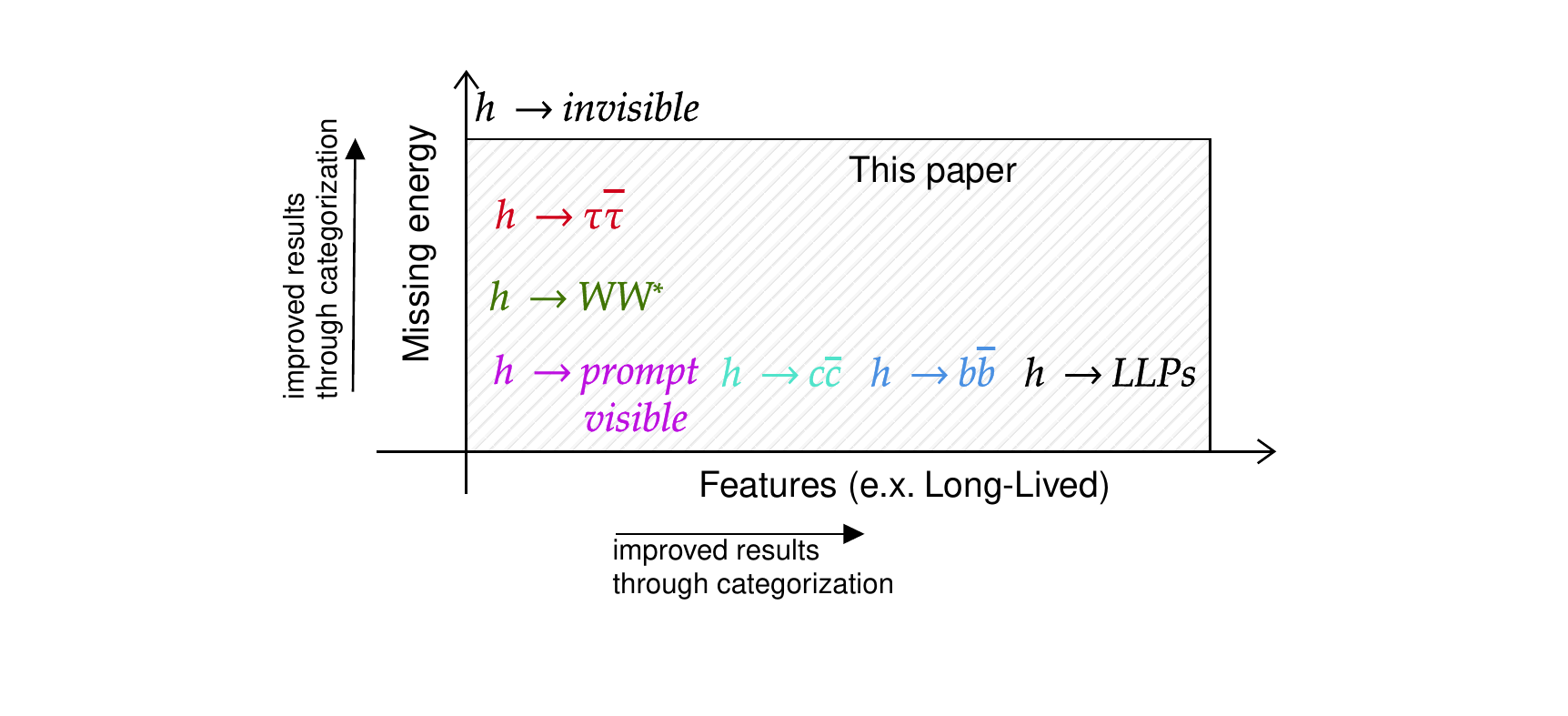}
\caption{\label{fig:strategydiag} Diagram depicting the strategy towards future categorization of the current analysis. The box represents the final states covered by the current selection. In particular, Long-Lived Higgs decays were not studied in the current analysis but we believe they are covered by this approach. Adding missing energy categories could make the analysis sensitive to Higgs to invisible decays. Other features such as secondary vertices or timing information can improve the sensitivity to long-lived final states. 
Finally, other axes can be added to this categorization, such as the number of leptons or number of prongs to recover sensitivity to semi-visible and visible final states.}
\end{figure}

The split of Higgs boson bounds by categories to enhance sensitivity to Higgs boson final states highlights an approach towards how to deal with the inclusive Higgs boson measurements in future analyses. While this paper has attempted to limit the amount of model dependence in the Higgs boson final state to focus on an inclusive category, it is possible to significantly improve specific final states by further categorization. Consider, for example, the missing transverse energy final state. The analysis quoted here, in the limit of large missing energy, could be made to approach the Higgs to invisible analysis, first, by opening up the categorization to include missing energy categories, second, by adding weak boson fusion categorization, and third, by adding events which pass a missing transverse energy trigger. A similar extension can be performed for long-lived models~\cite{Curtin:2013fra,Curtin:2014cca}. We view these additional final state analyses as further categorization towards a global all-encompassing Higgs boson measurement in all final states. The analysis presented here can be viewed as a catch-all category for events that cannot further be categorized. A diagram of this strategy is shown in Fig.~\ref{fig:strategydiag}. 

We would like to stress that we have aimed to keep assumptions as minimal as possible. By requiring a jet with a mass at the Higgs boson mass, we have only used Lorentz invariance as a means to select the Higgs boson. This property works for all particles that are reconstructed in the detector. By employing a selection using the combined jet and missing transverse energy mass, we have allowed for the potential of missing energy to be in the final state. Additionally, though the use of the mass ratios deep neural network, we have attempted to exploit the universal property that the Higgs boson is a color singlet. While these selections do add some model dependence in the Higgs boson final states, the amount of model dependence is minimal. Finally, by allowing a tagger to explicitly exploit the particle-based information within the jet, we can further enhance the sensitivity to the Higgs boson at the cost of more model dependence. This paper motivates a rich program of exploring the Higgs boson at high $p_{T}$ in all final states. We strongly encourage the community to consider this approach, and we hope that further developments can be made in this direction.

%% file: appendix.tex
\section{Fit procedure and signal extraction (35$fb^{-1}$}
\label{app:fit}
The upper bound on the Higgs boson total cross section quoted in this analysis is extracted from a likelihood fit that treats the Higgs boson as a signal.
We perform a simultaneous fit of the mass distribution in the range of 60 to 160~GeV and in 4 $p_{T}$ bins, corresponding to $500-600,600-750,750-900,900-2500$~GeV.
These 4 $p_{T}$ bins roughly match the binning present in the CMS $ggh \rightarrow b\bar{b}$ analysis~\cite{Sirunyan:2017dgc}.
The 1$\sigma$ confidence level CL$_{s}$~\cite{CLS2,CLS1} limit is then computed in a fit for the signal extraction where we have incorporated $W$, $Z$ boson, QCD multijet and $t\bar{t}$ backgrounds.

To ensure that our description of the backgrounds is not limited by per-bin statistical fluctuations, the $W$, $Z$, and $t\bar{t}$ backgrounds are smoothed.
The QCD multijet background shape is smoothed by fitting a 4th order Bernstein polynomial in each $p_{T}$ bin and subsequently sampling the MC predicted number of events from the fitted shape.

To perform the signal extraction fit, gaussian constraints on systematic uncertainties are added to best mimic the actual analysis strategy.
We apply a systematic uncertainty on the backgrounds from the luminosity uncertainty of 5\% and the boson tagging efficiency of 10\%.
For the QCD background, we apply a normalization uncertainty of 10\%.
A separate systematic uncertainty of 15\% is applied on each of the Z bosons and W boson normalization and a normalization uncertainty of 15\% is applied to the top quark, corresponding to the expected uncertainty attainable from a semileptonic top control region.
A mass scale and mass smearing uncertainty of 1\% and 5\% are applied to each of the resonant backgrounds, and the signal uncertainty.
These systematic uncertainties are treated as correlated across the different $p_{T}$ categories.

To estimate the QCD background contribution, we consider two possible approaches.
First, we consider an approach where we take the statistical precision of the QCD shape and perform a fixed template fit of the QCD multijet background.
Secondly, we perform a fit using the full fourth-order Bernstein polynomial.
The polynomial parameters are treated as fully unconstrained and allowed to float freely within the fit.
Finally, the likelihood is evaluated for an Asimov data set defined by the nominal model with the expected signal and background yields scaled to the integrated luminosity corresponding to LHC and HL-LHC and with the SM expectation for the signal strength.

\subsection{Motivating the order of extrapolation}
\label{app:extrap_order}
\begin{figure}[th]
\centering
\includegraphics[width=.7\textwidth]{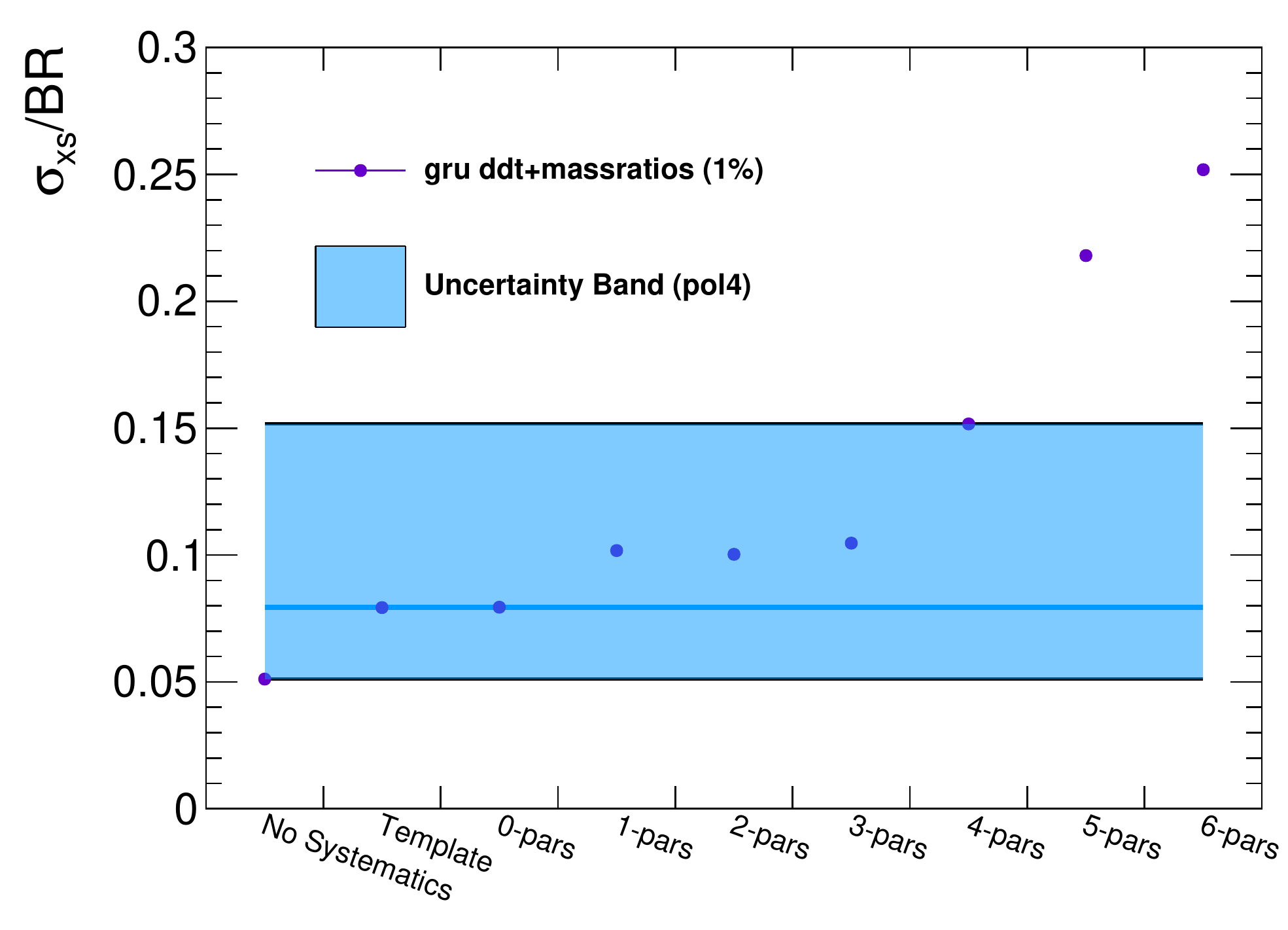}
\caption{\label{fig:polcomp} Comparison of the fit sensitivity for the GRU$^{\rm DDT}+\rm{mass-DNN}^{\rm DDT}$ observable selection at 1\% using different order polynomials to estimate the QCD multijet contribution. The band shown in this plot corresponds to the uncertainty bars shown in the fit results plots (e.g. Fig~\ref{fig:finalresult}) where the bottom part of the band corresponds to the result without any systematics applied, the middle part corresponds to a template fit, and the final top result corresponds to a fourth-order polynomial fit of the QCD background.}
\end{figure}

Our result at 35~fb$^{-1}$ has a sensitivity about 25\% worse than the existing $Z^{\prime}$ re-interpreted result. With existing background strategies, it is not clear that we will be able to continue to model the background with the same polynomial order. Figure~\ref{fig:polcomp} shows the variation in the sensitivity as the polynomial order of the background increases. As with the 35~$fb^{-1}$ fit, a separate polynomial is used for each of the four $p_{T}$ bin categories; thus, a 4th order polynomial corresponds to 16 floating parameters. When no systematics are present, we find for the benchmark discriminator a sensitivity to the inclusive cross section of $\delta\sigma = 0.055 \times \sigma_{\rm~SM}$. With a template fit the uncertainty is roughly doubled at $\delta\sigma = 0.095 \times \sigma_{\rm~SM}$ and this uncertainty progressively gets larger when higher-order polynomials are utilized for the background fit. Additional degradation of 60\% is present when extending from a 4th order to a 6th polynomial order (24 total freely floating parameters).

To improve the polynomial order, or at least preserve the order when more data is used, several approaches have been developed over the past years. In particular, the replacement of functional form fits the QCD multijet background with template-based approaches using scale factors from control regions. This strategy, in various forms, has been used for the CMS boosted $Z^{\prime}$ analyses, as well as the most recent QCD di-jet search~\cite{Sirunyan:2018ikr,CMS-PAS-EXO-17-026,CMS-PAS-EXO-19-012}; a related approach has been used for the modeling of single-boson processes in dark matter searches~\cite{Aaboud:2018xdl,Aaboud:2017phn,Aaboud:2018zpr,Sirunyan:2018owy,Sirunyan:2018dub,Sirunyan:2018gdw,Sirunyan:2018gka,Sirunyan:2017qfc,Sirunyan:2017jix}. In these approaches, the MC is used   to extrapolate the shape in data from a control region into the signal region. Corrections on this extrapolation factor are then applied through the use of polynomial functions or other functions, which allow the extrapolation factor to vary within the uncertainties of the extrapolation. Through this approach, the polynomial order has been reduced in modeling QCD processes. A notable example of this reduction occurred in the recent CMS di-jet search, which was able to reduce the fit function from four freely floating parameters to two freely floating parameters. To reduce the number of freely floating parameters in the fit function for this analysis, improved calculations of the QCD multijet jet soft drop mass are needed and these would have to be incorporated into the analysis through a modified Monte Carlo Parton shower models, or by other means.

\clearpage
\section{Identification of boosted Higgs bosons}

\subsection{MET regression}
\label{app:met_reg}
\begin{figure}[th]
\centering
\includegraphics[width=.325\textwidth]{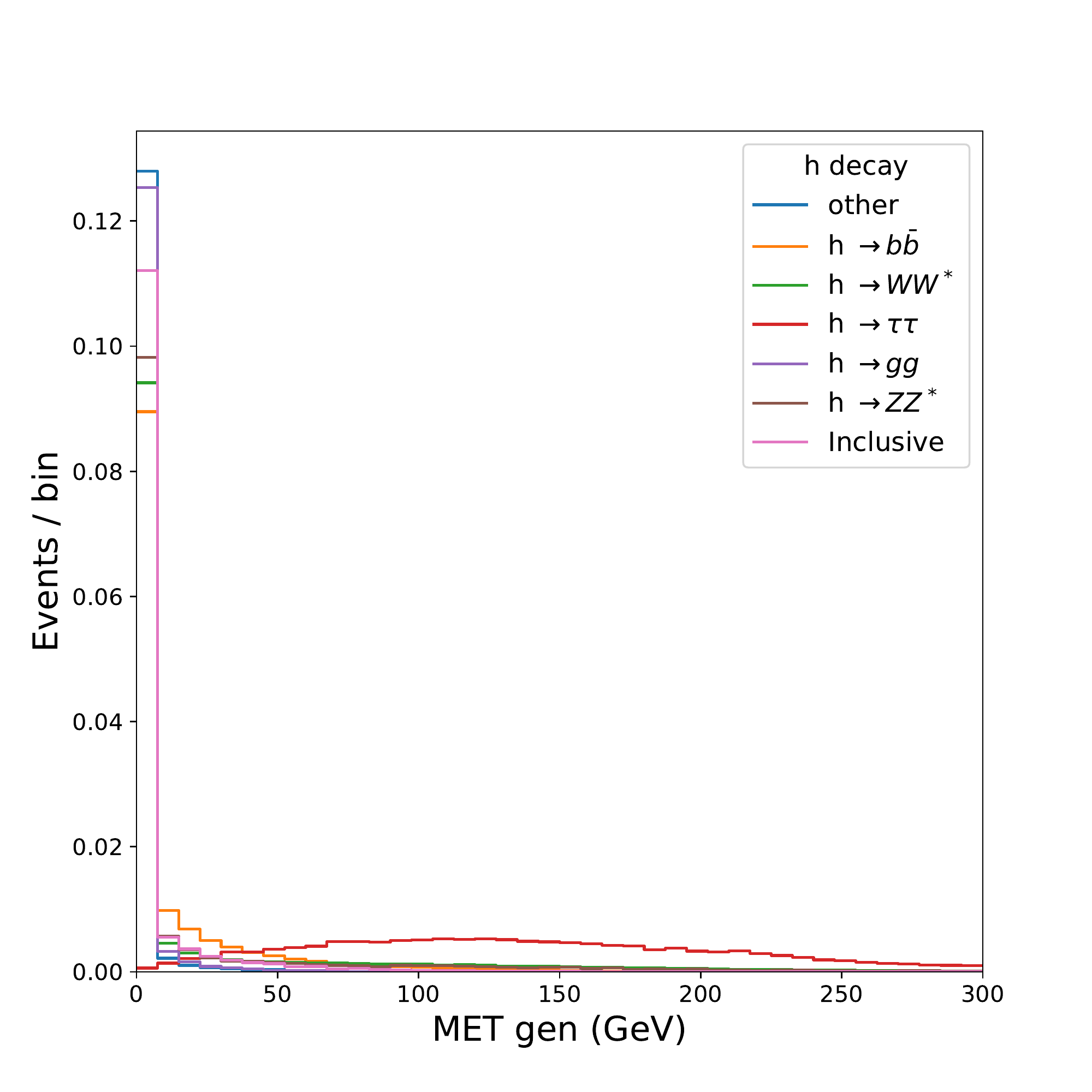}
\includegraphics[width=.325\textwidth]{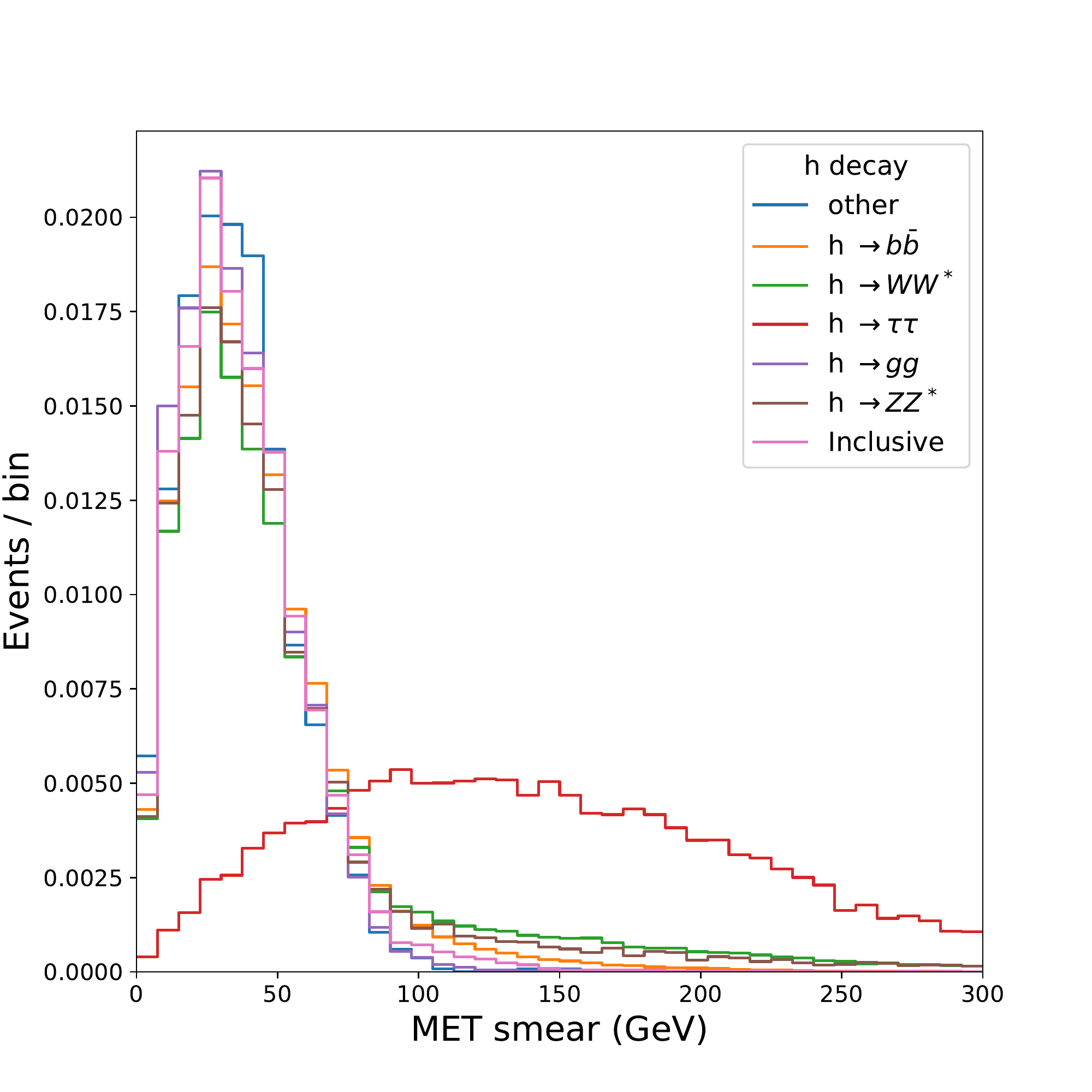}
\includegraphics[width=.325\textwidth]{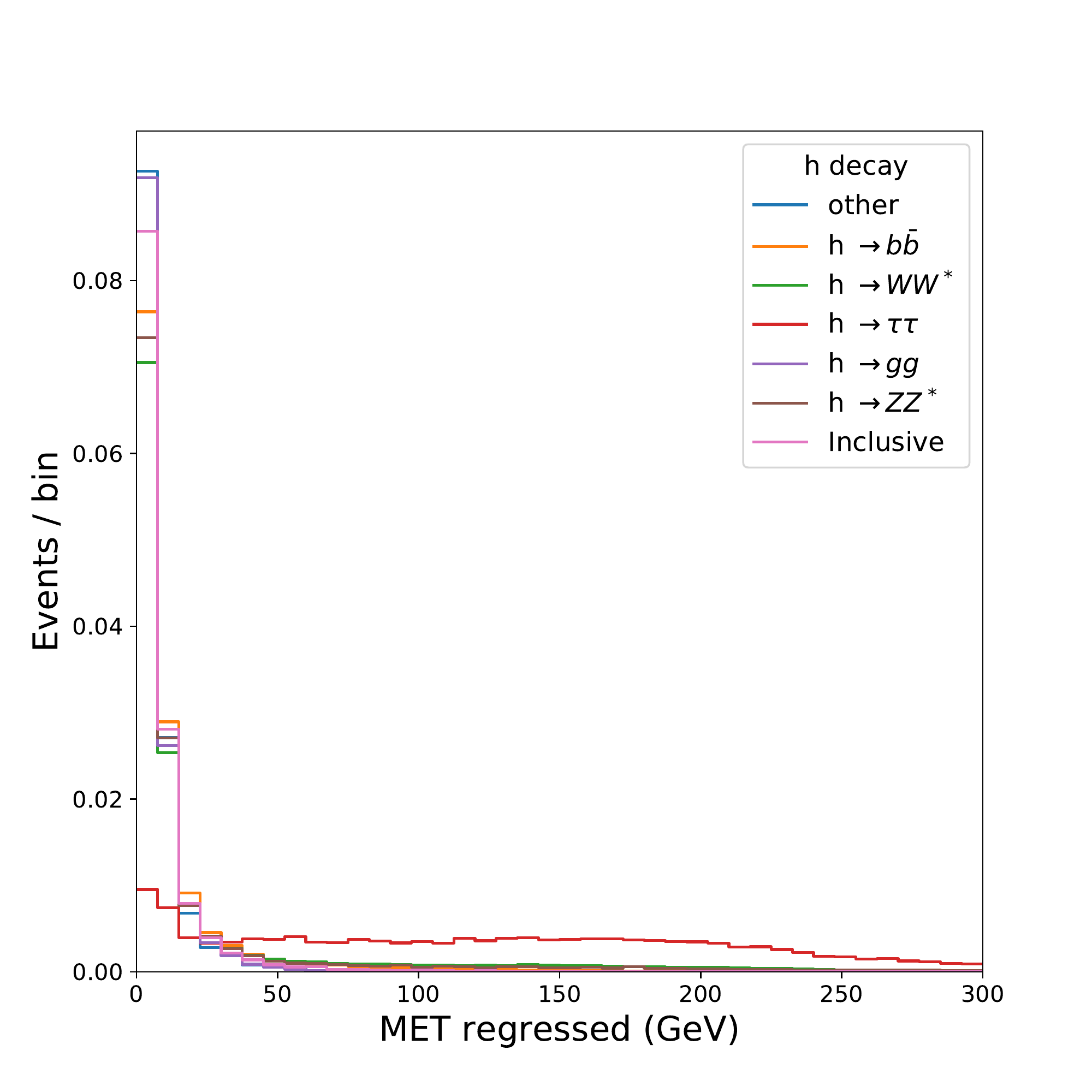}\\
\includegraphics[width=.325\textwidth]{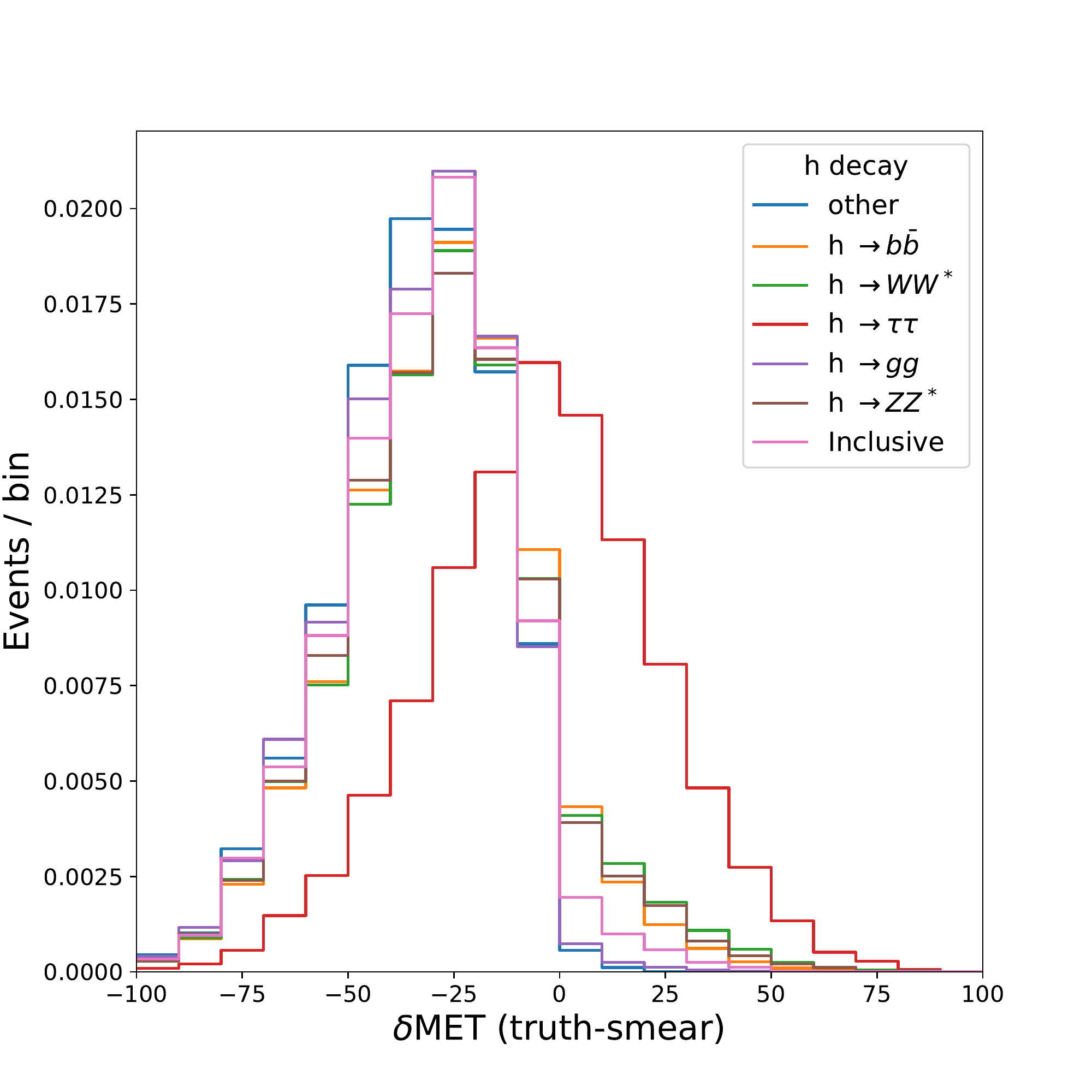}
\includegraphics[width=.325\textwidth]{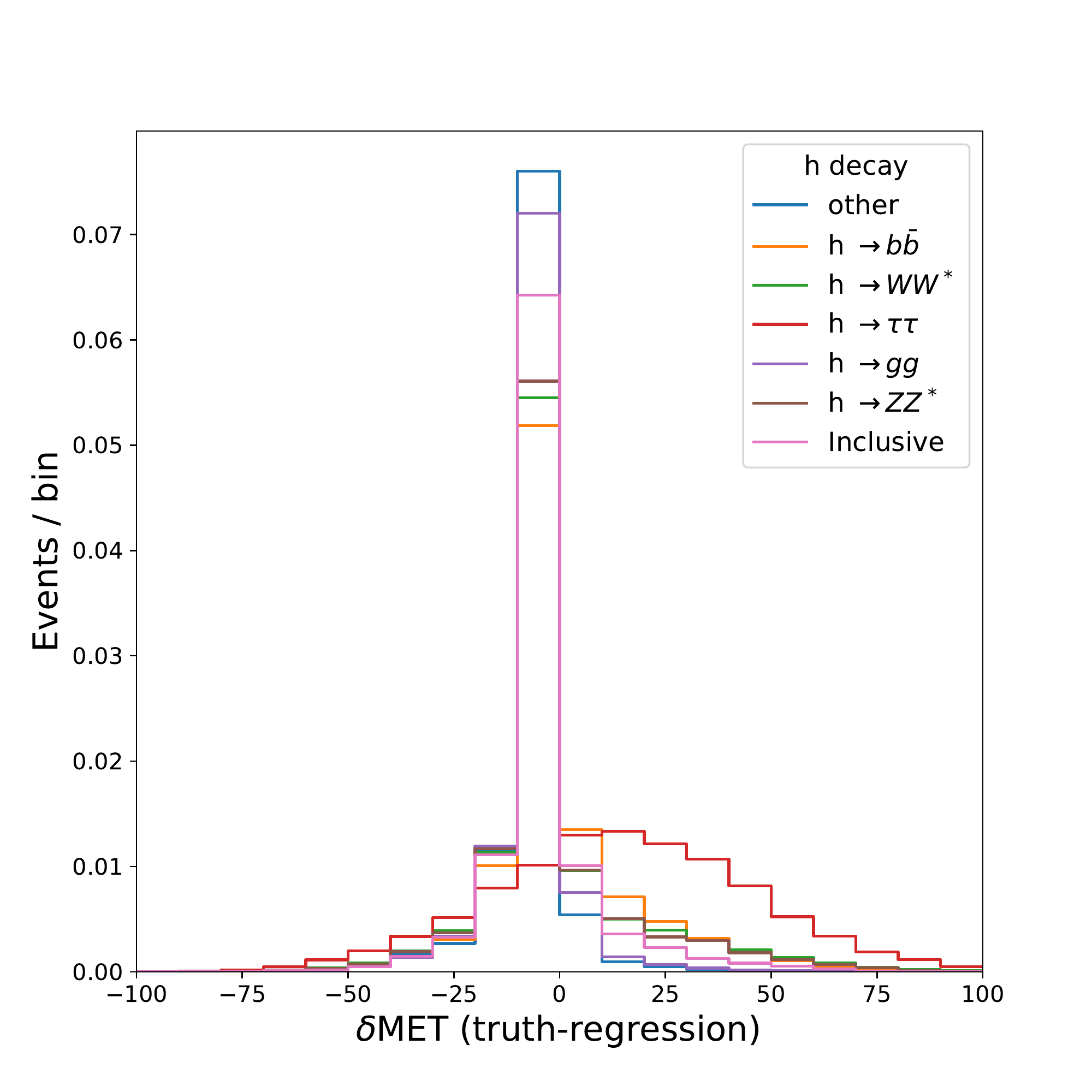}
\caption{\label{fig:met_regression} Performance of the regressed MET as compared to the default MET and the true MET, shown for different SM Higgs decays.}
\end{figure}

The MET regression is performed using a small fully connected neural network using inputs related only to the jet and MET kinematics and trained on Higgs signal events.
This method ensures that the regression does not introduce any additional model dependence on the result.
We find that this simple regression is capable of removing artificial MET very efficiently while predicting the true MET quite accurately in most cases.
The regressed MET as compared to the default MET and the true MET is shown in Fig.~\ref{fig:met_regression} for different SM Higgs decays.

\subsection{The GRU model for Higgs tagging}
\label{app:gru}
We employ a deep learning technique, that takes as inputs the four momenta of jet constituents and the particle type, to discriminate Higgs boson decays against the QCD background.
As suggested in~\cite{Louppe:2017ipp} we can embed jets of a variable number of constituents using a sequence-based recurrent neural network; we consider a gated recurrent unit (GRU)~\cite{DBLP:journals/corr/ChungGCB14} that takes the particle array sorted by decreasing $p_T$ as input.
We restrict our inputs to the first 20 jet constituents.
Increasing the number of constituents used in the GRU does not seem to significantly change the performance.
We use a small dense network on the outputs of the GRU to perform the final classification task of separating signal and background.
In the text, we refer to the combination of the GRU and dense network as the GRU classifier.

\clearpage
\subsection{Additional tagging performance plots}
\label{app:performance}

\begin{figure}[th]
\centering
\includegraphics[width=.45\textwidth]{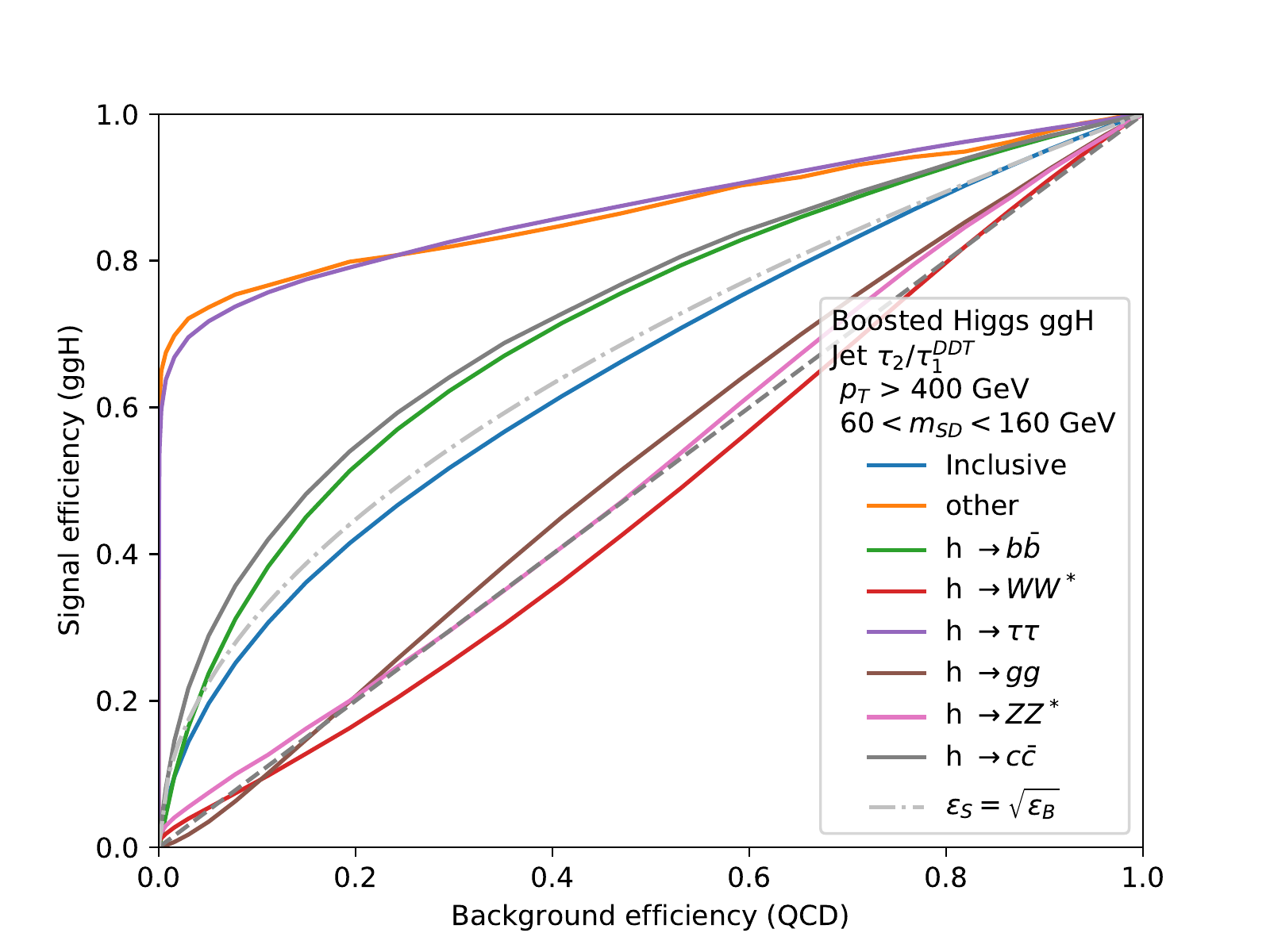}
\includegraphics[width=.45\textwidth]{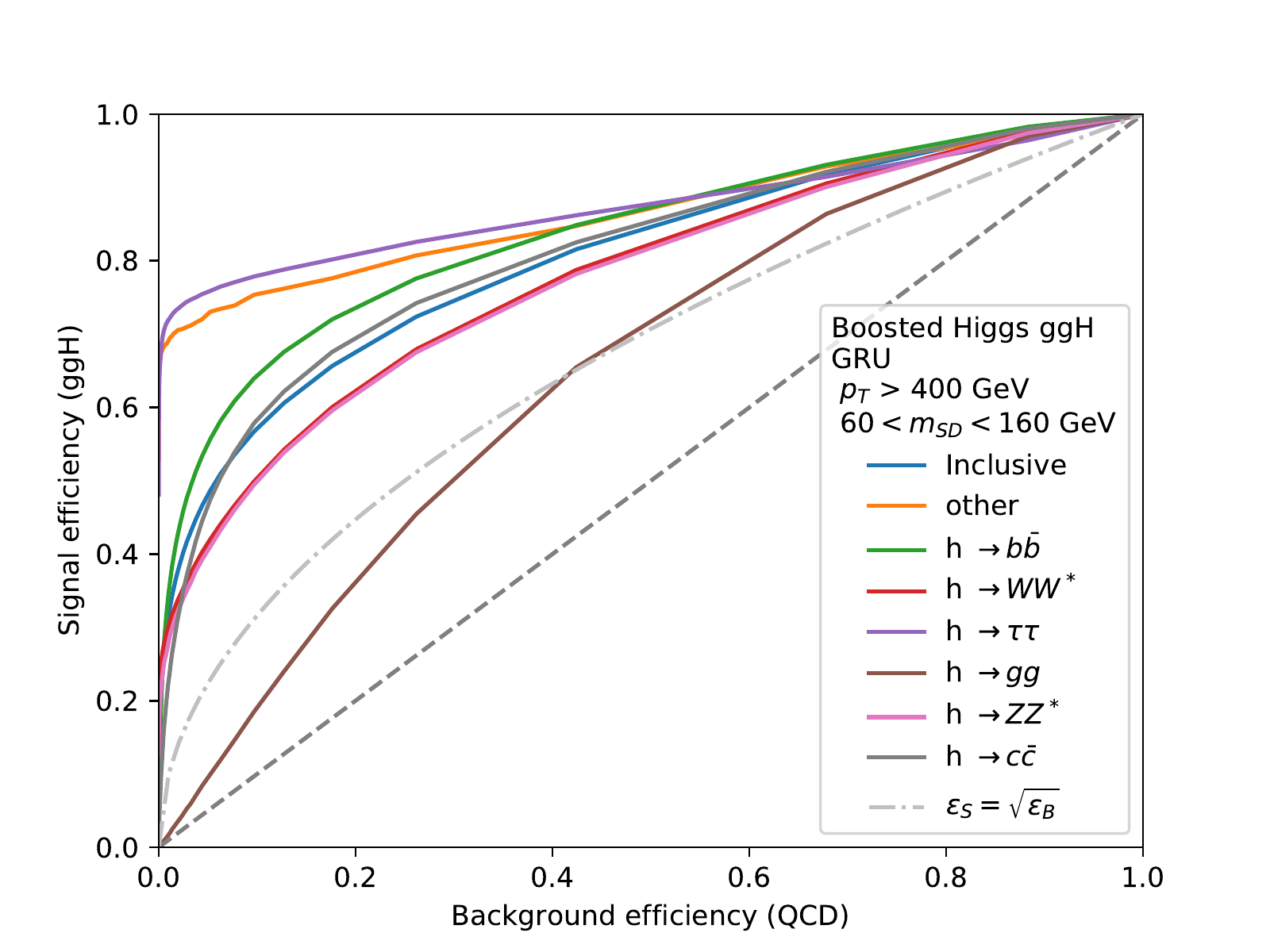}\\
\includegraphics[width=.45\textwidth]{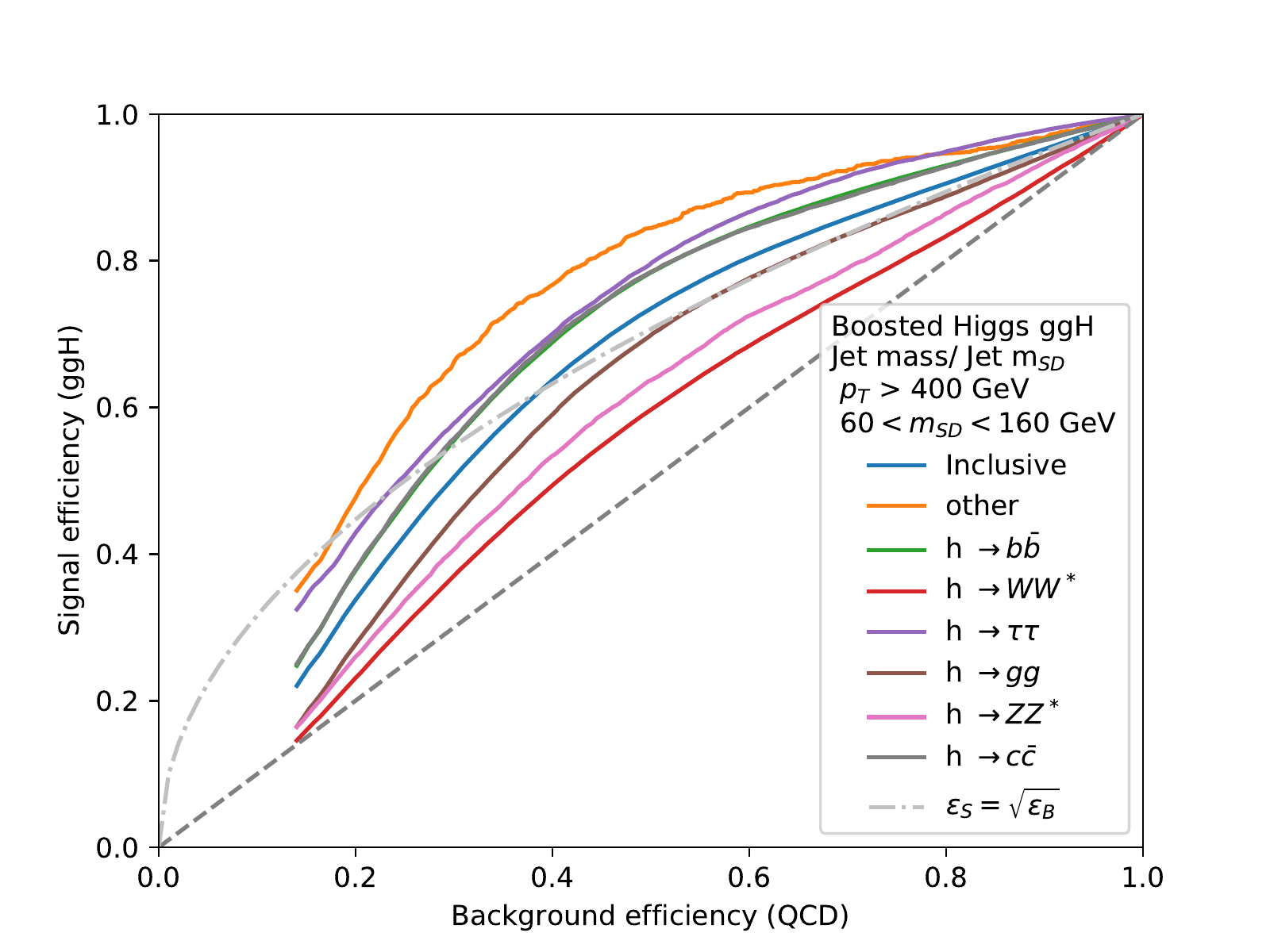}
\includegraphics[width=.45\textwidth]{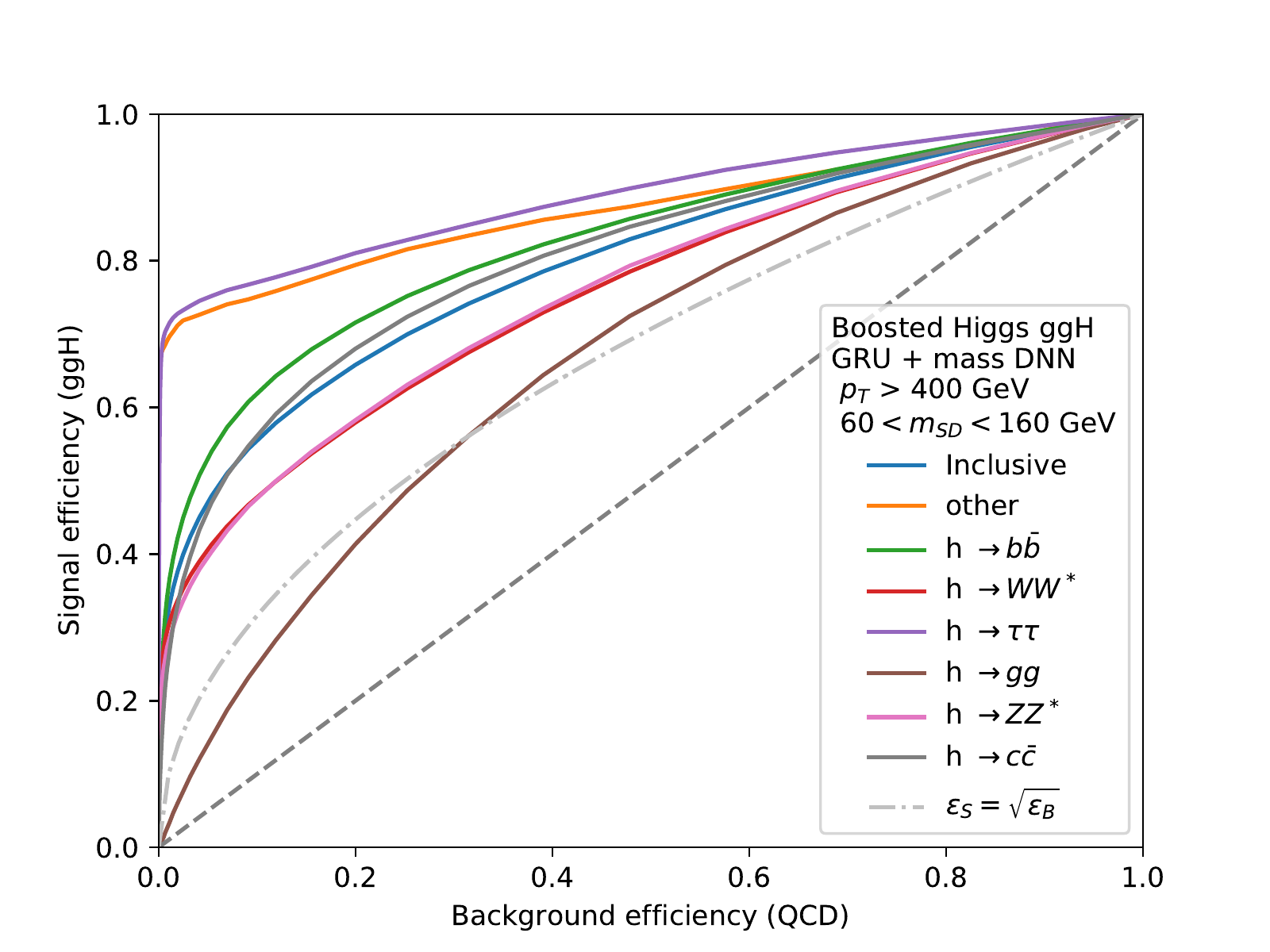}
\caption{\label{fig:roc_byvar} Comparison of the performance for the $\tau_2/\tau_1^{\rm DDT}$ (top-left), GRU (top-right), Jet mass/Jet m$_{\rm~SD}$ (bottom-left) and averaged discriminator of the GRU and the Mass-DNN (bottom-right).}
\end{figure}

\begin{figure}[th]
\centering
\includegraphics[width=.45\textwidth]{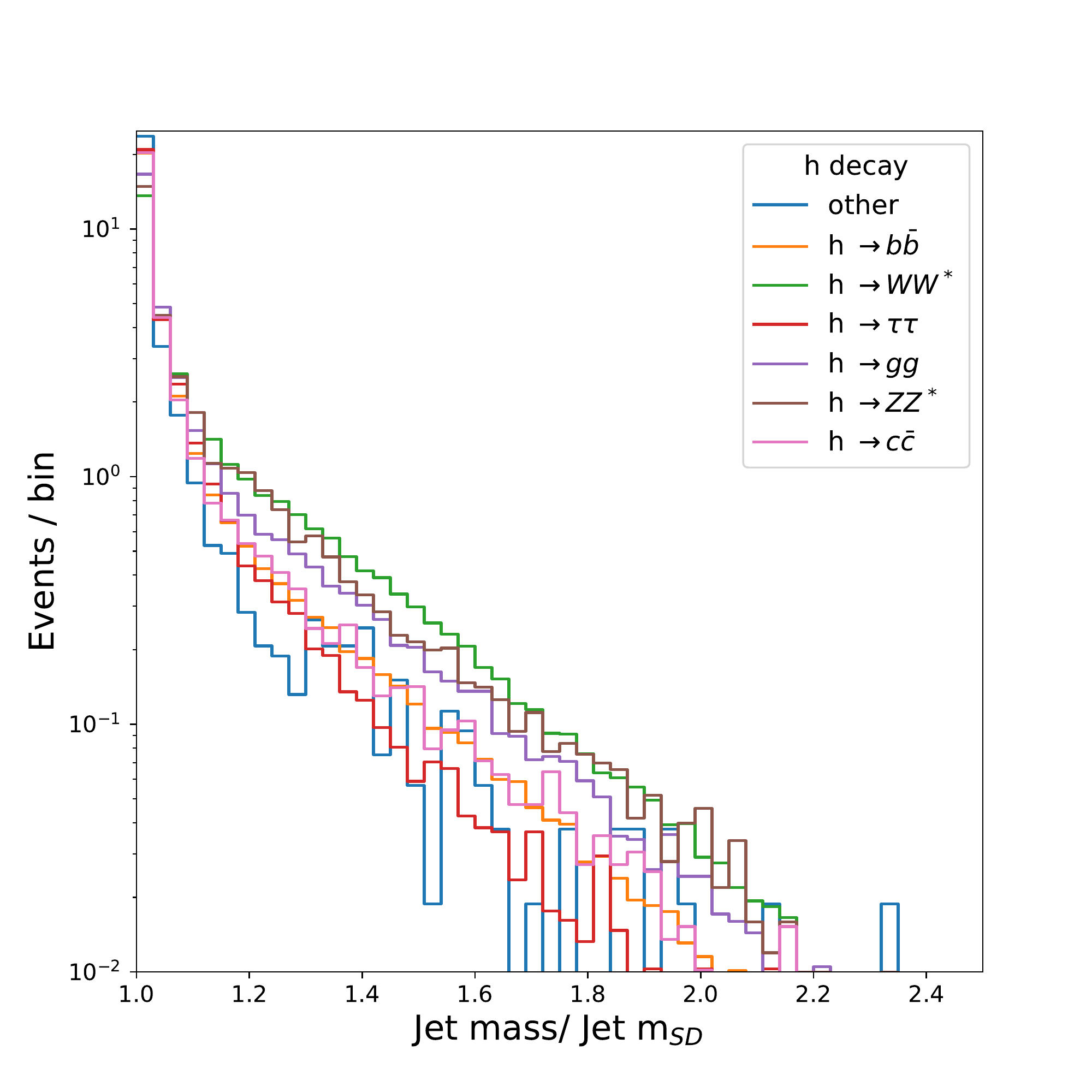}
\includegraphics[width=.45\textwidth]{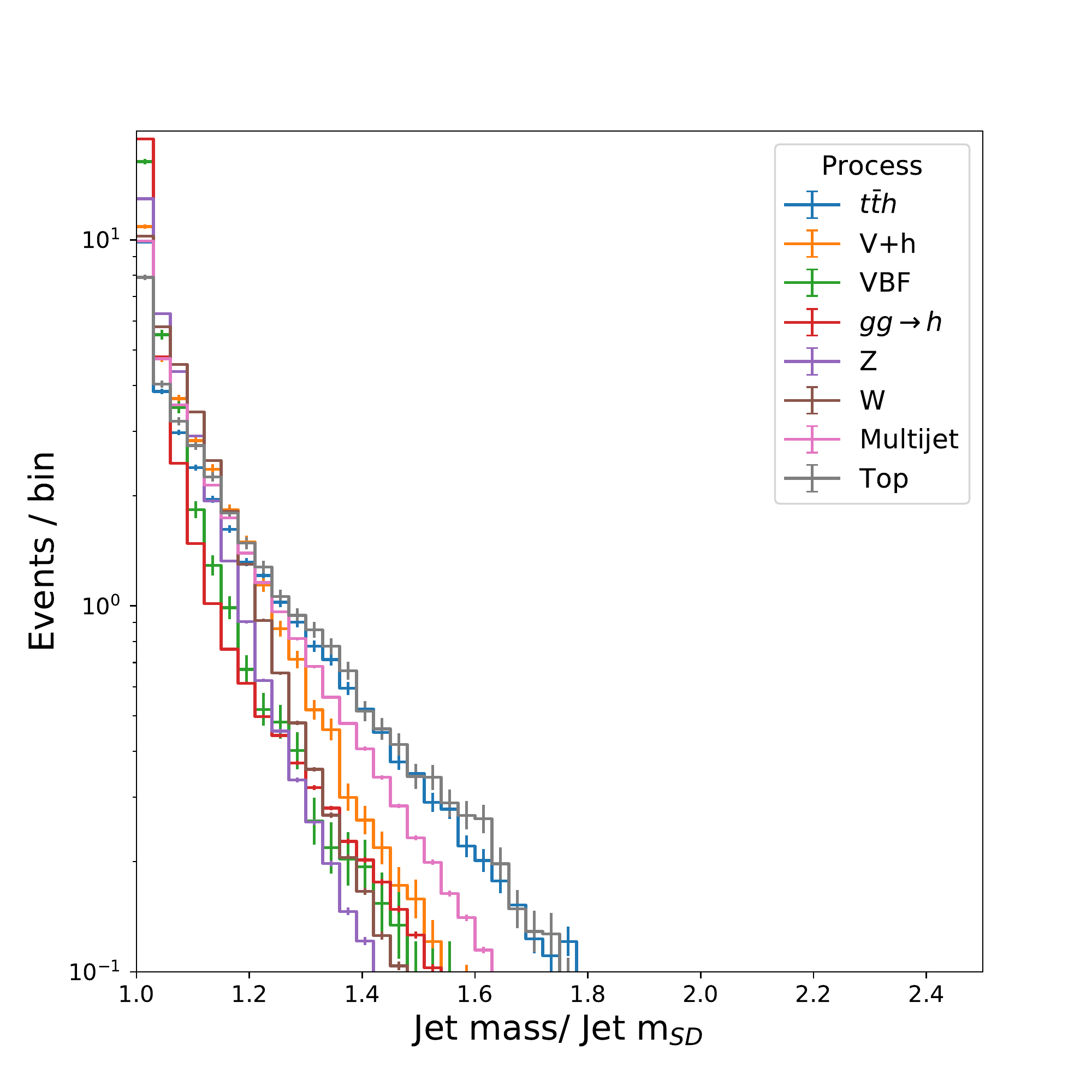}
\caption{\label{fig:massratio} Ratio of the ungroomed mass of the jet with respect to the groomed mass, using the soft-drop algorithm. The two jet masses differ more for QCD background decays leading to a discrimination favoring color-singlet jets.}
\end{figure}

\begin{figure}[th]
\centering
\includegraphics[width=.7\textwidth]{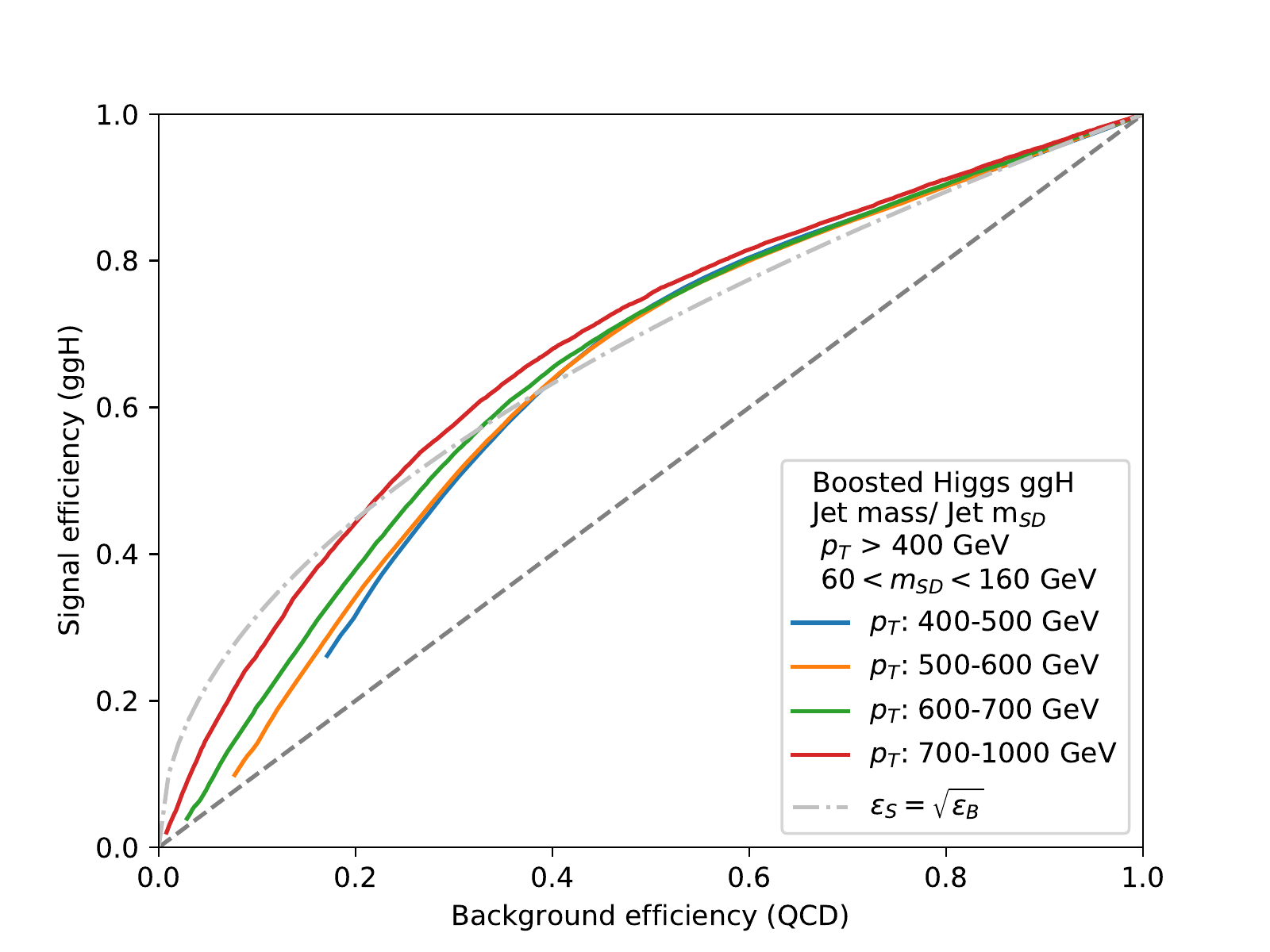}
\caption{\label{fig:rocmrbypt} Comparison of the performance of one of the mass ratios observables: $m_{\rm jet}/m_{\rm SD}$ for different $p_T$ ranges.}
\end{figure}

\begin{figure}[th]
\centering
\includegraphics[width=.7\textwidth]{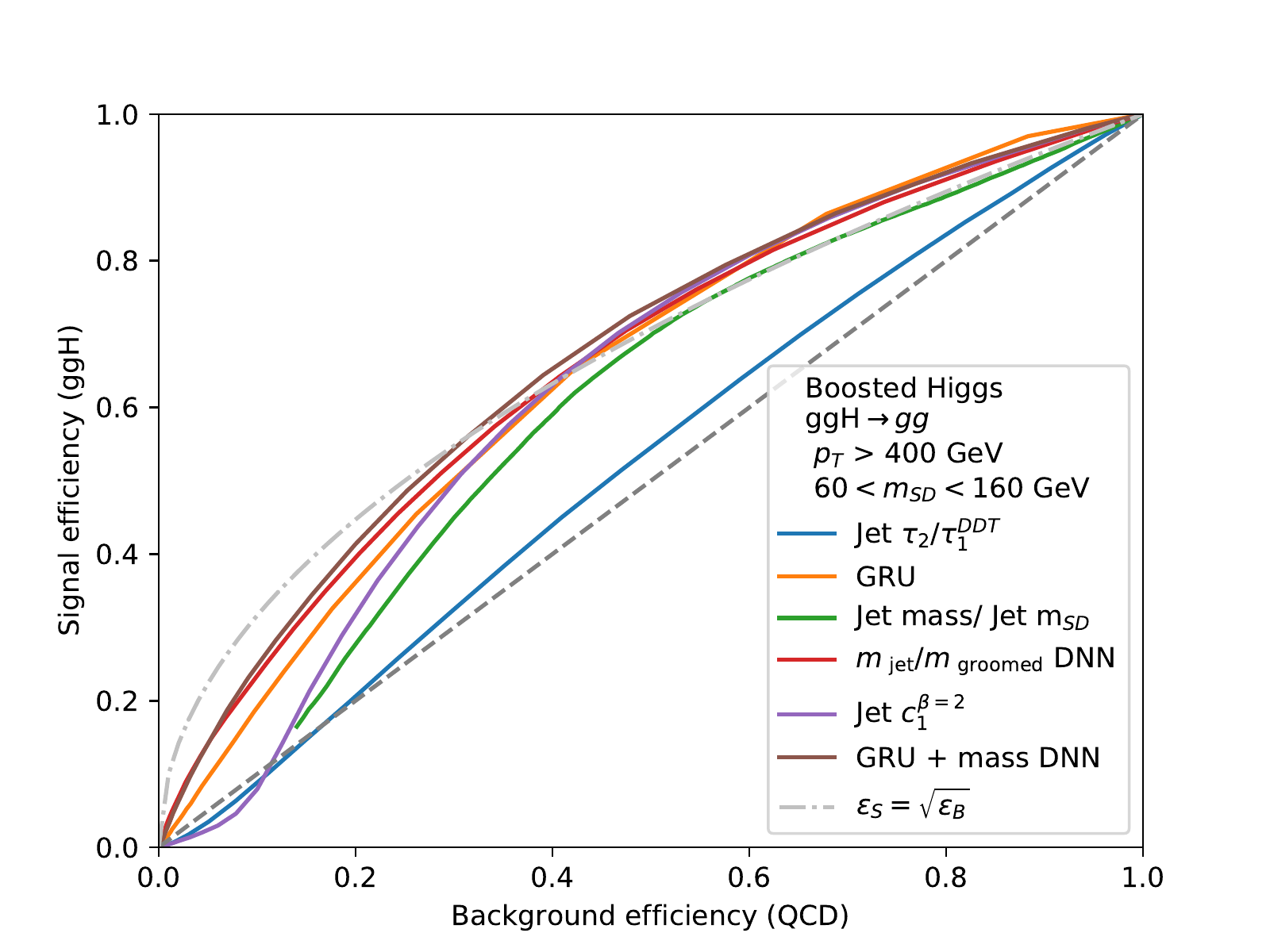}
\caption{\label{fig:rocs_gg} Comparison of the performance for the different algorithms for $h\rightarrow gg$ decays.}
\end{figure}

\clearpage
\subsection{Jet mass adversarial neural network}
\label{app:jmass_adv}
In addition to the DDT procedure, we studied other decorrelation techniques that alter the training of neural networks such they are not correlated with the jet mass and $p_{T}$.
In particular, we explored adversarial training~\cite{Shimmin:2017mfk,ATL-PHYS-PUB-2018-014,CMS-PAS-JME-18-002}, that prevents the classifier output from distorting the background mass distribution through penalty terms in the loss function, and a passive approach that constrains the number of events available in the training such that jet mass and $p_T$ distributions are identical for the background and signal samples~\cite{CMS-PAS-JME-18-002}; this prevents the network from making a simple selection on the jet mass and $p_T$ and forces it to attempt to learn more fundamental differences from the input variables.
Both approaches can reduce the mass bias but for very tight cuts we find a visible distortion present in the jet mass distributions.
Given that in our analysis we keep only 1\% or 10\% of the QCD multijet background, we only use the DDT approach in this paper, which is both easy to implement and successful in keeping the jet mass distribution unaffected at any given working point.

\clearpage

\section{Decay mode sensitivity}
\label{app:decay}
We studied the relative sensitivity of each decay mode channel to establish which decay mode limits the sensitivity of the inclusive Higgs boson cross section measurement.
For each of the taggers used in this analysis, we use the standard model Higgs boson decays as a proxy for the sensitivity to all Higgs boson decays.
We split the Higgs sample into the dominant SM decay modes and compute the individual cross section limit and correct it by the branching ratio as:
\begin{eqnarray}
\delta\sigma_{ggh} = \frac{\delta\sigma_{ggh \rightarrow XX}}{BR(h \rightarrow XX)}
\end{eqnarray}
where we assume branching ratios $BR(h\rightarrow XX)$ for a SM Higgs boson with $m_h=125$~GeV: $h \rightarrow b\bar{b}$ (BR=0.584), $h \rightarrow WW^*$ (BR=0.214), $h \rightarrow gg$ (BR=0.082),  $h \rightarrow \tau^{+}\tau^{-}$ (BR=0.062), $h \rightarrow c\bar{c}$ (BR=0.028) and $h \rightarrow ZZ^*$ (BR=0.026).
Other SM Higgs decays, such as $h \rightarrow \gamma\gamma$, $h \rightarrow \gamma Z$ and $h \rightarrow \mu^{+}\mu^{-}$ have much smaller predicted rates and thus we neglect them for now.

Three different scenarios are considered when computing the 1$\sigma$ limit on $\sigma_{ggh}$: a likelihood fit with no systematic uncertainties, a template fit for the signal extraction and a polynomial fit to the QCD background.

\begin{figure}[tbp]
\centering
\includegraphics[width=.7\textwidth]{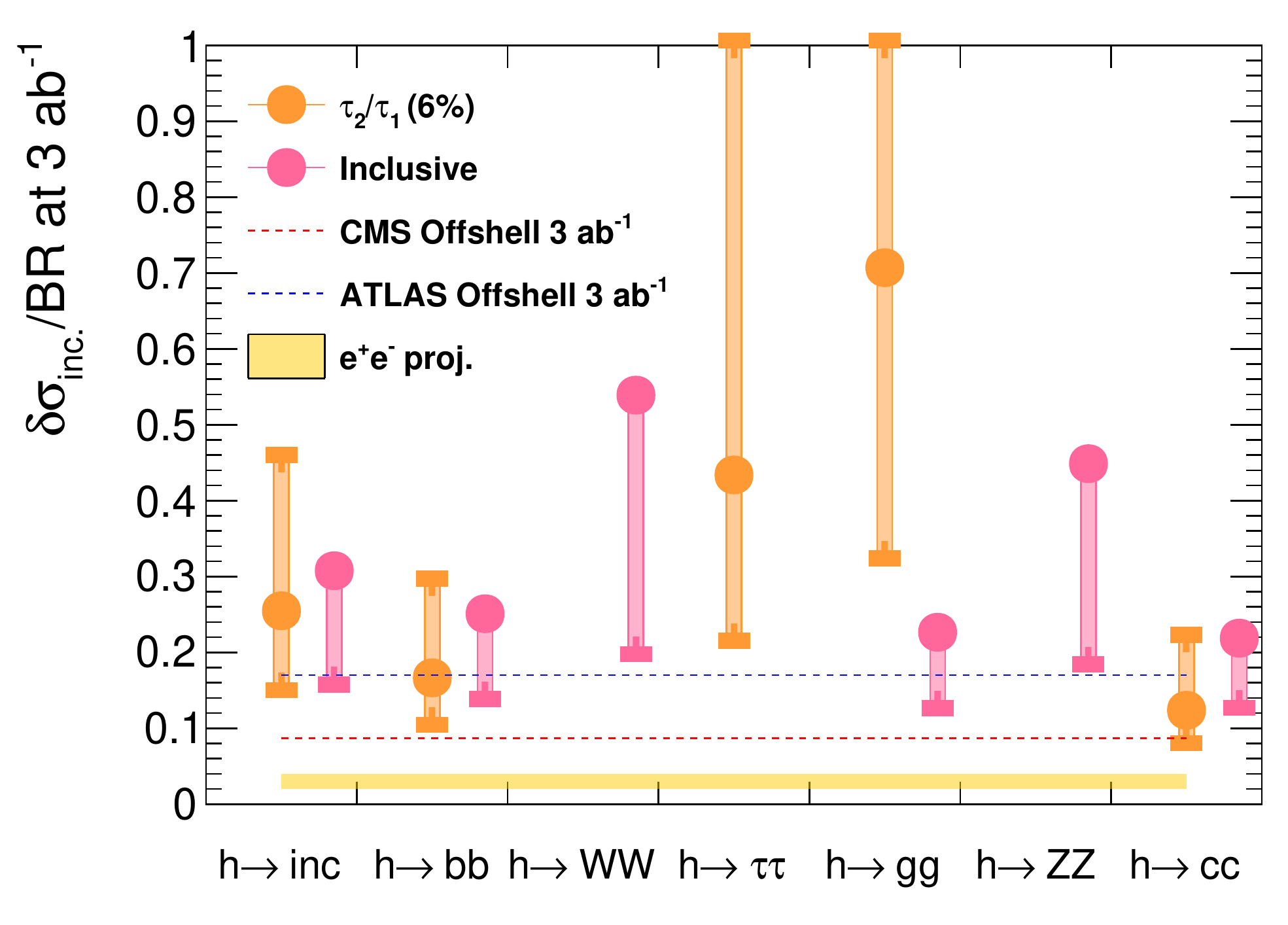}
\caption{\label{fig:inct21res} Estimated 1$\sigma$ sensitivity across different decay modes for a fit to the inclusive mass distribution (pink) and a selection on $\tau_{2}/\tau_{1}^{\rm DDT}$ (orange).
For each point, the upper bound of the band indicates the sensitivity using a fourth-order Bernstein polynomial to fit the background.
The lower bound of the band shows the result when no systematic uncertainties are included in the calculation and the point indicates the performance when a template fit is utilized for the signal extraction. 
The inclusive bin has a tadpole-like point, indicating that the Bernstein polynomial fit did not converge. Missing points imply the limit is outside the quoted sensitivity range in the plot.
The results are compared to CMS and ATLAS projections of constraints on the Higgs width, where their bounds have been extrapolated to a limit on the $gg \rightarrow h$ cross section.
}
\end{figure}

Figure~\ref{fig:inct21res} shows the results of extracting the Higgs boson peak from the inclusive distribution and the baseline $\tau_{2}/\tau_{1}^{\rm DDT}$ selection.
The position of the lower error bar shows the result with no systematic uncertainties, the circular marker shows the result obtained with a template fit for the signal extraction and the upper error bar position shows the limit obtained with the polynomial fit.
For the inclusive results, the variation of this distribution is largest across channels where missing energy is present and the mass distribution is therefore modified.
The polynomial based mass fit was found to be unstable when fitting a large number of events with the 3~ab$^{-1}$ simulated dataset.
Consequently, we do not quote a result for the inclusive fit using the polynomial fit.
When a $\tau_{2}/\tau_{1}^{\rm DDT}$ selection is applied, further model dependence is added and the sensitivity to non-two-pronged decays (gluon, $W$ boson, and $Z$ boson final states) significantly degrades.
On the contrary, the $(\tau_{2}/\tau_{1})^{DDT}$ observable discriminates very well two-pronged $\tau$ lepton decays but the Higgs mass is broader because of neutrino decays.

\begin{figure}[tbp]
\centering
\includegraphics[width=.7\textwidth]{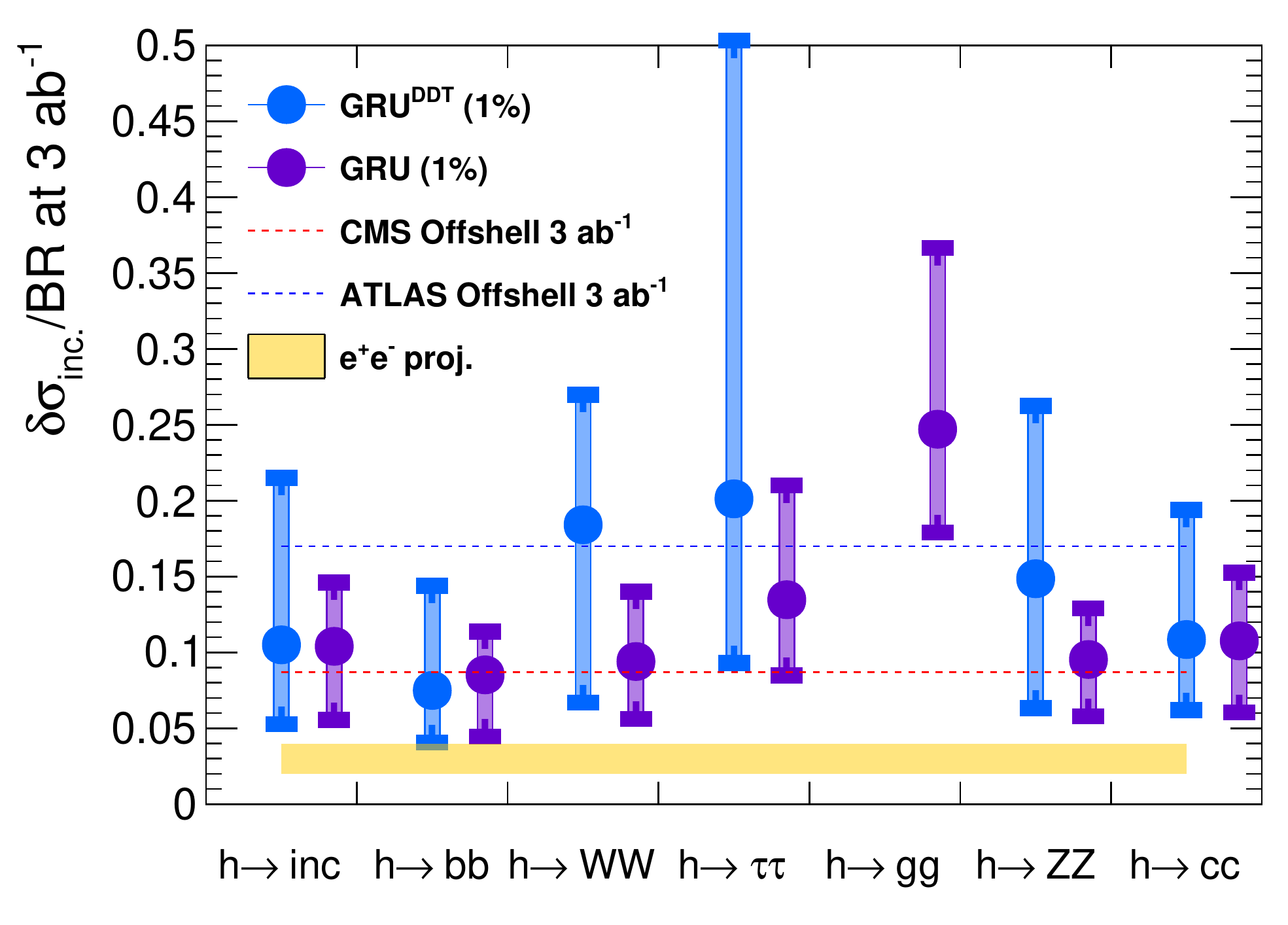}
\caption{\label{fig:grures} Estimated 1$\sigma$ sensitivity across different decay modes for a selection on the deep learning taggers: the GRU (violet) and GRU$^{\rm DDT}$ (blue) variables. For each point, the upper bound of the band indicates the sensitivity using a fourth-order Bernstein polynomial to fit the background. The lower bound of the band shows the result when no systematic uncertainties are included in the calculation and the point indicates the performance when a template fit is utilized for the signal extraction. Missing points imply the limit is outside the quoted sensitivity range in the plot. See the text for details.}
\end{figure}

Figure~\ref{fig:grures} shows the results of the signal extraction using the GRU, and GRU$^{\rm DDT}$ selections both using a 1\% background working point. 
In each case, a larger uniformity is present amongst the different decay modes when compared with that of either the aforementioned selections. 
The one notable exception is in the selection of Higgs to the di-gluon final state where the sensitivity significantly degrades for the GRU$^{\rm DDT}$. 
For the GRU, we find a comparable sensitivity to the inclusive result. 
However, upon further inspection, we find that this improvement is coming mostly from the $W$ boson peak at high $p_{T}$, through associated $W+H$ and $tt+H$ production. 
While this does hint that there is potential for further discrimination through the isolation of explicit Higgs boson decay modes, we do not think that the GRU based result for Higgs in the di-gluon final should be considered as definitive; the application of this strategy within a full data-based analysis would complicate many aspects of the measurement including the calibration through the use of the W and Z boson peaks. 
This is also true for the other GRU based final states. 
Consequently, we suggest that the GRU$^{DDT}$ be taken as more representative of the sensitivity across decay modes.

\begin{figure}[tbp]
\centering
\includegraphics[width=.7\textwidth]{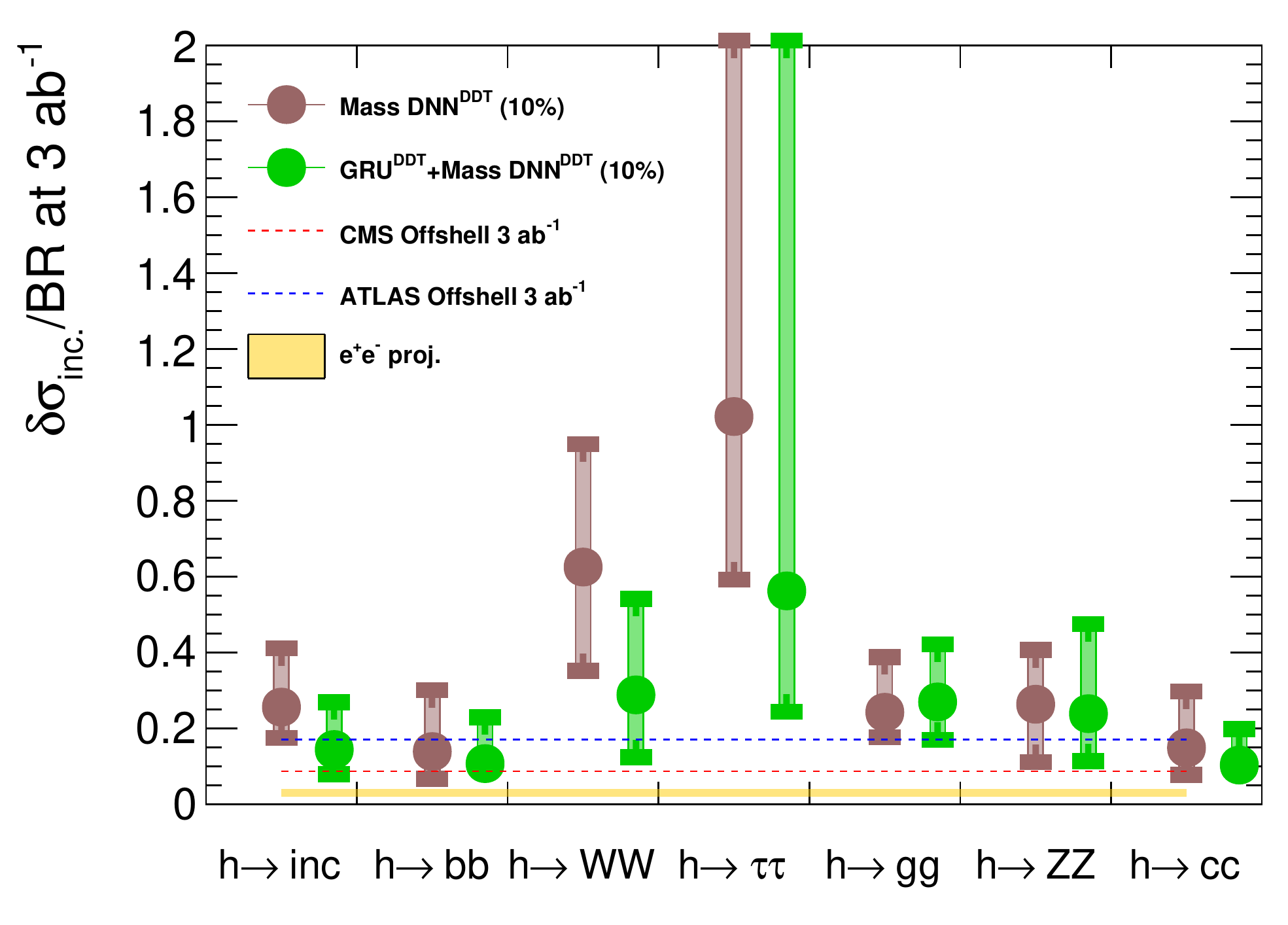}
\caption{\label{fig:massratiosresult} Estimated 1$\sigma$ sensitivity across different decay modes for a selection on the mass ratios DNN$^{\rm DDT}$ and on the average discriminator of the GRU$^{\rm DDT}$ and the mass ratios DNN$^{\rm DDT}$. For each point, the upper bound of the band indicates the sensitivity using a fourth-order Bernstein polynomial to fit the background. The lower bound of the band shows the result when no systematic uncertainties are included in the calculation and the point indicates the performance when a template fit is utilized for the signal extraction.}
\end{figure}

Figure~\ref{fig:massratiosresult} shows the results of extracting the Higgs boson peak after a selection on the mass ratio, and the combined mass ratio variables with GRU$^{\rm DDT}$. 
For the 10\% and 1\% working points the DDT is separately performed corresponding to their respective working points. 
We find that the mass ratio approach, upon decorrelation, gives a sensitivity similar to that of the inclusive analysis with the $h\rightarrow gg$ being one of the most sensitive analyses and the decays where missing energy is present as less sensitive. 
This reflects the behavior observed in the ROC curves where we find that this discriminator is largely independent of the decay mode. 
When combining this discriminator with the GRU$^{\rm DDT}$ discriminator, we find an improvement in sensitivity for all channels except for the Higgs boson decays to di-gluons where a small degradation is present. 
The quoted result on the mass ratios DNN at a 10\% working point is currently the best limit on the Higgs to a di-gluon final state.

